\documentclass[aps,prl,twocolumn,superscriptaddress,amsmath,amssymb,showpacs]{revtex4-1}
\pdfoutput=1 

\usepackage{graphicx}	
\usepackage{dcolumn}	
\usepackage{bm}		
\usepackage{verbatim} 	
\usepackage{braket}
\usepackage{xr}
\usepackage{hyperref}
\usepackage{booktabs}
\usepackage{upgreek}
\usepackage{notes2bib}
\usepackage{pdfpages}
\usepackage{lipsum}  
\usepackage{tikz}
\usepackage{moresize}
\usetikzlibrary{backgrounds}

\makeatletter
\AtBeginDocument{\let\LS@rot\@undefined}
\makeatother

\newcolumntype{L}[1]{>{\raggedright\let\newline\\\arraybackslash\hspace{0pt}}m{#1}}
\newcolumntype{C}[1]{>{\centering\let\newline\\\arraybackslash\hspace{0pt}}m{#1}}
\newcolumntype{R}[1]{>{\raggedleft\let\newline\\\arraybackslash\hspace{0pt}}m{#1}}

\newcommand{\appropto}{\mathrel{\vcenter{
  \offinterlineskip\halign{\hfil$##$\cr
    \propto\cr\noalign{\kern2pt}\sim\cr\noalign{\kern-2pt}}}}}
\newcommand{\resim}{\mathord{\sim}}
\newcommand{\RomanCaps}[1]{\MakeUppercase{\romannumeral #1}}
\begin{document}

\title{Fundamental thermal noise limits for optical microcavities}

\author{Christopher Panuski}
\email{cpanuski@mit.edu}
\affiliation{Research Laboratory of Electronics, MIT, Cambridge, MA 02139, USA}

\author{Dirk Englund}
\affiliation{Research Laboratory of Electronics, MIT, Cambridge, MA 02139, USA}

\author{Ryan Hamerly}
\affiliation{Research Laboratory of Electronics, MIT, Cambridge, MA 02139, USA}
\affiliation{NTT Research, Inc.\ Physics \& Informatics Laboratories, 1950 University Ave \#600, East Palo Alto, CA 94303, USA}

\begin{abstract}
We present a joint theoretical and experimental characterization of thermo-refractive noise in high quality factor ($Q$), small mode volume ($V$) optical microcavities. Analogous to well-studied stability limits imposed by Brownian motion in macroscopic Fabry-Perot resonators, microcavity thermo-refractive noise gives rise to a mode volume-dependent maximum effective quality factor. State-of-the-art fabricated microcavities are found to be within one order of magnitude of this bound. We confirm the assumptions of our theory by measuring the noise spectrum of high-$Q/V$ silicon photonic crystal cavities and apply our results to estimate the optimal performance of proposed room temperature, all-optical qubits using cavity-enhanced bulk material nonlinearities.
\end{abstract}

\maketitle

\begin{figure}
\includegraphics[width=0.48\textwidth]{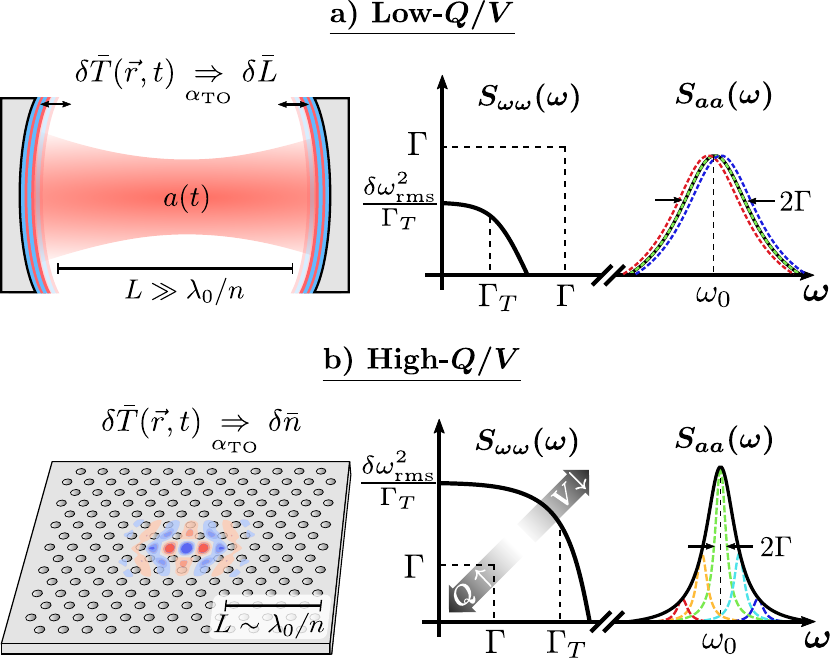}
\caption{Comparison of thermo-refractive noise (TRN) in macroscopic resonators and microcavities. Mode-averaged temperature fluctuations $\delta\bar{T}$ in large cavities induce refractive index noise $\delta\bar{n}$ (and thus pathlength changes $\delta\bar{L}$) due to the mirrors' non-zero thermo-optic coefficient $\alpha_\text{TO}=\mathrm{d}n/\mathrm{d}T$. The large mode volume $V$ reduces $\delta\bar{T}$, yielding a narrowband resonant frequency noise spectrum $S_{\omega\omega}(\omega)$ and an rms resonant frequency fluctuation $\delta\omega_\text{rms}\ll\Gamma$, the cavity half-linewidth. This non-dominant thermal noise inhomogeneously broadens the intracavity field spectrum $S_{aa}(\omega)$. Decreasing $V$ increases both the magnitude $\delta\omega_\text{rms}$ and bandwidth $\Gamma_T$ of TRN, while increasing the resonator quality factor $Q=\omega_0/2\Gamma$ causes both quantities to exceed $\Gamma$. TRN therefore becomes a dominant source of homogeneous broadening in wavelength-scale high-$Q/V$ microcavities, leading to thermal dephasing and reduced resonant excitation efficiency of the cavity field $a(t)$.}
\centering
\label{fig:overview}
\end{figure}

\paragraph*{Introduction.---} Room temperature, high-quality factor ($Q$) optical cavities enable the investigation of new physical phenomena by enhancing light-matter interaction \cite{Abbott2017a}, shaping electromagnetic modes \cite{Yu2018}, and modifying the vacuum photon density of states \cite{Delic2020}. However, these advantages come with an often forgotten cost --- interaction with a thermally equilibrated confining medium inherently injects noise into the optical mode in accordance with the fluctuation-dissipation theorem (FDT) \cite{Callen}. Macroscopic resonators (Fig.~\ref{fig:overview}a), such as those implemented in gravitational wave interferometers, minimize this interaction by supporting a large mode volume $\tilde{V} \gg 1$ in vacuum, where $\tilde{V}=V/(\lambda/n)^3$ for the volume $V$ of a $\lambda$-wavelength optical mode confined in a refractive index $n$. The surprising realization that the sensitivity of these $\resim$km-long cavities can still be limited by Brownian motion in few $\upmu$m-thick mirror coatings \cite{Levin1997,Martynov2016} has spurred interest in low-noise mirror coatings \cite{Cole2013}, grating-based mirrors \cite{Heinert2013,Kroker2017}, and the fundamental limits of macroscopic cavities in the presence of thermal fluctuations \cite{Numata,Kessler2011,Zhadnov2018}.

Here, we consider the opposite case: optical microcavities (Fig.~\ref{fig:overview}b) \cite{Vahala2003}, whose small mode volumes ($\tilde{V}\sim 1$) facilitate low-energy (even single-photon level) nonlinear interactions \cite{Li2019}, single molecule label-free sensing \cite{Armani2007}, and enhanced coupling for atom-photon interfaces \cite{Eichenfield2009a,Goban2014}. However, compared to macroscopic cavities, the fundamental temperature fluctuations $\langle\delta T^2\rangle \appropto 1/V$ \cite{Chui1992,Evans2008} are much larger in these near-diffraction-limited modes. The associated refractive index fluctuations, so called ``thermo-refractive noise" (TRN), become a principal source of resonant frequency noise in dielectric microcavities including microspheres \cite{Gorodetsky2004}, whispering-gallery mode resonators \cite{Matsko2007}, ring resonators \cite{Huang2019}, and photonic crystal (PhC) cavities \cite{Saurav}. TRN has also recently been shown to limit the stability of dielectric nanolasers \cite{Jiang2019} and microcavity frequency combs \cite{Drake2019}. 

To date, microcavity TRN has only been considered in a pertubative regime, where the resulting rms resonant frequency fluctuation $\delta\omega_\text{rms}\appropto 1/\sqrt{V}$ is much less than the loaded cavity linewidth $2\Gamma=\omega_0/Q$. For sufficiently high $Q$ and small $V$, this assumption becomes invalid. Continued improvements in microcavity performance --- yielding $Q>10^7$, $\tilde{V}\sim 1$ through fabrication advances \cite{Asano2017a} and $Q\sim 10^5$, $\tilde{V}\sim 10^{-3}$ using novel sub-wavelength dielectric features \cite{Hu2018} --- thus raises a simple question: when will fundamental thermal noise limit the performance of high-$Q/V$ microcavities?

In this Article, we answer this open question by deriving general bounds for optical microcavity performance in the presence of TRN and find that current devices are within one order of magnitude of this bound. We verify our theory by measuring TRN as the dominant noise source in high-$Q/V$ PhC cavities and demonstrate the ability to distinguish between sub-wavelength mode volumes ($\tilde{V}<1$) using fundamental noise spectra. As an example of the immediate impact of our formalism, we analyze the implications for an outstanding goal in quantum photonics: all-optical qubits using cavity-enhanced bulk material nonlinearities \cite{Mabuchi2012}. Our results highlight the importance of considering thermal noise in state-of-the-art optical resonators and also inform design choices to minimize its impact on device performance.

\paragraph*{Formalism.---} As schematically illustrated in Fig.~\ref{fig:overview}, fundamental stochastic temperature fluctuations $\delta T(\vec{r},t)$ within a cavity confining medium of refractive index $n$ and thermo-optic coefficient $\alpha_\text{TO}=\mathrm{d}n/\mathrm{d}T$ drive a mode-averaged refractive index change $\delta \bar{n}(t)=\alpha_\text{TO}\delta\bar{T}(t)$. For an optical mode completely confined in dielectric, the resulting resonance shift $\delta\omega(t)=\omega_0\alpha_\text{TO}\delta\bar{T}(t)/n$ follows from first-order perturbation theory \cite{Joannopoulos}. We neglect temperature-induced deformations of the cavity, as the thermo-elastic coefficient of typical dielectrics is two orders of magnitude smaller than $\alpha_\text{TO}$ \cite{Matsko2007}.

In the presence of TRN, the steady-state rotating-frame intra-cavity field amplitude is 
\begin{equation}
\tilde{a}(t) = i\sqrt{\Gamma}\tilde{s}_\text{in}\int_{-\infty}^t \mathrm{d}t' e^{-(i\Delta+\Gamma)(t-t')-i\int_{t'}^{t} \mathrm{d}t'' \delta\omega(t'')}
\label{eqn:cavityEvolution}
\end{equation}
for a loaded amplitude decay rate $\Gamma$ and critically-coupled static drive $\tilde{s}_\text{in}$ detuned by $\Delta$ from the cavity resonance. The associated statistical moments can be computed using the moment-generating properties of the characteristic functional $\langle e^{i\int_{t'}^t \delta\omega_0(t'')dt''}\rangle$, which in the case of zero-mean Gaussian noise only requires the autocorrelation $\langle\delta\omega(t)\delta\omega(t+\tau)\rangle=(\omega_0\alpha_\text{TO}/n)^2\langle\delta\bar{T}(t)\delta\bar{T}(t+\tau)\rangle$ \cite{Zhang2014a,Feynman1965}. The latter autocorrelation of temperature fluctuations can be computed from the heat equation
\begin{equation}
\frac{\partial }{\partial t}\delta T(\vec{r},t) + D_T\nabla^2\delta T(\vec{r},t)=F_T(\vec{r},t)
\label{eqn:generalHeat}
\end{equation}
in a medium of thermal diffusivity $D_T$ driven by a Langevin forcing term $F_T(\vec{r},t)$ which satisfies the FDT. As we will illustrate for slab PhC cavities, Eqn.~\ref{eqn:generalHeat} can be solved analytically for specific geometries; however, for generality we follow the approach of Ref.~\cite{Sun2017} and enforce a single-mode decay approximation by introducing a phenomenological thermal decay rate
\begin{equation}
\Gamma_T = D_T\frac{\int \left[\nabla\left( \epsilon(\vec{r}) |\vec{E}(\vec{r})|^2 \right) \right]^2 \mathrm{d}^3\vec{r}}{\int \epsilon(\vec{r})^2 |\vec{E}(\vec{r})|^4\mathrm{d}^3\vec{r}}
\label{eqn:GammaT}
\end{equation}
evaluated for the envelope of intracavity energy density. This form of $\Gamma_T$ is chosen for consistency with $\langle\delta\bar{T}^2\rangle=k_BT_0^2/c_VV_T$, the well-known statistical mechanics result for temperature fluctuations in a volume $V_T$ of specific heat capacity $c_V$ in thermal equilibrium with a bath temperature $T_0$  \cite{Land1980}. Averaging Eqn.~\ref{eqn:generalHeat} over the optical mode profile, we then find
\begin{equation}
\frac{\rm d}{{\rm d}t}\delta\bar{T}(t)+\Gamma_T\delta \bar{T}(t)=\bar{F}_T(t),
\label{eqn:modeHeat}
\end{equation}
leading to the solution
\begin{equation}
\langle \delta\omega(t)\delta\omega(t+\tau) \rangle = \underbrace{\left(\frac{\omega_0}{n}\alpha_\text{TO}\right)^2 \frac{k_BT_0^2}{c_V V_T}}_{\delta\omega_\text{rms}^2}e^{-\Gamma_T|\tau|},
\label{eqn:freqAutocorrelation}
\end{equation}
where the thermal mode volume
\begin{equation}
V_T = \frac{\left[ \int \epsilon(\vec{r}) |\vec{E}(\vec{r})|^2\mathrm{d}^3\vec{r} \right]^2}{\int \epsilon(\vec{r})^2 |\vec{E}(\vec{r})|^4\mathrm{d}^3\vec{r}}
\label{eqn:Vt}
\end{equation}
is the common Kerr nonlinear mode volume found by solving Eqn.~\ref{eqn:generalHeat} in a homogeneous medium \cite{Gorodetsky2004}. For a three-dimensional mode with a Gaussian energy density distribution, $V_T$ is larger than the standard Purcell mode volume $ V = \int \epsilon |\vec{E}|^2\mathrm{d}^3\vec{r}/\text{max}\lbrace \epsilon |\vec{E}|^2 \rbrace$ by a factor of $2\sqrt{2}.$

Combining Eqns.~\ref{eqn:cavityEvolution} and \ref{eqn:freqAutocorrelation}, we can solve for the statistical moments of $\tilde{a}(t)$ as a function of the primary parameters $V_T$ and $\Gamma_T$. Complete derivations are provided in Supplementary Section \RomanCaps{1}. If $V_T$ ($Q$) is sufficiently large (small) such that $\Gamma\gg (\delta\omega_\text{rms},~\Gamma_T)$ as in previous analyses, TRN can be treated perturbatively (Fig.~\ref{fig:overview}a). In this case, the cavity resonance is quasistatic over the photon decay period and shifts by much less than a cavity linewidth over time: TRN thus contributes to weak inhomogeneous broadening of the resonance. However, as the thermal mode volume $V_T$ shrinks, $\delta\omega_\text{rms}$ and $\Gamma_T$ increase until they eventually exceed $\Gamma$ (Fig.~\ref{fig:overview}b). The spectral density of the intracavity field is then homogeneously broadened to a linewidth $2\Gamma+2\delta\omega_\text{rms}^2/\Gamma_T\approx 2\delta\omega_\text{rms}^2/\Gamma_T$ by TRN. Our analysis focuses on the transition to this high-$Q/V$ limit. Specifically, we solve Eqn.~\ref{eqn:cavityEvolution} with non-perturbative TRN to calculate mode volume-dependent maximum ``effective" cavity quality factors $Q_\text{eff}$ to describe energy storage and dephasing in microcavities. 


\begin{figure*}
\includegraphics[width=\textwidth]{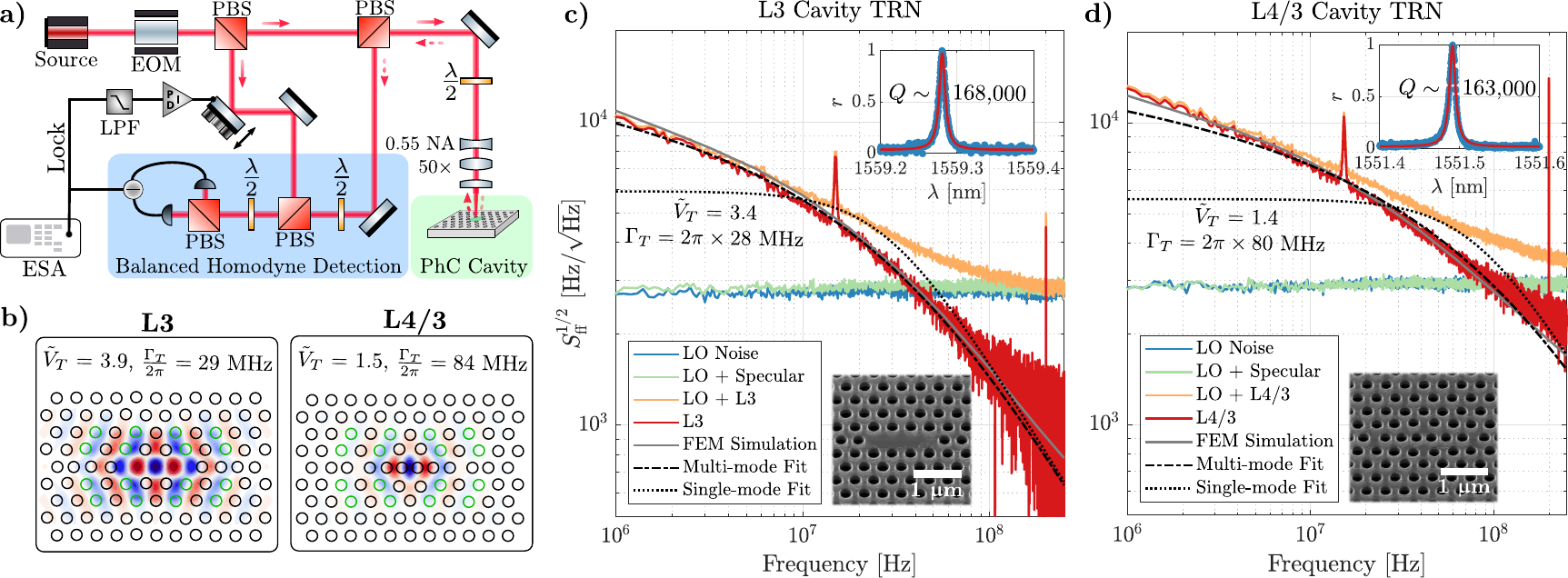}
\caption{Calibrated measurement of thermo-refractive noise (TRN) in high-$Q/V$ silicon PhC cavities. A shot noise-limited, balanced homodyne detector (a) is locked to the phase quadrature of the cavity reflection signal and records the spectrum of resonant frequency fluctuations. The simulated mode profiles, thermal mode volume $\tilde{V}_T$ (Eqn.~\ref{eqn:Vt}), and thermal decay rate $\Gamma_T$ (Eqn.~\ref{eqn:GammaT}) of the L3 and L4/3 devices tested are shown in (b). The radii of the green holes are increased by up to 5\% to form superimposed gratings which improve vertical coupling efficiency. The measured spectral density of cavity resonant frequency noise $S_\text{ff}(f)$ (red) for L3 (c) and L4/3 (d) cavities is compared to noise from a specular reflection off the sample surface, finite-element method (FEM) simulations of cavity TRN, as well as single- and multi-mode fits. The listed multi-mode fit parameters agree with the predicted values in (b). Inset reflection spectra of each device reveal quality factors on the order of $10^5$. Micrographs of the fabricated designs with enlarged holes relative to the optimal designs in (b) are also inset.}
\centering
\label{fig:expt}
\end{figure*}

\paragraph*{Measurement.---} Before pursuing these goals, we first experimentally verify the fundamental assumptions of our TRN model by measuring the noise spectrum of high-$Q/V$ PhC cavities. As shown in Fig.~\ref{fig:expt}, our setup uses a Mach-Zehnder interferometer to measure the phase of a cavity reflection signal via balanced homodyne detection. A variable beamsplitter separates the emission from an amplified tunable infrared laser into local oscillator (LO) and cavity input paths, which are passively balanced to minimize laser frequency noise coupling. A $\lambda/2$-plate rotates the input signal polarization by 45$^\circ$ relative to the dominant cavity polarization such that the cavity reflection can be isolated from any specular reflection from the sample using a polarizing beamsplitter (PBS) \cite{Galli2009}. The sample stage is temperature controlled to better than 10 mK using a Peltier plate and feedback temperature controller. A balanced, shot noise-limited photodetector measures the homodyne signal from the recombined cavity reflection and LO, and the result is recorded on an electronic spectrum analyzer (ESA). By actively locking to the phase quadrature of the homodyne signal with a piezo-controlled mirror, TRN-induced cavity frequency noise is detected as frequency-resolved voltage noise. To calibrate the spectrum, we inject a known phase noise with an electro-optic modulator (EOM) whose modulation efficiency is measured by sideband fitting \cite{Gorodetsky2010}.

Fig.~\ref{fig:expt} shows the resulting measurements for two released (air-clad) silicon PhC cavities: the common L3 cavity and the recently-proposed ``L4/3" cavity \cite{Minkov2014, Minkov2017}. Fabricated cavities yield high quality factors (up to $Q\approx$ 400,000 at $\lambda_0\approx 1550$~nm) with efficient vertical coupling. The variation of Purcell mode volume --- $\tilde{V} = (0.95,~0.32)$ for simulated L3 and L4/3 cavities, respectively --- also allows us to confirm the expected volume-dependence of TRN. Whereas a direct reflection from the sample surface (green trace) adds little additional noise to the LO background (blue), we observe broadband, \textit{input power-independent} noise from either cavity's reflection (orange). The calibration tone is visible at 200 MHz and we attribute the resonance at $\resim$15 MHz to optomechanical coupling from the fundamental flexural mode of the suspended membrane \cite{Gavartin}. In the corrected cavity noise curve (red), we have subtracted the LO shot noise and accounted for attenuation due to the finite cavity linewidth. As expected, the wavelength-scale mode volumes yield a spectral density of resonant frequency fluctuations $S_\text{ff}(f)$ with nearly two orders of magnitude larger amplitude and bandwidth compared to previous results in microspheres \cite{Gorodetsky2004} and ring resonators \cite{Huang2019}. Supplementary Section \RomanCaps{2} further details the experiments and provides TRN data as a function of cavity input power.

The measured noise spectra show excellent agreement with numerical simulations based on a modified version of the fluctuation-dissipation theorem for thermo-refractive noise \cite{Levin2008}. A noise model based on a multi-mode solution to Eqn.~\ref{eqn:generalHeat} in a thin slab is also well-fitted to the data, yielding the fit parameters $\lbrace \tilde{V}_T^\text{L3},\Gamma_T^\text{L3}/2\pi, \tilde{V}_T^\text{L4/3},\Gamma_T^\text{L4/3}/2\pi \rbrace=\lbrace 3.4\pm 0.3, 28\pm 1 \text{ MHz}, 1.4\pm 0.1, 80\pm 3\text{ MHz} \rbrace$ that compare favorably with the expected values ($\lbrace 3.9, 29\text{ MHz}, 1.5, 84\text{ MHz} \rbrace$) from Eqn.~\ref{eqn:Vt} (evaluated numerically from the simulated mode profiles) and Eqn.~\ref{eqn:GammaT}. In Eqn.~\ref{eqn:GammaT}, we assume a two-dimensional Gaussian mode and reduced thermal diffusivity $D_T=D(1-\phi)/(1+\phi)$ for the patterned slab with porosity $\phi$ compared to the unpatterned thin film diffusivity $D$ \cite{Jain2013,Cuffe2015}. As predicted, the reduced mode volume of the L4/3 cavity increases the bandwidth and spectral density of thermal fluctuations.

The noise spectra of the proposed single-mode approximation (Eqn.~\ref{eqn:freqAutocorrelation}) underestimates the measured noise of both devices at low frequencies $\omega\ll\Gamma_T$, but accurately approximate $S_\text{ff}$ in the range of frequencies of interest (near and above the cutoff frequency $\Gamma_T$) and conserve the integrated frequency noise $\langle\delta\omega_\text{rms}^2\rangle$. These results indicate that TRN is the dominant broadband noise source in high-$Q/V$ resonators, and validate the suitability of a single-mode approximation to describe the spectrum of frequency fluctuations in general microcavity geometries. More broadly, our demonstration unveils a new technique for evaluating the mode volume of fabricated optical resonators using fundamental quantities as opposed to complex invasive techniques, such as near-field scanning optical microscopy \cite{Hu2018}.

\paragraph*{$Q/V$ Limits.---} Given this experimental confirmation, we can apply our TRN model to estimate the fundamental performance limits of room temperature microcavities. We specifically consider the quality factor to mode volume ratio $Q/V$, which is proportional to the peak intracavity intensity and therefore of particular significance for single-photon nonlinearities \cite{Choi} and enhanced sensitivity to point-like defects \cite{Panuski2019}. 

\begin{figure}
\centering
\resizebox{.48\textwidth}{!}{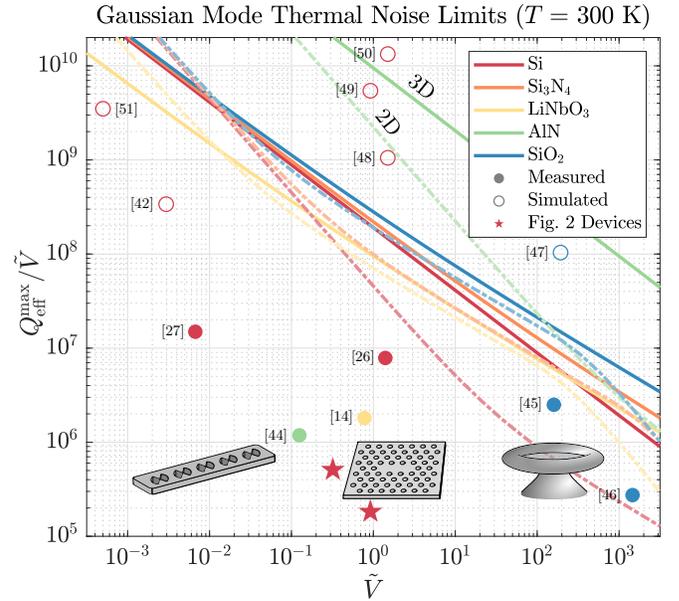}
\caption{Thermal noise-limited room temperature quality factor to mode volume ratios ($Q_\text{eff}^\text{max}/\tilde{V}$) for the materials considered in Supplementary Section \RomanCaps{3} assuming a Gaussian-shaped mode at $\lambda_0=1550$~nm admitting thermal diffusion in two or three dimensions (dash-dot and solid lines, respectively). These limits are compared our devices as well as other fabricated and proposed microcavities. Insets illustrate typical confinement geometries for the range of $\tilde{V}$ listed.}
\label{fig:limits}
\end{figure}

The effective quality factor $Q_\text{eff}=\omega_0\langle|\tilde{a}(t)|^2\rangle/2|\tilde{s}_\text{in}|^2$ of interest in this case is  the ratio of intracavity energy $\langle|\tilde{a}(t)|^2\rangle$ to energy input per cycle $2|\tilde{s}_\text{in}|^2/\omega_0$ in a resonantly excited, critically-coupled cavity. Under the same conditions, solving Eqn.~\ref{eqn:cavityEvolution} for $\langle|\tilde{a}(t)|^2\rangle$ subject to the noise autocorrelation of Eqn.~\ref{eqn:freqAutocorrelation} yields
\begin{equation}
\frac{Q_\text{eff}}{\tilde{V}} = \frac{\omega_0}{2\Gamma_T\tilde{V}}e^x x^{-s}\gamma_l(s,x) 
\label{eqn:GeneralQ-Energy}
\end{equation}
where $\gamma_l$ is the lower incomplete Gamma function, $x=(\delta\omega_\text{rms}/\Gamma_T)^2$, and $s = \Gamma/\Gamma_T+x$. Intuitively, decreasing the cavity linewidth $2\Gamma$ well below the broadened linewidth $2\delta
\omega_\text{rms}^2/\Gamma_T$ has little impact: the prolonged energy storage offsets the reduced excitation rate of the rapidly shifting resonance, leaving the intracavity energy unaltered. $Q_\text{eff}$ is maximized in this limiting case. The corresponding upper bound of Eqn.~\ref{eqn:GeneralQ-Energy} at $T=300~$K is plotted for various material systems in Fig.~\ref{fig:limits} as a function of $\tilde{V}$ assuming a three-dimensional Gaussian mode in a homogeneous three- or two-dimensional confining medium. In the latter case, the decay rate $\Gamma_T=3\pi D_T/V^{2/3}$ decreases by a factor of $3\sqrt{2}V^{1/3}$ to account for the restricted dimensionality of thermal diffusion.

In the joint limit $(\delta\omega_\text{RMS},~\Gamma) \ll \Gamma_T$, valid for sufficiently high-$Q$, low $V$ cavities, Eqn.~\ref{eqn:GeneralQ-Energy} simplifies to $Q_\text{eff}^\text{max}=\omega_0\Gamma_T/2\delta\omega_\text{rms}^2$, thereby recovering the broadened linewidth $2\delta\omega_\text{rms}^2/\Gamma_T$. $Q_\text{eff}^\text{max}$ then scales as $V^{1/3}$ in a homogeneous medium, indicating that larger mode volumes reduce the integrated thermo-optic noise, as expected. For this reason, recent ultra-high-$Q$ ($Q>10^8$) integrated resonators have been specifically designed with $\tilde{V}\gg 1$ to limit TRN \cite{Lee2012,YoulYang2018}. Alternatively, Fig.~\ref{fig:limits} illustrates the advantage of reducing $V$ to maximize $Q_\text{eff}^\text{max}/V$. Further optimization of sub-wavelength cavities \cite{Hu2018, Wang2018} could therefore improve the intensity enhancement achievable in room temperature devices. 

While our review of high-$Q/V$ cavities (Fig.~\ref{fig:limits}) in various materials shows that all fabricated cavities obey the projected bounds, silicon PhC slab cavities \cite{Asano2017a} and silica microtoroids \cite{Kippenberg2004} lie within an order of magnitude of the thermal noise limit. Furthermore, various simulated devices \cite{Spillane2005,Asano2018b,Alpeggiani2015,Quan2011} exceed the limit; their practical realization will thus require low-temperature operation or novel noise suppression techniques.

\begin{figure*}
\includegraphics[width=0.85\textwidth]{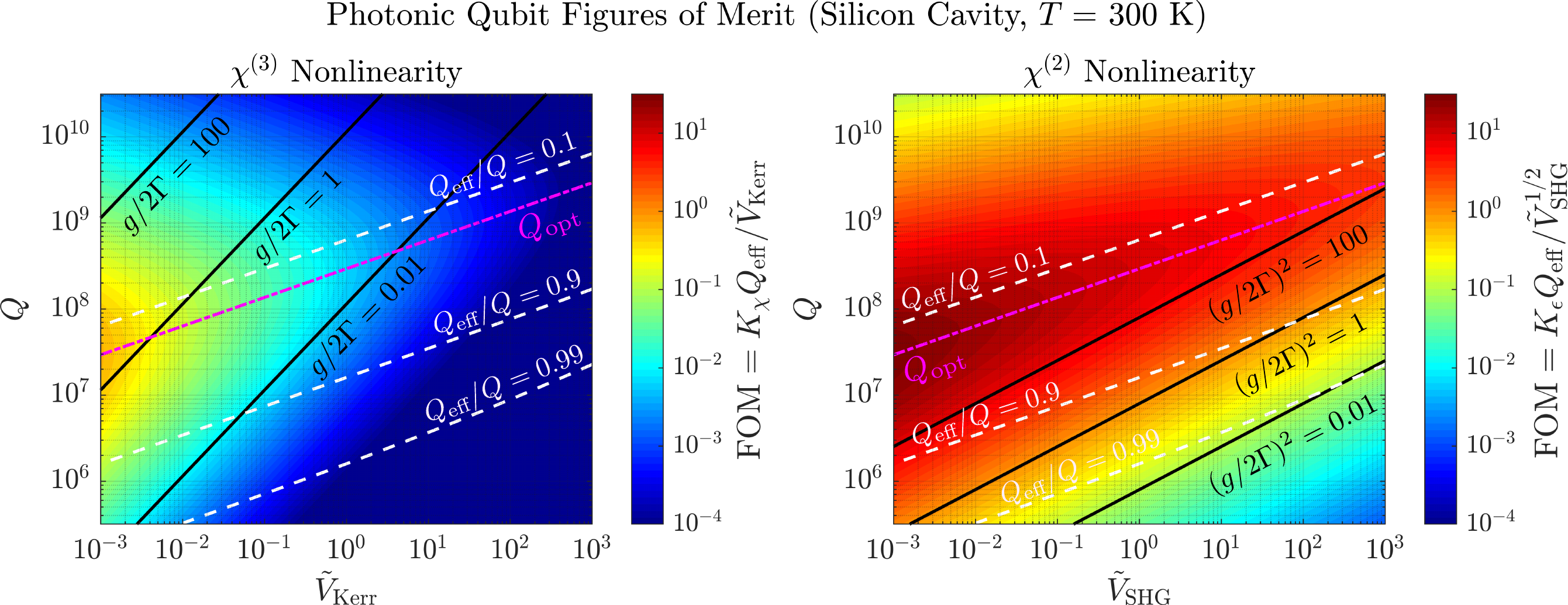}
\caption{Performance of room temperature all-optical qubits using bulk $\chi^{(3)}$ (left) or electric field-induced $\chi^{(2)}$ (right) nonlinearities in silicon microcavities as a function of loaded cavity quality factor $Q$ at $\lambda_0=2.3~\upmu$m and the relevant normalized nonlinear mode volume $\tilde{V}$ ($\tilde{V}_\text{Kerr}=\tilde{V}_T$ and $\tilde{V}_\text{SHG}$, assumed to be equal to $\tilde{V}_T$, for $\chi^{(3)}$ and $\chi^{(2)}$, respectively) . The figure of merit (FOM) --- the ratio of qubit coupling rate $g$ to the composite decay and thermal dephasing rate $2\Gamma_\text{eff}=\omega_0/Q_\text{eff}$ --- is largest for strong coupling ($g/2\Gamma,~(g/\Gamma)^2\gg 1$) and weak dephasing ($Q_\text{eff}\approx Q$). Three dimensional thermal diffusion in a homogenous medium is assumed.}
\centering
\label{fig:qubits}
\end{figure*}

\paragraph*{Implications for All-Optical Qubits.---} These proposed thermal noise limits have practical impact for future devices. Chief among the applications driving the pursuit for high-$Q/V$ cavities is quantum information. Recent photonic proposals \cite{Heuck2020,Krastanov2020} have explored the feasibility of reaching the qubit limit of cavity nonlinear optics \cite{Mabuchi2012} at room temperature by leveraging the relative immunity of optical photons to thermal noise. While this insensitivity is granted by Planck's Law, we have shown here that through the thermo-refractive effect, temperature fluctuations can significantly impact light in a high-$Q/V$ resonator. For coherent processes, TRN-induced dephasing of the field amplitude $\tilde{a}(t)$ must be considered in addition to the previously discussed intracavity energy limitations. We therefore define the effective quality factor $Q_\text{eff}=\omega_0|\langle\tilde{a}(t)\rangle|^2/2|\tilde{s}_\text{in}|^2$ based on the mean field amplitude of a resonant, critically-coupled cavity, yielding
\begin{equation}
Q_\text{eff} = \frac{\omega_0}{2\Gamma_\text{eff}} = Q\left(\frac{\Gamma}{\Gamma_T}\right)^2e^{2x} x^{-2s}\gamma_l^2(s,x) 
\end{equation}
for the previously defined $x$, $s$. 

We can compare $\Gamma_\text{eff}$ to a nonlinear coupling rate $g$ between qubit basis states with the simple figure of merit $\text{FOM} = g/\Gamma_\text{eff}$, which intuitively corresponds to the number of qubit operations that can be completed prior to decay or dephasing. For bulk $\chi^{(3)}$ and $\chi^{(2)}$ nonlinearities, the coupling rate $g$ is a function of material parameters and mode volumes.  For $\chi^{(3)}$, $g=2\Gamma K_\chi/\tilde{V}_\text{Kerr}\gg 2\Gamma$ is required to reach the strong coupling regime where the anharmonicity of Fock state energies decouples the qubit basis (zero and one photon states) from higher energy states \cite{Mabuchi2012}. Similarly, $g=2\Gamma K_\epsilon/\tilde{V}_\text{shg}^{1/2}$ is the coupling rate between doubly resonant first- and second-harmonic basis states using the bulk $\chi^{(2)}$ nonlinearity, which requires $(g/2\Gamma)^2 \gg 1$ for strong coupling. Derivations of the proportionality constants $K_\chi$, $K_\epsilon$ and mode volumes are included in the Supplementary Section \RomanCaps{4}. The resulting figures of merit
\begin{align}
\text{FOM}_{\chi^{(3)}} = K_\chi \frac{Q_\text{eff}}{\tilde{V}_\text{Kerr}} && \text{FOM}_{\chi^{(2)}} = K_\epsilon \frac{Q_\text{eff}}{\tilde{V}_\text{SHG}^{1/2}}
\end{align}
are plotted in Fig.~\ref{fig:qubits} for Gaussian modes in silicon, where we assume the intrinsic $\chi^{(3)}$ nonlinearity can create an electric-field induced $\chi^{(2)}=3\chi^{(3)}E_\text{dc}$ near the breakdown dc electric field $E_\text{dc}$ \cite{Timurdogan2017}.

An ideal qubit operates well within the strong coupling regime with minimal dephasing. In the presence of TRN, increasing $Q/V$ improves the former at the cost of the latter, leading to the observed mode volume-dependent optimum loaded quality factor $Q_\text{opt} \approx \omega_0\Gamma_T/2\delta\omega_{\text{rms}}^2$. Fig.~\ref{fig:qubits} also illustrates a relative performance advantage for $\chi^{(2)}$ devices in silicon, as strong coupling can be achieved at lower quality factors. For example, the peak $\text{FOM}_{\chi^{(2)}}\sim 10$ is three orders of magnitude greater than $\text{FOM}_{\chi^{(3)}}$ assuming $Q=Q_\text{opt}$ and $\tilde{V}_\text{Kerr}=\tilde{V}_\text{SHG}=1$. Although small $\tilde{V}_\text{SHG}$ --- which involves maximizing a nonlinear overlap function between two co-localized cavity modes --- is generally more difficult to achieve than small $\tilde{V}_\text{Kerr}$ \cite{Minkov2019}, $\text{FOM}_{\chi^{(2)}}\propto \tilde{V}_\text{SHG}^{-1/2}$ also demonstrates favorable scaling at larger mode volumes. 

\paragraph*{Conclusion.---} This brief example manifests the practical limitations imposed by fundamental thermal noise in microcavities while highlighting design choices that optimize device performance in its presence. Both outcomes rely on proper noise characterization. Towards this end, we have presented a general theory for thermo-refractive noise in optical microcavities and experimentally verified our model by measuring the effect of temperature fluctuations in high-$Q/V$ silicon PhC cavities. The results show that non-perturbative TRN ultimately limits the achievable quality factor in small mode volume cavities and that experimental devices have neared this fundamental bound. Violating the observed tradeoff between mode volume and thermo-optic noise stands as an exciting avenue for future investigation that we are currently pursuing. Ultimately, these improvements will be necessary to achieve the performance required for further advances in optical quantum information processing, cavity optomechanics, precision optical sensing, and beyond.

\begin{acknowledgments}
The authors thank M. Dykman, T. Kippenberg, and G.Huang for useful discussions. C.P. was supported by the Hertz Foundation Elizabeth and Stephen Fantone Family Fellowship. R.H. was supported by an IC Postdoctoral Fellowship at MIT, administered by ORISE through U.S. DoE/ODNI. Experiments were supported in part by AFOSR grant FA9550-16-1-0391, supervised by G. Pomrenke.
\end{acknowledgments}

\vspace{-10pt}

\bibliography{ThermorefractiveNoise}

\begin{thebibliography}{58}%
\makeatletter
\providecommand \@ifxundefined [1]{%
 \@ifx{#1\undefined}
}%
\providecommand \@ifnum [1]{%
 \ifnum #1\expandafter \@firstoftwo
 \else \expandafter \@secondoftwo
 \fi
}%
\providecommand \@ifx [1]{%
 \ifx #1\expandafter \@firstoftwo
 \else \expandafter \@secondoftwo
 \fi
}%
\providecommand \natexlab [1]{#1}%
\providecommand \enquote  [1]{``#1''}%
\providecommand \bibnamefont  [1]{#1}%
\providecommand \bibfnamefont [1]{#1}%
\providecommand \citenamefont [1]{#1}%
\providecommand \href@noop [0]{\@secondoftwo}%
\providecommand \href [0]{\begingroup \@sanitize@url \@href}%
\providecommand \@href[1]{\@@startlink{#1}\@@href}%
\providecommand \@@href[1]{\endgroup#1\@@endlink}%
\providecommand \@sanitize@url [0]{\catcode `\\12\catcode `\$12\catcode
  `\&12\catcode `\#12\catcode `\^12\catcode `\_12\catcode `\%12\relax}%
\providecommand \@@startlink[1]{}%
\providecommand \@@endlink[0]{}%
\providecommand \url  [0]{\begingroup\@sanitize@url \@url }%
\providecommand \@url [1]{\endgroup\@href {#1}{\urlprefix }}%
\providecommand \urlprefix  [0]{URL }%
\providecommand \Eprint [0]{\href }%
\providecommand \doibase [0]{http://dx.doi.org/}%
\providecommand \selectlanguage [0]{\@gobble}%
\providecommand \bibinfo  [0]{\@secondoftwo}%
\providecommand \bibfield  [0]{\@secondoftwo}%
\providecommand \translation [1]{[#1]}%
\providecommand \BibitemOpen [0]{}%
\providecommand \bibitemStop [0]{}%
\providecommand \bibitemNoStop [0]{.\EOS\space}%
\providecommand \EOS [0]{\spacefactor3000\relax}%
\providecommand \BibitemShut  [1]{\csname bibitem#1\endcsname}%
\let\auto@bib@innerbib\@empty
\bibitem [{\citenamefont {{The LIGO Scientific
  Collaboration}}(2017)}]{Abbott2017a}%
  \BibitemOpen
  \bibfield  {author} {\bibinfo {author} {\bibnamefont {{The LIGO Scientific
  Collaboration}}},\ }\href@noop {} {\bibfield  {journal} {\bibinfo  {journal}
  {Physical Review Letters}\ }\textbf {\bibinfo {volume} {119}},\ \bibinfo
  {pages} {161101} (\bibinfo {year} {2017})}\BibitemShut {NoStop}%
\bibitem [{\citenamefont {Yu}\ \emph {et~al.}(2019)\citenamefont {Yu},
  \citenamefont {Muniz}, \citenamefont {Hung},\ and\ \citenamefont
  {Kimble}}]{Yu2018}%
  \BibitemOpen
  \bibfield  {author} {\bibinfo {author} {\bibfnamefont {S.~P.}\ \bibnamefont
  {Yu}}, \bibinfo {author} {\bibfnamefont {J.~A.}\ \bibnamefont {Muniz}},
  \bibinfo {author} {\bibfnamefont {C.~L.}\ \bibnamefont {Hung}}, \ and\
  \bibinfo {author} {\bibfnamefont {H.~J.}\ \bibnamefont {Kimble}},\ }\href
  {https://arxiv.org/pdf/1812.08936.pdf} {\bibfield  {journal} {\bibinfo
  {journal} {Proceedings of the National Academy of Sciences of the United
  States of America}\ }\textbf {\bibinfo {volume} {116}},\ \bibinfo {pages}
  {12743} (\bibinfo {year} {2019})}\BibitemShut {NoStop}%
\bibitem [{\citenamefont {Deli{\'{c}}}\ \emph {et~al.}(2020)\citenamefont
  {Deli{\'{c}}}, \citenamefont {Reisenbauer}, \citenamefont {Dare},
  \citenamefont {Grass}, \citenamefont {Vuleti{\'{c}}}, \citenamefont
  {Kiesel},\ and\ \citenamefont {Aspelmeyer}}]{Delic2020}%
  \BibitemOpen
  \bibfield  {author} {\bibinfo {author} {\bibfnamefont {U.}~\bibnamefont
  {Deli{\'{c}}}}, \bibinfo {author} {\bibfnamefont {M.}~\bibnamefont
  {Reisenbauer}}, \bibinfo {author} {\bibfnamefont {K.}~\bibnamefont {Dare}},
  \bibinfo {author} {\bibfnamefont {D.}~\bibnamefont {Grass}}, \bibinfo
  {author} {\bibfnamefont {V.}~\bibnamefont {Vuleti{\'{c}}}}, \bibinfo {author}
  {\bibfnamefont {N.}~\bibnamefont {Kiesel}}, \ and\ \bibinfo {author}
  {\bibfnamefont {M.}~\bibnamefont {Aspelmeyer}},\ }\href
  {http://science.sciencemag.org/
  https://www.sciencemag.org/lookup/doi/10.1126/science.aba3993} {\bibfield
  {journal} {\bibinfo  {journal} {Science}\ }\textbf {\bibinfo {volume}
  {367}},\ \bibinfo {pages} {892} (\bibinfo {year} {2020})}\BibitemShut
  {NoStop}%
\bibitem [{\citenamefont {Callen}\ and\ \citenamefont {Welton}(1951)}]{Callen}%
  \BibitemOpen
  \bibfield  {author} {\bibinfo {author} {\bibfnamefont {H.~B.}\ \bibnamefont
  {Callen}}\ and\ \bibinfo {author} {\bibfnamefont {T.~A.}\ \bibnamefont
  {Welton}},\ }\href {\doibase 10.1103/PhysRev.83.34} {\bibfield  {journal}
  {\bibinfo  {journal} {Physical Review}\ }\textbf {\bibinfo {volume} {83}},\
  \bibinfo {pages} {34} (\bibinfo {year} {1951})}\BibitemShut {NoStop}%
\bibitem [{\citenamefont {Levin}(1998)}]{Levin1997}%
  \BibitemOpen
  \bibfield  {author} {\bibinfo {author} {\bibfnamefont {Y.}~\bibnamefont
  {Levin}},\ }\href {\doibase 10.1103/PhysRevE.67.046106} {\bibfield  {journal}
  {\bibinfo  {journal} {Physical Review D}\ }\textbf {\bibinfo {volume} {57}},\
  \bibinfo {pages} {659} (\bibinfo {year} {1998})}\BibitemShut {NoStop}%
\bibitem [{\citenamefont {{The LIGO Scientific
  Collaboration}}(2016)}]{Martynov2016}%
  \BibitemOpen
  \bibfield  {author} {\bibinfo {author} {\bibnamefont {{The LIGO Scientific
  Collaboration}}},\ }\href@noop {} {\bibfield  {journal} {\bibinfo  {journal}
  {Physical Review D}\ }\textbf {\bibinfo {volume} {93}},\ \bibinfo {pages}
  {112004} (\bibinfo {year} {2016})}\BibitemShut {NoStop}%
\bibitem [{\citenamefont {Cole}\ \emph {et~al.}(2013)\citenamefont {Cole},
  \citenamefont {Zhang}, \citenamefont {Martin}, \citenamefont {Ye},\ and\
  \citenamefont {Aspelmeyer}}]{Cole2013}%
  \BibitemOpen
  \bibfield  {author} {\bibinfo {author} {\bibfnamefont {G.~D.}\ \bibnamefont
  {Cole}}, \bibinfo {author} {\bibfnamefont {W.}~\bibnamefont {Zhang}},
  \bibinfo {author} {\bibfnamefont {M.~J.}\ \bibnamefont {Martin}}, \bibinfo
  {author} {\bibfnamefont {J.}~\bibnamefont {Ye}}, \ and\ \bibinfo {author}
  {\bibfnamefont {M.}~\bibnamefont {Aspelmeyer}},\ }\href {\doibase
  10.1038/nphoton.2013.174} {\bibfield  {journal} {\bibinfo  {journal} {Nature
  Photonics}\ }\textbf {\bibinfo {volume} {7}},\ \bibinfo {pages} {644}
  (\bibinfo {year} {2013})}\BibitemShut {NoStop}%
\bibitem [{\citenamefont {Heinert}\ \emph {et~al.}(2013)\citenamefont
  {Heinert}, \citenamefont {Kroker}, \citenamefont {Friedrich}, \citenamefont
  {Hild}, \citenamefont {Kley}, \citenamefont {Leavey}, \citenamefont {Martin},
  \citenamefont {Nawrodt}, \citenamefont {T{\"{u}}nnermann}, \citenamefont
  {Vyatchanin},\ and\ \citenamefont {Yamamoto}}]{Heinert2013}%
  \BibitemOpen
  \bibfield  {author} {\bibinfo {author} {\bibfnamefont {D.}~\bibnamefont
  {Heinert}}, \bibinfo {author} {\bibfnamefont {S.}~\bibnamefont {Kroker}},
  \bibinfo {author} {\bibfnamefont {D.}~\bibnamefont {Friedrich}}, \bibinfo
  {author} {\bibfnamefont {S.}~\bibnamefont {Hild}}, \bibinfo {author}
  {\bibfnamefont {E.~B.}\ \bibnamefont {Kley}}, \bibinfo {author}
  {\bibfnamefont {S.}~\bibnamefont {Leavey}}, \bibinfo {author} {\bibfnamefont
  {I.~W.}\ \bibnamefont {Martin}}, \bibinfo {author} {\bibfnamefont
  {R.}~\bibnamefont {Nawrodt}}, \bibinfo {author} {\bibfnamefont
  {A.}~\bibnamefont {T{\"{u}}nnermann}}, \bibinfo {author} {\bibfnamefont
  {S.~P.}\ \bibnamefont {Vyatchanin}}, \ and\ \bibinfo {author} {\bibfnamefont
  {K.}~\bibnamefont {Yamamoto}},\ }\href@noop {} {\bibfield  {journal}
  {\bibinfo  {journal} {Physical Review D}\ }\textbf {\bibinfo {volume} {88}},\
  \bibinfo {pages} {042001} (\bibinfo {year} {2013})}\BibitemShut {NoStop}%
\bibitem [{\citenamefont {Kroker}\ \emph {et~al.}(2017)\citenamefont {Kroker},
  \citenamefont {Dickmann}, \citenamefont {{Rojas Hurtado}}, \citenamefont
  {Heinert}, \citenamefont {Nawrodt}, \citenamefont {Levin},\ and\
  \citenamefont {Vyatchanin}}]{Kroker2017}%
  \BibitemOpen
  \bibfield  {author} {\bibinfo {author} {\bibfnamefont {S.}~\bibnamefont
  {Kroker}}, \bibinfo {author} {\bibfnamefont {J.}~\bibnamefont {Dickmann}},
  \bibinfo {author} {\bibfnamefont {C.~B.}\ \bibnamefont {{Rojas Hurtado}}},
  \bibinfo {author} {\bibfnamefont {D.}~\bibnamefont {Heinert}}, \bibinfo
  {author} {\bibfnamefont {R.}~\bibnamefont {Nawrodt}}, \bibinfo {author}
  {\bibfnamefont {Y.}~\bibnamefont {Levin}}, \ and\ \bibinfo {author}
  {\bibfnamefont {S.~P.}\ \bibnamefont {Vyatchanin}},\ }\href@noop {}
  {\bibfield  {journal} {\bibinfo  {journal} {Physical Review D}\ }\textbf
  {\bibinfo {volume} {96}},\ \bibinfo {pages} {022002} (\bibinfo {year}
  {2017})}\BibitemShut {NoStop}%
\bibitem [{\citenamefont {Numata}\ \emph {et~al.}(2004)\citenamefont {Numata},
  \citenamefont {Kemery},\ and\ \citenamefont {Camp}}]{Numata}%
  \BibitemOpen
  \bibfield  {author} {\bibinfo {author} {\bibfnamefont {K.}~\bibnamefont
  {Numata}}, \bibinfo {author} {\bibfnamefont {A.}~\bibnamefont {Kemery}}, \
  and\ \bibinfo {author} {\bibfnamefont {J.}~\bibnamefont {Camp}},\ }\href
  {https://journals.aps.org/prl/pdf/10.1103/PhysRevLett.93.250602} {\bibfield
  {journal} {\bibinfo  {journal} {Physical Review Letters}\ }\textbf {\bibinfo
  {volume} {93}},\ \bibinfo {pages} {250602} (\bibinfo {year}
  {2004})}\BibitemShut {NoStop}%
\bibitem [{\citenamefont {Kessler}\ \emph {et~al.}(2012)\citenamefont
  {Kessler}, \citenamefont {Legero},\ and\ \citenamefont
  {Sterr}}]{Kessler2011}%
  \BibitemOpen
  \bibfield  {author} {\bibinfo {author} {\bibfnamefont {T.}~\bibnamefont
  {Kessler}}, \bibinfo {author} {\bibfnamefont {T.}~\bibnamefont {Legero}}, \
  and\ \bibinfo {author} {\bibfnamefont {U.}~\bibnamefont {Sterr}},\ }\href
  {\doibase 10.1364/josab.29.000178} {\bibfield  {journal} {\bibinfo  {journal}
  {Journal of the Optical Society of America B}\ }\textbf {\bibinfo {volume}
  {29}},\ \bibinfo {pages} {178} (\bibinfo {year} {2012})}\BibitemShut
  {NoStop}%
\bibitem [{\citenamefont {Zhadnov}\ \emph {et~al.}(2018)\citenamefont
  {Zhadnov}, \citenamefont {Kudeyarov}, \citenamefont {Kryuchkov},
  \citenamefont {Semerikov}, \citenamefont {Khabarova},\ and\ \citenamefont
  {Kolachevsky}}]{Zhadnov2018}%
  \BibitemOpen
  \bibfield  {author} {\bibinfo {author} {\bibfnamefont {N.~O.}\ \bibnamefont
  {Zhadnov}}, \bibinfo {author} {\bibfnamefont {K.~S.}\ \bibnamefont
  {Kudeyarov}}, \bibinfo {author} {\bibfnamefont {D.~S.}\ \bibnamefont
  {Kryuchkov}}, \bibinfo {author} {\bibfnamefont {I.~A.}\ \bibnamefont
  {Semerikov}}, \bibinfo {author} {\bibfnamefont {K.~Y.}\ \bibnamefont
  {Khabarova}}, \ and\ \bibinfo {author} {\bibfnamefont {N.~N.}\ \bibnamefont
  {Kolachevsky}},\ }\href {\doibase 10.1070/QEL16654} {\bibfield  {journal}
  {\bibinfo  {journal} {Quantum Electronics}\ }\textbf {\bibinfo {volume}
  {48}},\ \bibinfo {pages} {425} (\bibinfo {year} {2018})}\BibitemShut
  {NoStop}%
\bibitem [{\citenamefont {Vahala}(2003)}]{Vahala2003}%
  \BibitemOpen
  \bibfield  {author} {\bibinfo {author} {\bibfnamefont {K.~J.}\ \bibnamefont
  {Vahala}},\ }\href {\doibase 10.1038/nature01939} {\bibfield  {journal}
  {\bibinfo  {journal} {Nature}\ }\textbf {\bibinfo {volume} {424}},\ \bibinfo
  {pages} {839} (\bibinfo {year} {2003})}\BibitemShut {NoStop}%
\bibitem [{\citenamefont {Li}\ \emph {et~al.}(2019)\citenamefont {Li},
  \citenamefont {Liang}, \citenamefont {Luo}, \citenamefont {He}, \citenamefont
  {Ling},\ and\ \citenamefont {Lin}}]{Li2019}%
  \BibitemOpen
  \bibfield  {author} {\bibinfo {author} {\bibfnamefont {M.}~\bibnamefont
  {Li}}, \bibinfo {author} {\bibfnamefont {H.}~\bibnamefont {Liang}}, \bibinfo
  {author} {\bibfnamefont {R.}~\bibnamefont {Luo}}, \bibinfo {author}
  {\bibfnamefont {Y.}~\bibnamefont {He}}, \bibinfo {author} {\bibfnamefont
  {J.}~\bibnamefont {Ling}}, \ and\ \bibinfo {author} {\bibfnamefont
  {Q.}~\bibnamefont {Lin}},\ }\href {\doibase 10.1364/optica.6.000860}
  {\bibfield  {journal} {\bibinfo  {journal} {Optica}\ }\textbf {\bibinfo
  {volume} {6}},\ \bibinfo {pages} {860} (\bibinfo {year} {2019})}\BibitemShut
  {NoStop}%
\bibitem [{\citenamefont {Armani}\ \emph {et~al.}(2007)\citenamefont {Armani},
  \citenamefont {Kulkarni}, \citenamefont {Fraser}, \citenamefont {Flagan},\
  and\ \citenamefont {Vahala}}]{Armani2007}%
  \BibitemOpen
  \bibfield  {author} {\bibinfo {author} {\bibfnamefont {A.~M.}\ \bibnamefont
  {Armani}}, \bibinfo {author} {\bibfnamefont {R.~P.}\ \bibnamefont
  {Kulkarni}}, \bibinfo {author} {\bibfnamefont {S.~E.}\ \bibnamefont
  {Fraser}}, \bibinfo {author} {\bibfnamefont {R.~C.}\ \bibnamefont {Flagan}},
  \ and\ \bibinfo {author} {\bibfnamefont {K.~J.}\ \bibnamefont {Vahala}},\
  }\href {\doibase 10.1126/science.1145002} {\bibfield  {journal} {\bibinfo
  {journal} {Science}\ }\textbf {\bibinfo {volume} {317}},\ \bibinfo {pages}
  {783} (\bibinfo {year} {2007})}\BibitemShut {NoStop}%
\bibitem [{\citenamefont {Eichenfield}\ \emph {et~al.}(2009)\citenamefont
  {Eichenfield}, \citenamefont {Camacho}, \citenamefont {Chan}, \citenamefont
  {Vahala},\ and\ \citenamefont {Painter}}]{Eichenfield2009a}%
  \BibitemOpen
  \bibfield  {author} {\bibinfo {author} {\bibfnamefont {M.}~\bibnamefont
  {Eichenfield}}, \bibinfo {author} {\bibfnamefont {R.}~\bibnamefont
  {Camacho}}, \bibinfo {author} {\bibfnamefont {J.}~\bibnamefont {Chan}},
  \bibinfo {author} {\bibfnamefont {K.~J.}\ \bibnamefont {Vahala}}, \ and\
  \bibinfo {author} {\bibfnamefont {O.}~\bibnamefont {Painter}},\ }\href
  {\doibase 10.1038/nature08061} {\bibfield  {journal} {\bibinfo  {journal}
  {Nature}\ }\textbf {\bibinfo {volume} {459}},\ \bibinfo {pages} {550}
  (\bibinfo {year} {2009})}\BibitemShut {NoStop}%
\bibitem [{\citenamefont {Goban}\ \emph {et~al.}(2014)\citenamefont {Goban},
  \citenamefont {Hung}, \citenamefont {Yu}, \citenamefont {Hood}, \citenamefont
  {Muniz}, \citenamefont {Lee}, \citenamefont {Martin}, \citenamefont
  {McClung}, \citenamefont {Choi}, \citenamefont {Chang}, \citenamefont
  {Painter},\ and\ \citenamefont {Kimble}}]{Goban2014}%
  \BibitemOpen
  \bibfield  {author} {\bibinfo {author} {\bibfnamefont {A.}~\bibnamefont
  {Goban}}, \bibinfo {author} {\bibfnamefont {C.~L.}\ \bibnamefont {Hung}},
  \bibinfo {author} {\bibfnamefont {S.~P.}\ \bibnamefont {Yu}}, \bibinfo
  {author} {\bibfnamefont {J.~D.}\ \bibnamefont {Hood}}, \bibinfo {author}
  {\bibfnamefont {J.~A.}\ \bibnamefont {Muniz}}, \bibinfo {author}
  {\bibfnamefont {J.~H.}\ \bibnamefont {Lee}}, \bibinfo {author} {\bibfnamefont
  {M.~J.}\ \bibnamefont {Martin}}, \bibinfo {author} {\bibfnamefont {A.~C.}\
  \bibnamefont {McClung}}, \bibinfo {author} {\bibfnamefont {K.~S.}\
  \bibnamefont {Choi}}, \bibinfo {author} {\bibfnamefont {D.~E.}\ \bibnamefont
  {Chang}}, \bibinfo {author} {\bibfnamefont {O.}~\bibnamefont {Painter}}, \
  and\ \bibinfo {author} {\bibfnamefont {H.~J.}\ \bibnamefont {Kimble}},\
  }\href {www.nature.com/naturecommunications} {\bibfield  {journal} {\bibinfo
  {journal} {Nature Communications}\ }\textbf {\bibinfo {volume} {5}},\
  \bibinfo {pages} {3808} (\bibinfo {year} {2014})}\BibitemShut {NoStop}%
\bibitem [{\citenamefont {Chui}\ \emph {et~al.}(1992)\citenamefont {Chui},
  \citenamefont {Swanson}, \citenamefont {Adriaans}, \citenamefont {Nissen},\
  and\ \citenamefont {Lipa}}]{Chui1992}%
  \BibitemOpen
  \bibfield  {author} {\bibinfo {author} {\bibfnamefont {T.~C.}\ \bibnamefont
  {Chui}}, \bibinfo {author} {\bibfnamefont {D.~R.}\ \bibnamefont {Swanson}},
  \bibinfo {author} {\bibfnamefont {M.~J.}\ \bibnamefont {Adriaans}}, \bibinfo
  {author} {\bibfnamefont {J.~A.}\ \bibnamefont {Nissen}}, \ and\ \bibinfo
  {author} {\bibfnamefont {J.~A.}\ \bibnamefont {Lipa}},\ }\href {\doibase
  10.1103/PhysRevLett.69.3005} {\bibfield  {journal} {\bibinfo  {journal}
  {Physical Review Letters}\ }\textbf {\bibinfo {volume} {69}},\ \bibinfo
  {pages} {3005} (\bibinfo {year} {1992})}\BibitemShut {NoStop}%
\bibitem [{\citenamefont {Evans}\ \emph {et~al.}(2008)\citenamefont {Evans},
  \citenamefont {Ballmer}, \citenamefont {Fejer}, \citenamefont {Fritschel},
  \citenamefont {Harry},\ and\ \citenamefont {Ogin}}]{Evans2008}%
  \BibitemOpen
  \bibfield  {author} {\bibinfo {author} {\bibfnamefont {M.}~\bibnamefont
  {Evans}}, \bibinfo {author} {\bibfnamefont {S.}~\bibnamefont {Ballmer}},
  \bibinfo {author} {\bibfnamefont {M.}~\bibnamefont {Fejer}}, \bibinfo
  {author} {\bibfnamefont {P.}~\bibnamefont {Fritschel}}, \bibinfo {author}
  {\bibfnamefont {G.}~\bibnamefont {Harry}}, \ and\ \bibinfo {author}
  {\bibfnamefont {G.}~\bibnamefont {Ogin}},\ }\href@noop {} {\bibfield
  {journal} {\bibinfo  {journal} {Physical Review D - Particles, Fields,
  Gravitation and Cosmology}\ }\textbf {\bibinfo {volume} {78}},\ \bibinfo
  {pages} {102003} (\bibinfo {year} {2008})}\BibitemShut {NoStop}%
\bibitem [{\citenamefont {Gorodetsky}\ and\ \citenamefont
  {Grudinin}(2004)}]{Gorodetsky2004}%
  \BibitemOpen
  \bibfield  {author} {\bibinfo {author} {\bibfnamefont {M.~L.}\ \bibnamefont
  {Gorodetsky}}\ and\ \bibinfo {author} {\bibfnamefont {I.~S.}\ \bibnamefont
  {Grudinin}},\ }\href {\doibase 10.1364/JOSAB.21.000697} {\bibfield  {journal}
  {\bibinfo  {journal} {Journal of the Optical Society of America B}\ }\textbf
  {\bibinfo {volume} {21}},\ \bibinfo {pages} {697} (\bibinfo {year}
  {2004})}\BibitemShut {NoStop}%
\bibitem [{\citenamefont {Savchenkov}\ \emph {et~al.}(2007)\citenamefont
  {Savchenkov}, \citenamefont {Matsko}, \citenamefont {Ilchenko}, \citenamefont
  {Yu},\ and\ \citenamefont {Maleki}}]{Matsko2007}%
  \BibitemOpen
  \bibfield  {author} {\bibinfo {author} {\bibfnamefont {A.~A.}\ \bibnamefont
  {Savchenkov}}, \bibinfo {author} {\bibfnamefont {A.~B.}\ \bibnamefont
  {Matsko}}, \bibinfo {author} {\bibfnamefont {V.~S.}\ \bibnamefont
  {Ilchenko}}, \bibinfo {author} {\bibfnamefont {N.}~\bibnamefont {Yu}}, \ and\
  \bibinfo {author} {\bibfnamefont {L.}~\bibnamefont {Maleki}},\ }\href
  {\doibase 10.1364/josab.24.002988} {\bibfield  {journal} {\bibinfo  {journal}
  {Journal of the Optical Society of America B}\ }\textbf {\bibinfo {volume}
  {24}},\ \bibinfo {pages} {2988} (\bibinfo {year} {2007})}\BibitemShut
  {NoStop}%
\bibitem [{\citenamefont {Huang}\ \emph {et~al.}(2019)\citenamefont {Huang},
  \citenamefont {Lucas}, \citenamefont {Liu}, \citenamefont {Raja},
  \citenamefont {Lihachev}, \citenamefont {Gorodetsky}, \citenamefont
  {Engelsen},\ and\ \citenamefont {Kippenberg}}]{Huang2019}%
  \BibitemOpen
  \bibfield  {author} {\bibinfo {author} {\bibfnamefont {G.}~\bibnamefont
  {Huang}}, \bibinfo {author} {\bibfnamefont {E.}~\bibnamefont {Lucas}},
  \bibinfo {author} {\bibfnamefont {J.}~\bibnamefont {Liu}}, \bibinfo {author}
  {\bibfnamefont {A.~S.}\ \bibnamefont {Raja}}, \bibinfo {author}
  {\bibfnamefont {G.}~\bibnamefont {Lihachev}}, \bibinfo {author}
  {\bibfnamefont {M.~L.}\ \bibnamefont {Gorodetsky}}, \bibinfo {author}
  {\bibfnamefont {N.~J.}\ \bibnamefont {Engelsen}}, \ and\ \bibinfo {author}
  {\bibfnamefont {T.~J.}\ \bibnamefont {Kippenberg}},\ }\href {\doibase
  10.1103/PhysRevA.99.061801} {\bibfield  {journal} {\bibinfo  {journal}
  {Physical Review A}\ }\textbf {\bibinfo {volume} {99}},\ \bibinfo {pages}
  {061801} (\bibinfo {year} {2019})}\BibitemShut {NoStop}%
\bibitem [{\citenamefont {Saurav}\ and\ \citenamefont {{Le
  Thomas}}(2017)}]{Saurav}%
  \BibitemOpen
  \bibfield  {author} {\bibinfo {author} {\bibfnamefont {K.}~\bibnamefont
  {Saurav}}\ and\ \bibinfo {author} {\bibfnamefont {N.}~\bibnamefont {{Le
  Thomas}}},\ }\href {\doibase 10.1364/OPTICA.4.000757} {\bibfield  {journal}
  {\bibinfo  {journal} {Optica}\ }\textbf {\bibinfo {volume} {4}},\ \bibinfo
  {pages} {757} (\bibinfo {year} {2017})}\BibitemShut {NoStop}%
\bibitem [{\citenamefont {Jiang}\ \emph {et~al.}(2019)\citenamefont {Jiang},
  \citenamefont {Pan}, \citenamefont {Deka}, \citenamefont {Fang},
  \citenamefont {Chen}, \citenamefont {Fainman},\ and\ \citenamefont {{El
  Amili}}}]{Jiang2019}%
  \BibitemOpen
  \bibfield  {author} {\bibinfo {author} {\bibfnamefont {S.}~\bibnamefont
  {Jiang}}, \bibinfo {author} {\bibfnamefont {S.~H.}\ \bibnamefont {Pan}},
  \bibinfo {author} {\bibfnamefont {S.~S.}\ \bibnamefont {Deka}}, \bibinfo
  {author} {\bibfnamefont {C.~Y.}\ \bibnamefont {Fang}}, \bibinfo {author}
  {\bibfnamefont {Z.}~\bibnamefont {Chen}}, \bibinfo {author} {\bibfnamefont
  {Y.}~\bibnamefont {Fainman}}, \ and\ \bibinfo {author} {\bibfnamefont
  {A.}~\bibnamefont {{El Amili}}},\ }\href@noop {} {\bibfield  {journal}
  {\bibinfo  {journal} {IEEE Journal of Quantum Electronics}\ }\textbf
  {\bibinfo {volume} {55}},\ \bibinfo {pages} {2000910} (\bibinfo {year}
  {2019})}\BibitemShut {NoStop}%
\bibitem [{\citenamefont {Drake}\ \emph {et~al.}(2019)\citenamefont {Drake},
  \citenamefont {Stone}, \citenamefont {Briles},\ and\ \citenamefont
  {Papp}}]{Drake2019}%
  \BibitemOpen
  \bibfield  {author} {\bibinfo {author} {\bibfnamefont {T.~E.}\ \bibnamefont
  {Drake}}, \bibinfo {author} {\bibfnamefont {J.~R.}\ \bibnamefont {Stone}},
  \bibinfo {author} {\bibfnamefont {T.~C.}\ \bibnamefont {Briles}}, \ and\
  \bibinfo {author} {\bibfnamefont {S.~B.}\ \bibnamefont {Papp}},\ }\href
  {https://arxiv.org/pdf/1903.00431.pdf} {\  (\bibinfo {year} {2019})},\
  \Eprint {http://arxiv.org/abs/1903.00431v1} {arXiv:1903.00431v1} \BibitemShut
  {NoStop}%
\bibitem [{\citenamefont {Asano}\ \emph {et~al.}(2017)\citenamefont {Asano},
  \citenamefont {Ochi}, \citenamefont {Takahashi}, \citenamefont {Kishimoto},\
  and\ \citenamefont {Noda}}]{Asano2017a}%
  \BibitemOpen
  \bibfield  {author} {\bibinfo {author} {\bibfnamefont {T.}~\bibnamefont
  {Asano}}, \bibinfo {author} {\bibfnamefont {Y.}~\bibnamefont {Ochi}},
  \bibinfo {author} {\bibfnamefont {Y.}~\bibnamefont {Takahashi}}, \bibinfo
  {author} {\bibfnamefont {K.}~\bibnamefont {Kishimoto}}, \ and\ \bibinfo
  {author} {\bibfnamefont {S.}~\bibnamefont {Noda}},\ }\href {\doibase
  10.1364/OE.25.001769} {\bibfield  {journal} {\bibinfo  {journal} {Optics
  Express}\ }\textbf {\bibinfo {volume} {25}},\ \bibinfo {pages} {1769}
  (\bibinfo {year} {2017})}\BibitemShut {NoStop}%
\bibitem [{\citenamefont {Hu}\ \emph {et~al.}(2018)\citenamefont {Hu},
  \citenamefont {Khater}, \citenamefont {Salas-Montiel}, \citenamefont
  {Kratschmer}, \citenamefont {Engelmann}, \citenamefont {Green},\ and\
  \citenamefont {Weiss}}]{Hu2018}%
  \BibitemOpen
  \bibfield  {author} {\bibinfo {author} {\bibfnamefont {S.}~\bibnamefont
  {Hu}}, \bibinfo {author} {\bibfnamefont {M.}~\bibnamefont {Khater}}, \bibinfo
  {author} {\bibfnamefont {R.}~\bibnamefont {Salas-Montiel}}, \bibinfo {author}
  {\bibfnamefont {E.}~\bibnamefont {Kratschmer}}, \bibinfo {author}
  {\bibfnamefont {S.}~\bibnamefont {Engelmann}}, \bibinfo {author}
  {\bibfnamefont {W.~M.}\ \bibnamefont {Green}}, \ and\ \bibinfo {author}
  {\bibfnamefont {S.~M.}\ \bibnamefont {Weiss}},\ }\href
  {http://advances.sciencemag.org/} {\bibfield  {journal} {\bibinfo  {journal}
  {Science Advances}\ }\textbf {\bibinfo {volume} {4}} (\bibinfo {year}
  {2018})}\BibitemShut {NoStop}%
\bibitem [{\citenamefont {Mabuchi}(2012)}]{Mabuchi2012}%
  \BibitemOpen
  \bibfield  {author} {\bibinfo {author} {\bibfnamefont {H.}~\bibnamefont
  {Mabuchi}},\ }\href {\doibase 10.1103/PhysRevA.85.015806} {\bibfield
  {journal} {\bibinfo  {journal} {Physical Review A}\ }\textbf {\bibinfo
  {volume} {85}},\ \bibinfo {pages} {015806} (\bibinfo {year}
  {2012})}\BibitemShut {NoStop}%
\bibitem [{\citenamefont {Joannopoulos}\ \emph {et~al.}(2008)\citenamefont
  {Joannopoulos}, \citenamefont {Johnson}, \citenamefont {Winn},\ and\
  \citenamefont {Meade}}]{Joannopoulos}%
  \BibitemOpen
  \bibfield  {author} {\bibinfo {author} {\bibfnamefont {J.~D.}\ \bibnamefont
  {Joannopoulos}}, \bibinfo {author} {\bibfnamefont {S.~G.}\ \bibnamefont
  {Johnson}}, \bibinfo {author} {\bibfnamefont {J.~N.}\ \bibnamefont {Winn}}, \
  and\ \bibinfo {author} {\bibfnamefont {R.~D.}\ \bibnamefont {Meade}},\ }\href
  {\doibase 10.1063/1.1586781} {\emph {\bibinfo {title} {{Photonic crystals:
  molding the flow of light}}}}\ (\bibinfo  {publisher} {Princeton University
  Press},\ \bibinfo {year} {2008})\BibitemShut {NoStop}%
\bibitem [{\citenamefont {Zhang}\ \emph {et~al.}(2014)\citenamefont {Zhang},
  \citenamefont {Moser}, \citenamefont {G{\"{u}}ttinger}, \citenamefont
  {Bachtold},\ and\ \citenamefont {Dykman}}]{Zhang2014a}%
  \BibitemOpen
  \bibfield  {author} {\bibinfo {author} {\bibfnamefont {Y.}~\bibnamefont
  {Zhang}}, \bibinfo {author} {\bibfnamefont {J.}~\bibnamefont {Moser}},
  \bibinfo {author} {\bibfnamefont {J.}~\bibnamefont {G{\"{u}}ttinger}},
  \bibinfo {author} {\bibfnamefont {A.}~\bibnamefont {Bachtold}}, \ and\
  \bibinfo {author} {\bibfnamefont {M.~I.}\ \bibnamefont {Dykman}},\ }\href
  {https://journals.aps.org/prl/pdf/10.1103/PhysRevLett.113.255502} {\bibfield
  {journal} {\bibinfo  {journal} {Physical Review Letters}\ }\textbf {\bibinfo
  {volume} {113}},\ \bibinfo {pages} {255502} (\bibinfo {year}
  {2014})}\BibitemShut {NoStop}%
\bibitem [{\citenamefont {Feynman}\ and\ \citenamefont
  {Hibbs}(1965)}]{Feynman1965}%
  \BibitemOpen
  \bibfield  {author} {\bibinfo {author} {\bibfnamefont {R.~P.}\ \bibnamefont
  {Feynman}}\ and\ \bibinfo {author} {\bibfnamefont {A.~R.}\ \bibnamefont
  {Hibbs}},\ }\href@noop {} {\emph {\bibinfo {title} {{Quantum Mechanics and
  Path Integrals}}}}\ (\bibinfo  {publisher} {McGraw-Hill},\ \bibinfo {address}
  {New York},\ \bibinfo {year} {1965})\BibitemShut {NoStop}%
\bibitem [{\citenamefont {Sun}\ \emph {et~al.}(2017)\citenamefont {Sun},
  \citenamefont {Luo}, \citenamefont {Zhang},\ and\ \citenamefont
  {Lin}}]{Sun2017}%
  \BibitemOpen
  \bibfield  {author} {\bibinfo {author} {\bibfnamefont {X.}~\bibnamefont
  {Sun}}, \bibinfo {author} {\bibfnamefont {R.}~\bibnamefont {Luo}}, \bibinfo
  {author} {\bibfnamefont {X.~C.}\ \bibnamefont {Zhang}}, \ and\ \bibinfo
  {author} {\bibfnamefont {Q.}~\bibnamefont {Lin}},\ }\href {\doibase
  10.1103/PhysRevA.95.023822} {\bibfield  {journal} {\bibinfo  {journal}
  {Physical Review A}\ }\textbf {\bibinfo {volume} {95}},\ \bibinfo {pages}
  {23822} (\bibinfo {year} {2017})}\BibitemShut {NoStop}%
\bibitem [{\citenamefont {Landau}\ and\ \citenamefont
  {Lifshitz}(1980)}]{Land1980}%
  \BibitemOpen
  \bibfield  {author} {\bibinfo {author} {\bibfnamefont {L.~D.}\ \bibnamefont
  {Landau}}\ and\ \bibinfo {author} {\bibfnamefont {E.~M.}\ \bibnamefont
  {Lifshitz}},\ }\href {\doibase 10.1016/C2009-0-24487-4} {\emph {\bibinfo
  {title} {{Statistical Physics}}}}\ (\bibinfo  {publisher} {Elsevier},\
  \bibinfo {year} {1980})\BibitemShut {NoStop}%
\bibitem [{\citenamefont {Galli}\ \emph {et~al.}(2009)\citenamefont {Galli},
  \citenamefont {Portalupi}, \citenamefont {Belotti}, \citenamefont {Andreani},
  \citenamefont {O'Faolain},\ and\ \citenamefont {Krauss}}]{Galli2009}%
  \BibitemOpen
  \bibfield  {author} {\bibinfo {author} {\bibfnamefont {M.}~\bibnamefont
  {Galli}}, \bibinfo {author} {\bibfnamefont {S.~L.}\ \bibnamefont
  {Portalupi}}, \bibinfo {author} {\bibfnamefont {M.}~\bibnamefont {Belotti}},
  \bibinfo {author} {\bibfnamefont {L.~C.}\ \bibnamefont {Andreani}}, \bibinfo
  {author} {\bibfnamefont {L.}~\bibnamefont {O'Faolain}}, \ and\ \bibinfo
  {author} {\bibfnamefont {T.~F.}\ \bibnamefont {Krauss}},\ }\href {\doibase
  10.1063/1.3080683} {\bibfield  {journal} {\bibinfo  {journal} {Applied
  Physics Letters}\ }\textbf {\bibinfo {volume} {94}},\ \bibinfo {pages}
  {71101} (\bibinfo {year} {2009})}\BibitemShut {NoStop}%
\bibitem [{\citenamefont {Gorodetsky}\ \emph {et~al.}(2010)\citenamefont
  {Gorodetsky}, \citenamefont {Schliesser}, \citenamefont {Anetsberger},
  \citenamefont {Deleglise},\ and\ \citenamefont
  {Kippenberg}}]{Gorodetsky2010}%
  \BibitemOpen
  \bibfield  {author} {\bibinfo {author} {\bibfnamefont {M.~L.}\ \bibnamefont
  {Gorodetsky}}, \bibinfo {author} {\bibfnamefont {A.}~\bibnamefont
  {Schliesser}}, \bibinfo {author} {\bibfnamefont {G.}~\bibnamefont
  {Anetsberger}}, \bibinfo {author} {\bibfnamefont {S.}~\bibnamefont
  {Deleglise}}, \ and\ \bibinfo {author} {\bibfnamefont {T.~J.}\ \bibnamefont
  {Kippenberg}},\ }\href {\doibase 10.1364/OE.18.023236} {\bibfield  {journal}
  {\bibinfo  {journal} {Optics Express}\ }\textbf {\bibinfo {volume} {18}},\
  \bibinfo {pages} {23236} (\bibinfo {year} {2010})}\BibitemShut {NoStop}%
\bibitem [{\citenamefont {Minkov}\ and\ \citenamefont
  {Savona}(2014)}]{Minkov2014}%
  \BibitemOpen
  \bibfield  {author} {\bibinfo {author} {\bibfnamefont {M.}~\bibnamefont
  {Minkov}}\ and\ \bibinfo {author} {\bibfnamefont {V.}~\bibnamefont
  {Savona}},\ }\href {www.nature.com/scientificreports} {\bibfield  {journal}
  {\bibinfo  {journal} {Scientific Reports}\ }\textbf {\bibinfo {volume} {4}},\
  \bibinfo {pages} {5124} (\bibinfo {year} {2014})}\BibitemShut {NoStop}%
\bibitem [{\citenamefont {Minkov}\ \emph {et~al.}(2017)\citenamefont {Minkov},
  \citenamefont {Savona},\ and\ \citenamefont {Gerace}}]{Minkov2017}%
  \BibitemOpen
  \bibfield  {author} {\bibinfo {author} {\bibfnamefont {M.}~\bibnamefont
  {Minkov}}, \bibinfo {author} {\bibfnamefont {V.}~\bibnamefont {Savona}}, \
  and\ \bibinfo {author} {\bibfnamefont {D.}~\bibnamefont {Gerace}},\ }\href
  {\doibase 10.1063/1.4991416} {\bibfield  {journal} {\bibinfo  {journal}
  {Applied Physics Letters}\ }\textbf {\bibinfo {volume} {111}},\ \bibinfo
  {pages} {131104} (\bibinfo {year} {2017})}\BibitemShut {NoStop}%
\bibitem [{\citenamefont {Gavartin}\ \emph {et~al.}(2011)\citenamefont
  {Gavartin}, \citenamefont {Braive}, \citenamefont {Sagnes}, \citenamefont
  {Arcizet}, \citenamefont {Beveratos}, \citenamefont {Kippenberg},\ and\
  \citenamefont {Robert-Philip}}]{Gavartin}%
  \BibitemOpen
  \bibfield  {author} {\bibinfo {author} {\bibfnamefont {E.}~\bibnamefont
  {Gavartin}}, \bibinfo {author} {\bibfnamefont {R.}~\bibnamefont {Braive}},
  \bibinfo {author} {\bibfnamefont {I.}~\bibnamefont {Sagnes}}, \bibinfo
  {author} {\bibfnamefont {O.}~\bibnamefont {Arcizet}}, \bibinfo {author}
  {\bibfnamefont {A.}~\bibnamefont {Beveratos}}, \bibinfo {author}
  {\bibfnamefont {T.~J.}\ \bibnamefont {Kippenberg}}, \ and\ \bibinfo {author}
  {\bibfnamefont {I.}~\bibnamefont {Robert-Philip}},\ }\href@noop {} {\bibfield
   {journal} {\bibinfo  {journal} {Physical Review Letters}\ }\textbf {\bibinfo
  {volume} {106}},\ \bibinfo {pages} {203902} (\bibinfo {year}
  {2011})}\BibitemShut {NoStop}%
\bibitem [{\citenamefont {Levin}(2008)}]{Levin2008}%
  \BibitemOpen
  \bibfield  {author} {\bibinfo {author} {\bibfnamefont {Y.}~\bibnamefont
  {Levin}},\ }\href {\doibase 10.1016/j.physleta.2007.11.007} {\bibfield
  {journal} {\bibinfo  {journal} {Physics Letters, Section A: General, Atomic
  and Solid State Physics}\ }\textbf {\bibinfo {volume} {372}},\ \bibinfo
  {pages} {1941} (\bibinfo {year} {2008})}\BibitemShut {NoStop}%
\bibitem [{\citenamefont {Jain}\ \emph {et~al.}(2013)\citenamefont {Jain},
  \citenamefont {Yu},\ and\ \citenamefont {McGaughey}}]{Jain2013}%
  \BibitemOpen
  \bibfield  {author} {\bibinfo {author} {\bibfnamefont {A.}~\bibnamefont
  {Jain}}, \bibinfo {author} {\bibfnamefont {Y.~J.}\ \bibnamefont {Yu}}, \ and\
  \bibinfo {author} {\bibfnamefont {A.~J.}\ \bibnamefont {McGaughey}},\ }\href
  {\doibase 10.1103/PhysRevB.87.195301} {\bibfield  {journal} {\bibinfo
  {journal} {Physical Review B}\ }\textbf {\bibinfo {volume} {87}},\ \bibinfo
  {pages} {195301} (\bibinfo {year} {2013})}\BibitemShut {NoStop}%
\bibitem [{\citenamefont {Cuffe}\ \emph {et~al.}(2015)\citenamefont {Cuffe},
  \citenamefont {Eliason}, \citenamefont {Maznev}, \citenamefont {Collins},
  \citenamefont {Johnson}, \citenamefont {Shchepetov}, \citenamefont
  {Prunnila}, \citenamefont {Ahopelto}, \citenamefont {{Sotomayor Torres}},
  \citenamefont {Chen},\ and\ \citenamefont {Nelson}}]{Cuffe2015}%
  \BibitemOpen
  \bibfield  {author} {\bibinfo {author} {\bibfnamefont {J.}~\bibnamefont
  {Cuffe}}, \bibinfo {author} {\bibfnamefont {J.~K.}\ \bibnamefont {Eliason}},
  \bibinfo {author} {\bibfnamefont {A.~A.}\ \bibnamefont {Maznev}}, \bibinfo
  {author} {\bibfnamefont {K.~C.}\ \bibnamefont {Collins}}, \bibinfo {author}
  {\bibfnamefont {J.~A.}\ \bibnamefont {Johnson}}, \bibinfo {author}
  {\bibfnamefont {A.}~\bibnamefont {Shchepetov}}, \bibinfo {author}
  {\bibfnamefont {M.}~\bibnamefont {Prunnila}}, \bibinfo {author}
  {\bibfnamefont {J.}~\bibnamefont {Ahopelto}}, \bibinfo {author}
  {\bibfnamefont {C.~M.}\ \bibnamefont {{Sotomayor Torres}}}, \bibinfo {author}
  {\bibfnamefont {G.}~\bibnamefont {Chen}}, \ and\ \bibinfo {author}
  {\bibfnamefont {K.~A.}\ \bibnamefont {Nelson}},\ }\href {\doibase
  10.1103/PhysRevB.91.245423} {\bibfield  {journal} {\bibinfo  {journal}
  {Physical Review B}\ }\textbf {\bibinfo {volume} {91}},\ \bibinfo {pages}
  {245423} (\bibinfo {year} {2015})}\BibitemShut {NoStop}%
\bibitem [{\citenamefont {Choi}\ \emph {et~al.}(2017)\citenamefont {Choi},
  \citenamefont {Heuck},\ and\ \citenamefont {Englund}}]{Choi}%
  \BibitemOpen
  \bibfield  {author} {\bibinfo {author} {\bibfnamefont {H.}~\bibnamefont
  {Choi}}, \bibinfo {author} {\bibfnamefont {M.}~\bibnamefont {Heuck}}, \ and\
  \bibinfo {author} {\bibfnamefont {D.}~\bibnamefont {Englund}},\ }\href
  {https://journals.aps.org/prl/pdf/10.1103/PhysRevLett.118.223605} {\bibfield
  {journal} {\bibinfo  {journal} {Physical Review Letters}\ }\textbf {\bibinfo
  {volume} {118}},\ \bibinfo {pages} {223605} (\bibinfo {year}
  {2017})}\BibitemShut {NoStop}%
\bibitem [{\citenamefont {Panuski}\ \emph {et~al.}(2019)\citenamefont
  {Panuski}, \citenamefont {Pant}, \citenamefont {Heuck}, \citenamefont
  {Hamerly},\ and\ \citenamefont {Englund}}]{Panuski2019}%
  \BibitemOpen
  \bibfield  {author} {\bibinfo {author} {\bibfnamefont {C.}~\bibnamefont
  {Panuski}}, \bibinfo {author} {\bibfnamefont {M.}~\bibnamefont {Pant}},
  \bibinfo {author} {\bibfnamefont {M.}~\bibnamefont {Heuck}}, \bibinfo
  {author} {\bibfnamefont {R.}~\bibnamefont {Hamerly}}, \ and\ \bibinfo
  {author} {\bibfnamefont {D.}~\bibnamefont {Englund}},\ }\href {\doibase
  10.1103/PhysRevB.99.205303} {\bibfield  {journal} {\bibinfo  {journal}
  {Physical Review B}\ }\textbf {\bibinfo {volume} {99}},\ \bibinfo {pages}
  {205303} (\bibinfo {year} {2019})}\BibitemShut {NoStop}%
\bibitem [{\citenamefont {Pernice}\ \emph {et~al.}(2012)\citenamefont
  {Pernice}, \citenamefont {Xiong}, \citenamefont {Schuck},\ and\ \citenamefont
  {Tang}}]{Pernice2012}%
  \BibitemOpen
  \bibfield  {author} {\bibinfo {author} {\bibfnamefont {W.~H.}\ \bibnamefont
  {Pernice}}, \bibinfo {author} {\bibfnamefont {C.}~\bibnamefont {Xiong}},
  \bibinfo {author} {\bibfnamefont {C.}~\bibnamefont {Schuck}}, \ and\ \bibinfo
  {author} {\bibfnamefont {H.~X.}\ \bibnamefont {Tang}},\ }\href@noop {}
  {\bibfield  {journal} {\bibinfo  {journal} {Applied Physics Letters}\
  }\textbf {\bibinfo {volume} {100}},\ \bibinfo {pages} {091105} (\bibinfo
  {year} {2012})}\BibitemShut {NoStop}%
\bibitem [{\citenamefont {Kippenberg}\ \emph {et~al.}(2004)\citenamefont
  {Kippenberg}, \citenamefont {Spillane},\ and\ \citenamefont
  {Vahala}}]{Kippenberg2004}%
  \BibitemOpen
  \bibfield  {author} {\bibinfo {author} {\bibfnamefont {T.~J.}\ \bibnamefont
  {Kippenberg}}, \bibinfo {author} {\bibfnamefont {S.~M.}\ \bibnamefont
  {Spillane}}, \ and\ \bibinfo {author} {\bibfnamefont {K.~J.}\ \bibnamefont
  {Vahala}},\ }\href@noop {} {\bibfield  {journal} {\bibinfo  {journal}
  {Applied Physics Letters}\ }\textbf {\bibinfo {volume} {85}},\ \bibinfo
  {pages} {6113} (\bibinfo {year} {2004})}\BibitemShut {NoStop}%
\bibitem [{\citenamefont {Zhu}\ \emph {et~al.}(2010)\citenamefont {Zhu},
  \citenamefont {Ozdemir}, \citenamefont {Xiao}, \citenamefont {Li},
  \citenamefont {He}, \citenamefont {Chen},\ and\ \citenamefont
  {Yang}}]{Zhu2010}%
  \BibitemOpen
  \bibfield  {author} {\bibinfo {author} {\bibfnamefont {J.}~\bibnamefont
  {Zhu}}, \bibinfo {author} {\bibfnamefont {S.~K.}\ \bibnamefont {Ozdemir}},
  \bibinfo {author} {\bibfnamefont {Y.~F.}\ \bibnamefont {Xiao}}, \bibinfo
  {author} {\bibfnamefont {L.}~\bibnamefont {Li}}, \bibinfo {author}
  {\bibfnamefont {L.}~\bibnamefont {He}}, \bibinfo {author} {\bibfnamefont
  {D.~R.}\ \bibnamefont {Chen}}, \ and\ \bibinfo {author} {\bibfnamefont
  {L.}~\bibnamefont {Yang}},\ }\href {\doibase 10.1038/nphoton.2009.237}
  {\bibfield  {journal} {\bibinfo  {journal} {Nature Photonics}\ }\textbf
  {\bibinfo {volume} {4}},\ \bibinfo {pages} {46} (\bibinfo {year}
  {2010})}\BibitemShut {NoStop}%
\bibitem [{\citenamefont {Spillane}\ \emph {et~al.}(2005)\citenamefont
  {Spillane}, \citenamefont {Kippenberg}, \citenamefont {Vahala}, \citenamefont
  {Goh}, \citenamefont {Wilcut},\ and\ \citenamefont {Kimble}}]{Spillane2005}%
  \BibitemOpen
  \bibfield  {author} {\bibinfo {author} {\bibfnamefont {S.~M.}\ \bibnamefont
  {Spillane}}, \bibinfo {author} {\bibfnamefont {T.~J.}\ \bibnamefont
  {Kippenberg}}, \bibinfo {author} {\bibfnamefont {K.~J.}\ \bibnamefont
  {Vahala}}, \bibinfo {author} {\bibfnamefont {K.~W.}\ \bibnamefont {Goh}},
  \bibinfo {author} {\bibfnamefont {E.}~\bibnamefont {Wilcut}}, \ and\ \bibinfo
  {author} {\bibfnamefont {H.~J.}\ \bibnamefont {Kimble}},\ }\href@noop {}
  {\bibfield  {journal} {\bibinfo  {journal} {Physical Review A}\ }\textbf
  {\bibinfo {volume} {71}},\ \bibinfo {pages} {013817} (\bibinfo {year}
  {2005})}\BibitemShut {NoStop}%
\bibitem [{\citenamefont {Asano}\ and\ \citenamefont
  {Noda}(2018)}]{Asano2018b}%
  \BibitemOpen
  \bibfield  {author} {\bibinfo {author} {\bibfnamefont {T.}~\bibnamefont
  {Asano}}\ and\ \bibinfo {author} {\bibfnamefont {S.}~\bibnamefont {Noda}},\
  }\href {\doibase 10.1364/oe.26.032704} {\bibfield  {journal} {\bibinfo
  {journal} {Optics Express}\ }\textbf {\bibinfo {volume} {26}},\ \bibinfo
  {pages} {32704} (\bibinfo {year} {2018})}\BibitemShut {NoStop}%
\bibitem [{\citenamefont {Alpeggiani}\ \emph {et~al.}(2015)\citenamefont
  {Alpeggiani}, \citenamefont {Andreani},\ and\ \citenamefont
  {Gerace}}]{Alpeggiani2015}%
  \BibitemOpen
  \bibfield  {author} {\bibinfo {author} {\bibfnamefont {F.}~\bibnamefont
  {Alpeggiani}}, \bibinfo {author} {\bibfnamefont {L.~C.}\ \bibnamefont
  {Andreani}}, \ and\ \bibinfo {author} {\bibfnamefont {D.}~\bibnamefont
  {Gerace}},\ }\href {\doibase 10.1063/1.4938395} {\bibfield  {journal}
  {\bibinfo  {journal} {Applied Physics Letters}\ }\textbf {\bibinfo {volume}
  {107}},\ \bibinfo {pages} {261110} (\bibinfo {year} {2015})}\BibitemShut
  {NoStop}%
\bibitem [{\citenamefont {Quan}\ and\ \citenamefont {Loncar}(2011)}]{Quan2011}%
  \BibitemOpen
  \bibfield  {author} {\bibinfo {author} {\bibfnamefont {Q.}~\bibnamefont
  {Quan}}\ and\ \bibinfo {author} {\bibfnamefont {M.}~\bibnamefont {Loncar}},\
  }\href {\doibase 10.1364/OE.19.018529} {\bibfield  {journal} {\bibinfo
  {journal} {Optics Express}\ }\textbf {\bibinfo {volume} {19}},\ \bibinfo
  {pages} {18529} (\bibinfo {year} {2011})}\BibitemShut {NoStop}%
\bibitem [{\citenamefont {Hu}\ and\ \citenamefont {Weiss}(2016)}]{Hu2016}%
  \BibitemOpen
  \bibfield  {author} {\bibinfo {author} {\bibfnamefont {S.}~\bibnamefont
  {Hu}}\ and\ \bibinfo {author} {\bibfnamefont {S.~M.}\ \bibnamefont {Weiss}},\
  }\href {\doibase 10.1021/acsphotonics.6b00219} {\bibfield  {journal}
  {\bibinfo  {journal} {ACS Photonics}\ }\textbf {\bibinfo {volume} {3}},\
  \bibinfo {pages} {1647} (\bibinfo {year} {2016})}\BibitemShut {NoStop}%
\bibitem [{\citenamefont {Lee}\ \emph {et~al.}(2012)\citenamefont {Lee},
  \citenamefont {Chen}, \citenamefont {Li}, \citenamefont {Yang}, \citenamefont
  {Jeon}, \citenamefont {Painter},\ and\ \citenamefont {Vahala}}]{Lee2012}%
  \BibitemOpen
  \bibfield  {author} {\bibinfo {author} {\bibfnamefont {H.}~\bibnamefont
  {Lee}}, \bibinfo {author} {\bibfnamefont {T.}~\bibnamefont {Chen}}, \bibinfo
  {author} {\bibfnamefont {J.}~\bibnamefont {Li}}, \bibinfo {author}
  {\bibfnamefont {K.~Y.}\ \bibnamefont {Yang}}, \bibinfo {author}
  {\bibfnamefont {S.}~\bibnamefont {Jeon}}, \bibinfo {author} {\bibfnamefont
  {O.}~\bibnamefont {Painter}}, \ and\ \bibinfo {author} {\bibfnamefont
  {K.~J.}\ \bibnamefont {Vahala}},\ }\href {\doibase 10.1038/nphoton.2012.109}
  {\bibfield  {journal} {\bibinfo  {journal} {Nature Photonics}\ }\textbf
  {\bibinfo {volume} {6}},\ \bibinfo {pages} {369} (\bibinfo {year}
  {2012})}\BibitemShut {NoStop}%
\bibitem [{\citenamefont {Yang}\ \emph {et~al.}(2018)\citenamefont {Yang},
  \citenamefont {Oh}, \citenamefont {Lee}, \citenamefont {Yang}, \citenamefont
  {Yi}, \citenamefont {Shen}, \citenamefont {Wang},\ and\ \citenamefont
  {Vahala}}]{YoulYang2018}%
  \BibitemOpen
  \bibfield  {author} {\bibinfo {author} {\bibfnamefont {K.~Y.}\ \bibnamefont
  {Yang}}, \bibinfo {author} {\bibfnamefont {D.~Y.}\ \bibnamefont {Oh}},
  \bibinfo {author} {\bibfnamefont {S.~H.}\ \bibnamefont {Lee}}, \bibinfo
  {author} {\bibfnamefont {Q.~F.}\ \bibnamefont {Yang}}, \bibinfo {author}
  {\bibfnamefont {X.}~\bibnamefont {Yi}}, \bibinfo {author} {\bibfnamefont
  {B.}~\bibnamefont {Shen}}, \bibinfo {author} {\bibfnamefont {H.}~\bibnamefont
  {Wang}}, \ and\ \bibinfo {author} {\bibfnamefont {K.}~\bibnamefont
  {Vahala}},\ }\href {\doibase 10.1038/s41566-018-0132-5} {\bibfield  {journal}
  {\bibinfo  {journal} {Nature Photonics}\ }\textbf {\bibinfo {volume} {12}},\
  \bibinfo {pages} {297} (\bibinfo {year} {2018})}\BibitemShut {NoStop}%
\bibitem [{\citenamefont {Wang}\ \emph {et~al.}(2018)\citenamefont {Wang},
  \citenamefont {Christiansen}, \citenamefont {Yu}, \citenamefont {M{\o}rk},\
  and\ \citenamefont {Sigmund}}]{Wang2018}%
  \BibitemOpen
  \bibfield  {author} {\bibinfo {author} {\bibfnamefont {F.}~\bibnamefont
  {Wang}}, \bibinfo {author} {\bibfnamefont {R.~E.}\ \bibnamefont
  {Christiansen}}, \bibinfo {author} {\bibfnamefont {Y.}~\bibnamefont {Yu}},
  \bibinfo {author} {\bibfnamefont {J.}~\bibnamefont {M{\o}rk}}, \ and\
  \bibinfo {author} {\bibfnamefont {O.}~\bibnamefont {Sigmund}},\ }\href
  {https://doi.org/10.1063/1.5064468} {\bibfield  {journal} {\bibinfo
  {journal} {Applied Physics Letters}\ }\textbf {\bibinfo {volume} {113}},\
  \bibinfo {pages} {241101} (\bibinfo {year} {2018})}\BibitemShut {NoStop}%
\bibitem [{\citenamefont {Heuck}\ \emph {et~al.}(2020)\citenamefont {Heuck},
  \citenamefont {Jacobs},\ and\ \citenamefont {Englund}}]{Heuck2020}%
  \BibitemOpen
  \bibfield  {author} {\bibinfo {author} {\bibfnamefont {M.}~\bibnamefont
  {Heuck}}, \bibinfo {author} {\bibfnamefont {K.}~\bibnamefont {Jacobs}}, \
  and\ \bibinfo {author} {\bibfnamefont {D.~R.}\ \bibnamefont {Englund}},\
  }\href {https://link.aps.org/doi/10.1103/PhysRevLett.124.160501} {\bibfield
  {journal} {\bibinfo  {journal} {Physical Review Letters}\ }\textbf {\bibinfo
  {volume} {124}},\ \bibinfo {pages} {160501} (\bibinfo {year}
  {2020})}\BibitemShut {NoStop}%
\bibitem [{\citenamefont {Krastanov}\ \emph {et~al.}(2020)\citenamefont
  {Krastanov}, \citenamefont {Heuck}, \citenamefont {Shapiro}, \citenamefont
  {Narang}, \citenamefont {Englund},\ and\ \citenamefont
  {Jacobs}}]{Krastanov2020}%
  \BibitemOpen
  \bibfield  {author} {\bibinfo {author} {\bibfnamefont {S.}~\bibnamefont
  {Krastanov}}, \bibinfo {author} {\bibfnamefont {M.}~\bibnamefont {Heuck}},
  \bibinfo {author} {\bibfnamefont {J.~H.}\ \bibnamefont {Shapiro}}, \bibinfo
  {author} {\bibfnamefont {P.}~\bibnamefont {Narang}}, \bibinfo {author}
  {\bibfnamefont {D.~R.}\ \bibnamefont {Englund}}, \ and\ \bibinfo {author}
  {\bibfnamefont {K.}~\bibnamefont {Jacobs}},\ }\href
  {http://arxiv.org/abs/2002.07193} {\  (\bibinfo {year} {2020})},\ \Eprint
  {http://arxiv.org/abs/2002.07193} {arXiv:2002.07193} \BibitemShut {NoStop}%
\bibitem [{\citenamefont {Timurdogan}\ \emph {et~al.}(2017)\citenamefont
  {Timurdogan}, \citenamefont {Poulton}, \citenamefont {Byrd},\ and\
  \citenamefont {Watts}}]{Timurdogan2017}%
  \BibitemOpen
  \bibfield  {author} {\bibinfo {author} {\bibfnamefont {E.}~\bibnamefont
  {Timurdogan}}, \bibinfo {author} {\bibfnamefont {C.~V.}\ \bibnamefont
  {Poulton}}, \bibinfo {author} {\bibfnamefont {M.~J.}\ \bibnamefont {Byrd}}, \
  and\ \bibinfo {author} {\bibfnamefont {M.~R.}\ \bibnamefont {Watts}},\ }\href
  {\doibase 10.1038/nphoton.2017.14} {\bibfield  {journal} {\bibinfo  {journal}
  {Nature Photonics}\ }\textbf {\bibinfo {volume} {11}},\ \bibinfo {pages}
  {200} (\bibinfo {year} {2017})}\BibitemShut {NoStop}%
\bibitem [{\citenamefont {Minkov}\ \emph {et~al.}(2019)\citenamefont {Minkov},
  \citenamefont {Gerace},\ and\ \citenamefont {Fan}}]{Minkov2019}%
  \BibitemOpen
  \bibfield  {author} {\bibinfo {author} {\bibfnamefont {M.}~\bibnamefont
  {Minkov}}, \bibinfo {author} {\bibfnamefont {D.}~\bibnamefont {Gerace}}, \
  and\ \bibinfo {author} {\bibfnamefont {S.}~\bibnamefont {Fan}},\ }\href
  {\doibase 10.1364/optica.6.001039} {\bibfield  {journal} {\bibinfo  {journal}
  {Optica}\ }\textbf {\bibinfo {volume} {6}},\ \bibinfo {pages} {1039}
  (\bibinfo {year} {2019})}\BibitemShut {NoStop}%
\end{thebibliography}%


\begin{thebibliography}{33}%
\makeatletter
\providecommand \@ifxundefined [1]{%
 \@ifx{#1\undefined}
}%
\providecommand \@ifnum [1]{%
 \ifnum #1\expandafter \@firstoftwo
 \else \expandafter \@secondoftwo
 \fi
}%
\providecommand \@ifx [1]{%
 \ifx #1\expandafter \@firstoftwo
 \else \expandafter \@secondoftwo
 \fi
}%
\providecommand \natexlab [1]{#1}%
\providecommand \enquote  [1]{``#1''}%
\providecommand \bibnamefont  [1]{#1}%
\providecommand \bibfnamefont [1]{#1}%
\providecommand \citenamefont [1]{#1}%
\providecommand \href@noop [0]{\@secondoftwo}%
\providecommand \href [0]{\begingroup \@sanitize@url \@href}%
\providecommand \@href[1]{\@@startlink{#1}\@@href}%
\providecommand \@@href[1]{\endgroup#1\@@endlink}%
\providecommand \@sanitize@url [0]{\catcode `\\12\catcode `\$12\catcode
  `\&12\catcode `\#12\catcode `\^12\catcode `\_12\catcode `\%12\relax}%
\providecommand \@@startlink[1]{}%
\providecommand \@@endlink[0]{}%
\providecommand \url  [0]{\begingroup\@sanitize@url \@url }%
\providecommand \@url [1]{\endgroup\@href {#1}{\urlprefix }}%
\providecommand \urlprefix  [0]{URL }%
\providecommand \Eprint [0]{\href }%
\providecommand \doibase [0]{https://doi.org/}%
\providecommand \selectlanguage [0]{\@gobble}%
\providecommand \bibinfo  [0]{\@secondoftwo}%
\providecommand \bibfield  [0]{\@secondoftwo}%
\providecommand \translation [1]{[#1]}%
\providecommand \BibitemOpen [0]{}%
\providecommand \bibitemStop [0]{}%
\providecommand \bibitemNoStop [0]{.\EOS\space}%
\providecommand \EOS [0]{\spacefactor3000\relax}%
\providecommand \BibitemShut  [1]{\csname bibitem#1\endcsname}%
\let\auto@bib@innerbib\@empty
\bibitem [{\citenamefont {Landau}\ and\ \citenamefont
  {Lifshitz}(1980)}]{Land1980}%
  \BibitemOpen
  \bibfield  {author} {\bibinfo {author} {\bibfnamefont {L.~D.}\ \bibnamefont
  {Landau}}\ and\ \bibinfo {author} {\bibfnamefont {E.~M.}\ \bibnamefont
  {Lifshitz}},\ }\href {https://doi.org/10.1016/C2009-0-24487-4} {\emph
  {\bibinfo {title} {{Statistical Physics}}}}\ (\bibinfo  {publisher}
  {Elsevier},\ \bibinfo {year} {1980})\BibitemShut {NoStop}%
\bibitem [{\citenamefont {Braginsky}\ \emph {et~al.}(1999)\citenamefont
  {Braginsky}, \citenamefont {Gorodetsky},\ and\ \citenamefont
  {Vyatchanin}}]{Braginsky1999}%
  \BibitemOpen
  \bibfield  {author} {\bibinfo {author} {\bibfnamefont {V.~B.}\ \bibnamefont
  {Braginsky}}, \bibinfo {author} {\bibfnamefont {M.~L.}\ \bibnamefont
  {Gorodetsky}},\ and\ \bibinfo {author} {\bibfnamefont {S.~P.}\ \bibnamefont
  {Vyatchanin}},\ }\bibfield  {title} {\bibinfo {title} {{Thermodynamical
  fluctuations and photo-thermal shot noise in gravitational wave antennae}},\
  }\href {https://doi.org/10.1016/S0375-9601(99)00785-9} {\bibfield  {journal}
  {\bibinfo  {journal} {Physics Letters, Section A: General, Atomic and Solid
  State Physics}\ }\textbf {\bibinfo {volume} {264}},\ \bibinfo {pages} {1}
  (\bibinfo {year} {1999})}\BibitemShut {NoStop}%
\bibitem [{Note1()}]{Note1}%
  \BibitemOpen
  \bibinfo {note} {The implications of this definition are especially important
  for the ultra-small move volume cavities proposed in recent studies \cite
  {Robinson2005a, Choi, Hu2016}. In these cavities, we found that $V_\protect
  \text {eff}^{(2)} < V_\protect \text {eff}$ \cite {Panuski2019}, implying
  that thermodynamic fluctuations in these cavities are significantly
  suppressed.}\BibitemShut {Stop}%
\bibitem [{\citenamefont {Gorodetsky}\ and\ \citenamefont
  {Grudinin}(2004)}]{Gorodetsky2004}%
  \BibitemOpen
  \bibfield  {author} {\bibinfo {author} {\bibfnamefont {M.~L.}\ \bibnamefont
  {Gorodetsky}}\ and\ \bibinfo {author} {\bibfnamefont {I.~S.}\ \bibnamefont
  {Grudinin}},\ }\bibfield  {title} {\bibinfo {title} {{Fundamental thermal
  fluctuations in microspheres}},\ }\href
  {https://doi.org/10.1364/JOSAB.21.000697} {\bibfield  {journal} {\bibinfo
  {journal} {Journal of the Optical Society of America B}\ }\textbf {\bibinfo
  {volume} {21}},\ \bibinfo {pages} {697} (\bibinfo {year} {2004})}\BibitemShut
  {NoStop}%
\bibitem [{\citenamefont {Braginsky}\ \emph {et~al.}(2000)\citenamefont
  {Braginsky}, \citenamefont {Gorodetsky},\ and\ \citenamefont
  {Vyatchanin}}]{Braginsky2000}%
  \BibitemOpen
  \bibfield  {author} {\bibinfo {author} {\bibfnamefont {V.~B.}\ \bibnamefont
  {Braginsky}}, \bibinfo {author} {\bibfnamefont {M.~L.}\ \bibnamefont
  {Gorodetsky}},\ and\ \bibinfo {author} {\bibfnamefont {S.~P.}\ \bibnamefont
  {Vyatchanin}},\ }\bibfield  {title} {\bibinfo {title} {{Thermo-refractive
  noise in gravitational wave antennae}},\ }\href
  {https://doi.org/10.1016/S0375-9601(00)00389-3} {\bibfield  {journal}
  {\bibinfo  {journal} {Physics Letters, Section A: General, Atomic and Solid
  State Physics}\ }\textbf {\bibinfo {volume} {271}},\ \bibinfo {pages} {303}
  (\bibinfo {year} {2000})}\BibitemShut {NoStop}%
\bibitem [{\citenamefont {{The LIGO Scientific
  Collaboration}}(2016)}]{Martynov2016}%
  \BibitemOpen
  \bibfield  {author} {\bibinfo {author} {\bibnamefont {{The LIGO Scientific
  Collaboration}}},\ }\bibfield  {title} {\bibinfo {title} {{Sensitivity of the
  Advanced LIGO detectors at the beginning of gravitational wave astronomy}},\
  }\href@noop {} {\bibfield  {journal} {\bibinfo  {journal} {Physical Review
  D}\ }\textbf {\bibinfo {volume} {93}},\ \bibinfo {pages} {112004} (\bibinfo
  {year} {2016})}\BibitemShut {NoStop}%
\bibitem [{\citenamefont {Lim}\ \emph {et~al.}(2017)\citenamefont {Lim},
  \citenamefont {Savchenkov}, \citenamefont {Dale}, \citenamefont {Liang},
  \citenamefont {Eliyahu}, \citenamefont {Ilchenko}, \citenamefont {Matsko},
  \citenamefont {Maleki},\ and\ \citenamefont {Wong}}]{Lim}%
  \BibitemOpen
  \bibfield  {author} {\bibinfo {author} {\bibfnamefont {J.}~\bibnamefont
  {Lim}}, \bibinfo {author} {\bibfnamefont {A.~A.}\ \bibnamefont {Savchenkov}},
  \bibinfo {author} {\bibfnamefont {E.}~\bibnamefont {Dale}}, \bibinfo {author}
  {\bibfnamefont {W.}~\bibnamefont {Liang}}, \bibinfo {author} {\bibfnamefont
  {D.}~\bibnamefont {Eliyahu}}, \bibinfo {author} {\bibfnamefont
  {V.}~\bibnamefont {Ilchenko}}, \bibinfo {author} {\bibfnamefont {A.~B.}\
  \bibnamefont {Matsko}}, \bibinfo {author} {\bibfnamefont {L.}~\bibnamefont
  {Maleki}},\ and\ \bibinfo {author} {\bibfnamefont {C.~W.}\ \bibnamefont
  {Wong}},\ }\bibfield  {title} {\bibinfo {title} {{Chasing the thermodynamical
  noise limit in whispering-gallery-mode resonators for ultrastable laser
  frequency stabilization}},\ }\href {www.nature.com/naturecommunications}
  {\bibfield  {journal} {\bibinfo  {journal} {Nature Communications}\ }\textbf
  {\bibinfo {volume} {8}} (\bibinfo {year} {2017})}\BibitemShut {NoStop}%
\bibitem [{\citenamefont {Zhang}\ \emph {et~al.}(2019)\citenamefont {Zhang},
  \citenamefont {Baynes}, \citenamefont {Diddams},\ and\ \citenamefont
  {Papp}}]{Zhang2019}%
  \BibitemOpen
  \bibfield  {author} {\bibinfo {author} {\bibfnamefont {W.}~\bibnamefont
  {Zhang}}, \bibinfo {author} {\bibfnamefont {F.}~\bibnamefont {Baynes}},
  \bibinfo {author} {\bibfnamefont {S.~A.}\ \bibnamefont {Diddams}},\ and\
  \bibinfo {author} {\bibfnamefont {S.~B.}\ \bibnamefont {Papp}},\ }\bibfield
  {title} {\bibinfo {title} {{Microrod Optical Frequency Reference in the
  Ambient Environment}},\ }\href
  {https://doi.org/10.1103/PhysRevApplied.12.024010} {\bibfield  {journal}
  {\bibinfo  {journal} {Physical Review Applied}\ }\textbf {\bibinfo {volume}
  {12}},\ \bibinfo {pages} {1} (\bibinfo {year} {2019})}\BibitemShut {NoStop}%
\bibitem [{\citenamefont {Feynman}\ and\ \citenamefont
  {Hibbs}(1965)}]{Feynman1965}%
  \BibitemOpen
  \bibfield  {author} {\bibinfo {author} {\bibfnamefont {R.~P.}\ \bibnamefont
  {Feynman}}\ and\ \bibinfo {author} {\bibfnamefont {A.~R.}\ \bibnamefont
  {Hibbs}},\ }\href@noop {} {\emph {\bibinfo {title} {{Quantum Mechanics and
  Path Integrals}}}}\ (\bibinfo  {publisher} {McGraw-Hill},\ \bibinfo {address}
  {New York},\ \bibinfo {year} {1965})\BibitemShut {NoStop}%
\bibitem [{\citenamefont {Zhang}\ \emph {et~al.}(2014)\citenamefont {Zhang},
  \citenamefont {Moser}, \citenamefont {G{\"{u}}ttinger}, \citenamefont
  {Bachtold},\ and\ \citenamefont {Dykman}}]{Zhang2014a}%
  \BibitemOpen
  \bibfield  {author} {\bibinfo {author} {\bibfnamefont {Y.}~\bibnamefont
  {Zhang}}, \bibinfo {author} {\bibfnamefont {J.}~\bibnamefont {Moser}},
  \bibinfo {author} {\bibfnamefont {J.}~\bibnamefont {G{\"{u}}ttinger}},
  \bibinfo {author} {\bibfnamefont {A.}~\bibnamefont {Bachtold}},\ and\
  \bibinfo {author} {\bibfnamefont {M.~I.}\ \bibnamefont {Dykman}},\ }\bibfield
   {title} {\bibinfo {title} {{Interplay of driving and frequency noise in the
  spectra of vibrational systems}},\ }\href
  {https://journals.aps.org/prl/pdf/10.1103/PhysRevLett.113.255502} {\bibfield
  {journal} {\bibinfo  {journal} {Physical Review Letters}\ }\textbf {\bibinfo
  {volume} {113}},\ \bibinfo {pages} {255502} (\bibinfo {year}
  {2014})}\BibitemShut {NoStop}%
\bibitem [{\citenamefont {Gardiner}(1984)}]{Gardiner1984}%
  \BibitemOpen
  \bibfield  {author} {\bibinfo {author} {\bibfnamefont {C.~W.}\ \bibnamefont
  {Gardiner}},\ }\bibfield  {title} {\bibinfo {title} {{Adiabatic elimination
  in stochastic systems. I. Formulation of methods and application to
  few-variable systems}},\ }\href {https://doi.org/10.1103/PhysRevA.29.2814}
  {\bibfield  {journal} {\bibinfo  {journal} {Physical Review A}\ }\textbf
  {\bibinfo {volume} {29}},\ \bibinfo {pages} {2814} (\bibinfo {year}
  {1984})}\BibitemShut {NoStop}%
\bibitem [{\citenamefont {Chui}\ \emph {et~al.}(1992)\citenamefont {Chui},
  \citenamefont {Swanson}, \citenamefont {Adriaans}, \citenamefont {Nissen},\
  and\ \citenamefont {Lipa}}]{Chui1992}%
  \BibitemOpen
  \bibfield  {author} {\bibinfo {author} {\bibfnamefont {T.~C.}\ \bibnamefont
  {Chui}}, \bibinfo {author} {\bibfnamefont {D.~R.}\ \bibnamefont {Swanson}},
  \bibinfo {author} {\bibfnamefont {M.~J.}\ \bibnamefont {Adriaans}}, \bibinfo
  {author} {\bibfnamefont {J.~A.}\ \bibnamefont {Nissen}},\ and\ \bibinfo
  {author} {\bibfnamefont {J.~A.}\ \bibnamefont {Lipa}},\ }\bibfield  {title}
  {\bibinfo {title} {{Temperature fluctuations in the canonical ensemble}},\
  }\href {https://doi.org/10.1103/PhysRevLett.69.3005} {\bibfield  {journal}
  {\bibinfo  {journal} {Physical Review Letters}\ }\textbf {\bibinfo {volume}
  {69}},\ \bibinfo {pages} {3005} (\bibinfo {year} {1992})}\BibitemShut
  {NoStop}%
\bibitem [{\citenamefont {Sun}\ \emph {et~al.}(2017)\citenamefont {Sun},
  \citenamefont {Luo}, \citenamefont {Zhang},\ and\ \citenamefont
  {Lin}}]{Sun2017}%
  \BibitemOpen
  \bibfield  {author} {\bibinfo {author} {\bibfnamefont {X.}~\bibnamefont
  {Sun}}, \bibinfo {author} {\bibfnamefont {R.}~\bibnamefont {Luo}}, \bibinfo
  {author} {\bibfnamefont {X.~C.}\ \bibnamefont {Zhang}},\ and\ \bibinfo
  {author} {\bibfnamefont {Q.}~\bibnamefont {Lin}},\ }\bibfield  {title}
  {\bibinfo {title} {{Squeezing the fundamental temperature fluctuations of a
  high- Q microresonator}},\ }\href
  {https://doi.org/10.1103/PhysRevA.95.023822} {\bibfield  {journal} {\bibinfo
  {journal} {Physical Review A}\ }\textbf {\bibinfo {volume} {95}},\ \bibinfo
  {pages} {23822} (\bibinfo {year} {2017})}\BibitemShut {NoStop}%
\bibitem [{\citenamefont {Cuffe}\ \emph {et~al.}(2015)\citenamefont {Cuffe},
  \citenamefont {Eliason}, \citenamefont {Maznev}, \citenamefont {Collins},
  \citenamefont {Johnson}, \citenamefont {Shchepetov}, \citenamefont
  {Prunnila}, \citenamefont {Ahopelto}, \citenamefont {{Sotomayor Torres}},
  \citenamefont {Chen},\ and\ \citenamefont {Nelson}}]{Cuffe2015}%
  \BibitemOpen
  \bibfield  {author} {\bibinfo {author} {\bibfnamefont {J.}~\bibnamefont
  {Cuffe}}, \bibinfo {author} {\bibfnamefont {J.~K.}\ \bibnamefont {Eliason}},
  \bibinfo {author} {\bibfnamefont {A.~A.}\ \bibnamefont {Maznev}}, \bibinfo
  {author} {\bibfnamefont {K.~C.}\ \bibnamefont {Collins}}, \bibinfo {author}
  {\bibfnamefont {J.~A.}\ \bibnamefont {Johnson}}, \bibinfo {author}
  {\bibfnamefont {A.}~\bibnamefont {Shchepetov}}, \bibinfo {author}
  {\bibfnamefont {M.}~\bibnamefont {Prunnila}}, \bibinfo {author}
  {\bibfnamefont {J.}~\bibnamefont {Ahopelto}}, \bibinfo {author}
  {\bibfnamefont {C.~M.}\ \bibnamefont {{Sotomayor Torres}}}, \bibinfo {author}
  {\bibfnamefont {G.}~\bibnamefont {Chen}},\ and\ \bibinfo {author}
  {\bibfnamefont {K.~A.}\ \bibnamefont {Nelson}},\ }\bibfield  {title}
  {\bibinfo {title} {{Reconstructing phonon mean-free-path contributions to
  thermal conductivity using nanoscale membranes}},\ }\href
  {https://doi.org/10.1103/PhysRevB.91.245423} {\bibfield  {journal} {\bibinfo
  {journal} {Physical Review B}\ }\textbf {\bibinfo {volume} {91}},\ \bibinfo
  {pages} {245423} (\bibinfo {year} {2015})}\BibitemShut {NoStop}%
\bibitem [{\citenamefont {Minkov}\ and\ \citenamefont
  {Savona}(2014)}]{Minkov2014}%
  \BibitemOpen
  \bibfield  {author} {\bibinfo {author} {\bibfnamefont {M.}~\bibnamefont
  {Minkov}}\ and\ \bibinfo {author} {\bibfnamefont {V.}~\bibnamefont
  {Savona}},\ }\bibfield  {title} {\bibinfo {title} {{Automated optimization of
  photonic crystal slab cavities}},\ }\href {www.nature.com/scientificreports}
  {\bibfield  {journal} {\bibinfo  {journal} {Scientific Reports}\ }\textbf
  {\bibinfo {volume} {4}},\ \bibinfo {pages} {5124} (\bibinfo {year}
  {2014})}\BibitemShut {NoStop}%
\bibitem [{\citenamefont {Minkov}\ \emph {et~al.}(2017)\citenamefont {Minkov},
  \citenamefont {Savona},\ and\ \citenamefont {Gerace}}]{Minkov2017}%
  \BibitemOpen
  \bibfield  {author} {\bibinfo {author} {\bibfnamefont {M.}~\bibnamefont
  {Minkov}}, \bibinfo {author} {\bibfnamefont {V.}~\bibnamefont {Savona}},\
  and\ \bibinfo {author} {\bibfnamefont {D.}~\bibnamefont {Gerace}},\
  }\bibfield  {title} {\bibinfo {title} {{Photonic crystal slab cavity
  simultaneously optimized for ultra-high Q/V and vertical radiation
  coupling}},\ }\href {https://doi.org/10.1063/1.4991416} {\bibfield  {journal}
  {\bibinfo  {journal} {Applied Physics Letters}\ }\textbf {\bibinfo {volume}
  {111}},\ \bibinfo {pages} {131104} (\bibinfo {year} {2017})}\BibitemShut
  {NoStop}%
\bibitem [{\citenamefont {Schliesser}\ \emph {et~al.}(2008)\citenamefont
  {Schliesser}, \citenamefont {Rivi{\`{e}}re}, \citenamefont {Anetsberger},
  \citenamefont {Arcizet},\ and\ \citenamefont {Kippenberg}}]{Schliesser2008}%
  \BibitemOpen
  \bibfield  {author} {\bibinfo {author} {\bibfnamefont {A.}~\bibnamefont
  {Schliesser}}, \bibinfo {author} {\bibfnamefont {R.}~\bibnamefont
  {Rivi{\`{e}}re}}, \bibinfo {author} {\bibfnamefont {G.}~\bibnamefont
  {Anetsberger}}, \bibinfo {author} {\bibfnamefont {O.}~\bibnamefont
  {Arcizet}},\ and\ \bibinfo {author} {\bibfnamefont {T.~J.}\ \bibnamefont
  {Kippenberg}},\ }\bibfield  {title} {\bibinfo {title} {{Resolved-sideband
  cooling of a micromechanical oscillator}},\ }\href
  {https://doi.org/10.1038/nphys939} {\bibfield  {journal} {\bibinfo  {journal}
  {Nature Physics}\ }\textbf {\bibinfo {volume} {4}},\ \bibinfo {pages} {415}
  (\bibinfo {year} {2008})}\BibitemShut {NoStop}%
\bibitem [{\citenamefont {Gorodetsky}\ \emph {et~al.}(2010)\citenamefont
  {Gorodetsky}, \citenamefont {Schliesser}, \citenamefont {Anetsberger},
  \citenamefont {Deleglise},\ and\ \citenamefont
  {Kippenberg}}]{Gorodetsky2010}%
  \BibitemOpen
  \bibfield  {author} {\bibinfo {author} {\bibfnamefont {M.~L.}\ \bibnamefont
  {Gorodetsky}}, \bibinfo {author} {\bibfnamefont {A.}~\bibnamefont
  {Schliesser}}, \bibinfo {author} {\bibfnamefont {G.}~\bibnamefont
  {Anetsberger}}, \bibinfo {author} {\bibfnamefont {S.}~\bibnamefont
  {Deleglise}},\ and\ \bibinfo {author} {\bibfnamefont {T.~J.}\ \bibnamefont
  {Kippenberg}},\ }\bibfield  {title} {\bibinfo {title} {{Determination of the
  vacuum optomechanical coupling rate using frequency noise calibration}},\
  }\href {https://doi.org/10.1364/OE.18.023236} {\bibfield  {journal} {\bibinfo
   {journal} {Optics Express}\ }\textbf {\bibinfo {volume} {18}},\ \bibinfo
  {pages} {23236} (\bibinfo {year} {2010})}\BibitemShut {NoStop}%
\bibitem [{\citenamefont {{Keysight
  Technologies}}(2017)}]{KeysightTechnologiesA}%
  \BibitemOpen
  \bibfield  {author} {\bibinfo {author} {\bibnamefont {{Keysight
  Technologies}}},\ }\bibfield  {title} {\bibinfo {title} {{Spectrum and Signal
  Analyzer Measurements and Noise}},\ }\href
  {https://literature.cdn.keysight.com/litweb/pdf/5966-4008E.pdf} {\bibfield
  {journal} {\bibinfo  {journal} {Application Note 5966}\ } (\bibinfo {year}
  {2017})}\BibitemShut {NoStop}%
\bibitem [{\citenamefont {Ludvigsen}\ \emph {et~al.}(1998)\citenamefont
  {Ludvigsen}, \citenamefont {Tossavainen},\ and\ \citenamefont
  {Kaivola}}]{Ludvigsen1998}%
  \BibitemOpen
  \bibfield  {author} {\bibinfo {author} {\bibfnamefont {H.}~\bibnamefont
  {Ludvigsen}}, \bibinfo {author} {\bibfnamefont {M.}~\bibnamefont
  {Tossavainen}},\ and\ \bibinfo {author} {\bibfnamefont {M.}~\bibnamefont
  {Kaivola}},\ }\bibfield  {title} {\bibinfo {title} {{Laser linewidth
  measurements using self-homodyne detection with short delay}},\ }\href
  {https://doi.org/10.1016/S0030-4018(98)00355-1} {\bibfield  {journal}
  {\bibinfo  {journal} {Optics Communications}\ }\textbf {\bibinfo {volume}
  {155}},\ \bibinfo {pages} {180} (\bibinfo {year} {1998})}\BibitemShut
  {NoStop}%
\bibitem [{\citenamefont {Jain}\ \emph {et~al.}(2013)\citenamefont {Jain},
  \citenamefont {Yu},\ and\ \citenamefont {McGaughey}}]{Jain2013}%
  \BibitemOpen
  \bibfield  {author} {\bibinfo {author} {\bibfnamefont {A.}~\bibnamefont
  {Jain}}, \bibinfo {author} {\bibfnamefont {Y.~J.}\ \bibnamefont {Yu}},\ and\
  \bibinfo {author} {\bibfnamefont {A.~J.}\ \bibnamefont {McGaughey}},\
  }\bibfield  {title} {\bibinfo {title} {{Phonon transport in periodic silicon
  nanoporous films with feature sizes greater than 100 nm}},\ }\href
  {https://doi.org/10.1103/PhysRevB.87.195301} {\bibfield  {journal} {\bibinfo
  {journal} {Physical Review B}\ }\textbf {\bibinfo {volume} {87}},\ \bibinfo
  {pages} {195301} (\bibinfo {year} {2013})}\BibitemShut {NoStop}%
\bibitem [{\citenamefont {Komma}\ \emph {et~al.}(2012)\citenamefont {Komma},
  \citenamefont {Schwarz}, \citenamefont {Hofmann}, \citenamefont {Heinert},\
  and\ \citenamefont {Nawrodt}}]{Komma2012}%
  \BibitemOpen
  \bibfield  {author} {\bibinfo {author} {\bibfnamefont {J.}~\bibnamefont
  {Komma}}, \bibinfo {author} {\bibfnamefont {C.}~\bibnamefont {Schwarz}},
  \bibinfo {author} {\bibfnamefont {G.}~\bibnamefont {Hofmann}}, \bibinfo
  {author} {\bibfnamefont {D.}~\bibnamefont {Heinert}},\ and\ \bibinfo {author}
  {\bibfnamefont {R.}~\bibnamefont {Nawrodt}},\ }\bibfield  {title} {\bibinfo
  {title} {{Thermo-optic coefficient of silicon at 1550 nm and cryogenic
  temperatures}},\ }\href {https://doi.org/10.1063/1.4738989} {\bibfield
  {journal} {\bibinfo  {journal} {Applied Physics Letters}\ }\textbf {\bibinfo
  {volume} {101}},\ \bibinfo {pages} {041905} (\bibinfo {year}
  {2012})}\BibitemShut {NoStop}%
\bibitem [{\citenamefont {Levin}(1998)}]{Levin1997}%
  \BibitemOpen
  \bibfield  {author} {\bibinfo {author} {\bibfnamefont {Y.}~\bibnamefont
  {Levin}},\ }\bibfield  {title} {\bibinfo {title} {{Internal thermal noise in
  the LIGO test masses: A direct approach}},\ }\href
  {https://doi.org/10.1103/PhysRevE.67.046106} {\bibfield  {journal} {\bibinfo
  {journal} {Physical Review D}\ }\textbf {\bibinfo {volume} {57}},\ \bibinfo
  {pages} {659} (\bibinfo {year} {1998})}\BibitemShut {NoStop}%
\bibitem [{\citenamefont {Levin}(2008)}]{Levin2008}%
  \BibitemOpen
  \bibfield  {author} {\bibinfo {author} {\bibfnamefont {Y.}~\bibnamefont
  {Levin}},\ }\bibfield  {title} {\bibinfo {title} {{Fluctuation-dissipation
  theorem for thermo-refractive noise}},\ }\href
  {https://doi.org/10.1016/j.physleta.2007.11.007} {\bibfield  {journal}
  {\bibinfo  {journal} {Physics Letters, Section A: General, Atomic and Solid
  State Physics}\ }\textbf {\bibinfo {volume} {372}},\ \bibinfo {pages} {1941}
  (\bibinfo {year} {2008})}\BibitemShut {NoStop}%
\bibitem [{\citenamefont {Panuski}\ \emph {et~al.}(2019)\citenamefont
  {Panuski}, \citenamefont {Pant}, \citenamefont {Heuck}, \citenamefont
  {Hamerly},\ and\ \citenamefont {Englund}}]{Panuski2019}%
  \BibitemOpen
  \bibfield  {author} {\bibinfo {author} {\bibfnamefont {C.}~\bibnamefont
  {Panuski}}, \bibinfo {author} {\bibfnamefont {M.}~\bibnamefont {Pant}},
  \bibinfo {author} {\bibfnamefont {M.}~\bibnamefont {Heuck}}, \bibinfo
  {author} {\bibfnamefont {R.}~\bibnamefont {Hamerly}},\ and\ \bibinfo {author}
  {\bibfnamefont {D.}~\bibnamefont {Englund}},\ }\bibfield  {title} {\bibinfo
  {title} {{Single photon detection by cavity-assisted all-optical gain}},\
  }\href {https://doi.org/10.1103/PhysRevB.99.205303} {\bibfield  {journal}
  {\bibinfo  {journal} {Physical Review B}\ }\textbf {\bibinfo {volume} {99}},\
  \bibinfo {pages} {205303} (\bibinfo {year} {2019})}\BibitemShut {NoStop}%
\bibitem [{\citenamefont {Choi}\ \emph {et~al.}(2017)\citenamefont {Choi},
  \citenamefont {Heuck},\ and\ \citenamefont {Englund}}]{Choi}%
  \BibitemOpen
  \bibfield  {author} {\bibinfo {author} {\bibfnamefont {H.}~\bibnamefont
  {Choi}}, \bibinfo {author} {\bibfnamefont {M.}~\bibnamefont {Heuck}},\ and\
  \bibinfo {author} {\bibfnamefont {D.}~\bibnamefont {Englund}},\ }\bibfield
  {title} {\bibinfo {title} {{Self-Similar Nanocavity Design with Ultrasmall
  Mode Volume for Single-Photon Nonlinearities}},\ }\href
  {https://journals.aps.org/prl/pdf/10.1103/PhysRevLett.118.223605} {\bibfield
  {journal} {\bibinfo  {journal} {Physical Review Letters}\ }\textbf {\bibinfo
  {volume} {118}},\ \bibinfo {pages} {223605} (\bibinfo {year}
  {2017})}\BibitemShut {NoStop}%
\bibitem [{\citenamefont {Mabuchi}(2012)}]{Mabuchi2012}%
  \BibitemOpen
  \bibfield  {author} {\bibinfo {author} {\bibfnamefont {H.}~\bibnamefont
  {Mabuchi}},\ }\bibfield  {title} {\bibinfo {title} {{Qubit limit of cavity
  nonlinear optics}},\ }\href {https://doi.org/10.1103/PhysRevA.85.015806}
  {\bibfield  {journal} {\bibinfo  {journal} {Physical Review A}\ }\textbf
  {\bibinfo {volume} {85}},\ \bibinfo {pages} {015806} (\bibinfo {year}
  {2012})}\BibitemShut {NoStop}%
\bibitem [{\citenamefont {Krastanov}\ \emph {et~al.}(2020)\citenamefont
  {Krastanov}, \citenamefont {Heuck}, \citenamefont {Shapiro}, \citenamefont
  {Narang}, \citenamefont {Englund},\ and\ \citenamefont
  {Jacobs}}]{Krastanov2020}%
  \BibitemOpen
  \bibfield  {author} {\bibinfo {author} {\bibfnamefont {S.}~\bibnamefont
  {Krastanov}}, \bibinfo {author} {\bibfnamefont {M.}~\bibnamefont {Heuck}},
  \bibinfo {author} {\bibfnamefont {J.~H.}\ \bibnamefont {Shapiro}}, \bibinfo
  {author} {\bibfnamefont {P.}~\bibnamefont {Narang}}, \bibinfo {author}
  {\bibfnamefont {D.~R.}\ \bibnamefont {Englund}},\ and\ \bibinfo {author}
  {\bibfnamefont {K.}~\bibnamefont {Jacobs}},\ }\bibfield  {title} {\bibinfo
  {title} {{Room-Temperature Photonic Logical Qubits via Second-Order
  Nonlinearities}},\ }\href {http://arxiv.org/abs/2002.07193} {\  (\bibinfo
  {year} {2020})},\ \Eprint {https://arxiv.org/abs/2002.07193}
  {arXiv:2002.07193} \BibitemShut {NoStop}%
\bibitem [{\citenamefont {Boyd}(2019)}]{BoydBook}%
  \BibitemOpen
  \bibfield  {author} {\bibinfo {author} {\bibfnamefont {R.~W.}\ \bibnamefont
  {Boyd}},\ }\href@noop {} {\emph {\bibinfo {title} {Nonlinear optics}}}\
  (\bibinfo  {publisher} {Academic press},\ \bibinfo {year} {2019})\BibitemShut
  {NoStop}%
\bibitem [{\citenamefont {Hamerly}(2016)}]{HamerlyThesis}%
  \BibitemOpen
  \bibfield  {author} {\bibinfo {author} {\bibfnamefont {R.}~\bibnamefont
  {Hamerly}},\ }\bibfield  {title} {\bibinfo {title} {{Coherent LQG control,
  Free-Carrier Oscillations, Optical Ising Machines and Pulsed OPO Dynamics}},\
  }\href {https://arxiv.org/abs/1608.07551} {\bibfield  {journal} {\bibinfo
  {journal} {arXiv}\ } (\bibinfo {year} {2016})}\BibitemShut {NoStop}%
\bibitem [{\citenamefont {Timurdogan}\ \emph {et~al.}(2017)\citenamefont
  {Timurdogan}, \citenamefont {Poulton}, \citenamefont {Byrd},\ and\
  \citenamefont {Watts}}]{Timurdogan2017}%
  \BibitemOpen
  \bibfield  {author} {\bibinfo {author} {\bibfnamefont {E.}~\bibnamefont
  {Timurdogan}}, \bibinfo {author} {\bibfnamefont {C.~V.}\ \bibnamefont
  {Poulton}}, \bibinfo {author} {\bibfnamefont {M.~J.}\ \bibnamefont {Byrd}},\
  and\ \bibinfo {author} {\bibfnamefont {M.~R.}\ \bibnamefont {Watts}},\
  }\bibfield  {title} {\bibinfo {title} {{Electric field-induced second-order
  nonlinear optical effects in silicon waveguides}},\ }\href
  {https://doi.org/10.1038/nphoton.2017.14} {\bibfield  {journal} {\bibinfo
  {journal} {Nature Photonics}\ }\textbf {\bibinfo {volume} {11}},\ \bibinfo
  {pages} {200} (\bibinfo {year} {2017})}\BibitemShut {NoStop}%
\bibitem [{\citenamefont {Robinson}\ \emph {et~al.}(2005)\citenamefont
  {Robinson}, \citenamefont {Manolatou}, \citenamefont {Chen},\ and\
  \citenamefont {Lipson}}]{Robinson2005a}%
  \BibitemOpen
  \bibfield  {author} {\bibinfo {author} {\bibfnamefont {J.~T.}\ \bibnamefont
  {Robinson}}, \bibinfo {author} {\bibfnamefont {C.}~\bibnamefont {Manolatou}},
  \bibinfo {author} {\bibfnamefont {L.}~\bibnamefont {Chen}},\ and\ \bibinfo
  {author} {\bibfnamefont {M.}~\bibnamefont {Lipson}},\ }\bibfield  {title}
  {\bibinfo {title} {{Ultrasmall mode volumes in dielectric optical
  microcavities}},\ }\href@noop {} {\bibfield  {journal} {\bibinfo  {journal}
  {Physical Review Letters}\ }\textbf {\bibinfo {volume} {95}} (\bibinfo {year}
  {2005})}\BibitemShut {NoStop}%
\bibitem [{\citenamefont {Hu}\ and\ \citenamefont {Weiss}(2016)}]{Hu2016}%
  \BibitemOpen
  \bibfield  {author} {\bibinfo {author} {\bibfnamefont {S.}~\bibnamefont
  {Hu}}\ and\ \bibinfo {author} {\bibfnamefont {S.~M.}\ \bibnamefont {Weiss}},\
  }\bibfield  {title} {\bibinfo {title} {{Design of Photonic Crystal Cavities
  for Extreme Light Concentration}},\ }\href
  {https://doi.org/10.1021/acsphotonics.6b00219} {\bibfield  {journal}
  {\bibinfo  {journal} {ACS Photonics}\ }\textbf {\bibinfo {volume} {3}},\
  \bibinfo {pages} {1647} (\bibinfo {year} {2016})}\BibitemShut {NoStop}%
\end{thebibliography}%

\end{document}


\title{Supplemental Materials: Fundamental thermal noise limits for optical microcavities}

\author{Christopher Panuski}
\email{cpanuski@mit.edu}
\affiliation{Research Laboratory of Electronics, MIT, Cambridge, MA 02139, USA}

\author{Dirk Englund}
\affiliation{Research Laboratory of Electronics, MIT, Cambridge, MA 02139, USA}

\author{Ryan Hamerly}
\affiliation{Research Laboratory of Electronics, MIT, Cambridge, MA 02139, USA}
\affiliation{NTT Research, Inc.\ Physics \& Informatics Laboratories, 1950 University Ave Ste.\ 600, East Palo Alto, CA 94303, USA}

\maketitle

\renewcommand{\theequation}{S\arabic{equation}}
\renewcommand{\thefigure}{S\arabic{figure}}
\renewcommand{\citenumfont}[1]{S#1}
\renewcommand{\bibnumfmt}[1]{[S#1]}
\renewcommand{\bibnumfmt}[1]{[S#1]}
\renewcommand{\citenumfont}[1]{S#1}

\twocolumngrid
\tableofcontents
\onecolumngrid

\section{Thermo-Refractive Noise (TRN) Theory}

While the development of ultrahigh-performance optical resonators has only recently warranted its study within optical systems, stochastic temperature fluctuations are a fundamental concept in thermodynamics \cite{Land1980}. Assuming Boltzmann statistics within a finite volume $V$ with specific heat capacity $c_V$ at temperature $T$, we find
\begin{equation}
\langle \delta T^2 \rangle = \frac{k_BT^2}{c_V V}.
\label{thermoBasic}
\end{equation}
In optical microcavities, $V$ approaches diffraction limited volumes leading to temperature fluctuations that significantly impact the resonance stability in materials with a temperature-dependent refractive index.

Here, we derive the associated thermo-refractive noise (TRN) spectrum in an optical microcavity under the single-mode approximation described in the main text. Using this approximation, the intracavity field statistics are derived. In the typical perturbative limit where the rms frequency fluctuation is much smaller than the loaded cavity linewidth ($\delta\omega_\text{rms} \ll 2\Gamma_l$), we use perturbation theory to solve for the evolution of the cavity field $a(t)$ and the associated noise spectrum $S_{aa}(t)$. We also provide general solutions for the first and second statistical moments of $a(t)$, which are used in the main text to derive ``effective" quality factors in the presence of thermal noise. The solution for $S_{aa}(t)$ in the limiting case of high-$Q$ cavities --- where the thermal decay rate $\Gamma_T\gg\Gamma_l$ and the frequency noise can be assumed to be white --- is also provided. Finally, we compare the single-mode noise spectrum to that derived from a formal solution to heat diffusion in an infinite two dimensional slab, which we found to most accurately model the specific geometry of the photonic crystal microcavities in our experiments.

A few notes on convention: 1) we derive two-sided angular frequency noise spectra $S_{\omega\omega}(\omega)$, but plot one-sided frequency spectra $S_\text{ff}(f)=2 S_{\omega\omega}(2\pi f)/2\pi = S_{\omega\omega}(2\pi f)/\pi$ for experimental measurements to conform with the common conventions of the gravitational wave community; 2) temporal coupled mode theory decay rates $\Gamma_i$ are amplitude decay rates; the associated quality factors are therefore defined as $Q_i=\omega_0/2\Gamma_i$.

\subsection{Statistics of Microcavity TRN}
\label{sec:TRNstatistics}
To first order, the change in resonant frequency under a permittivity perturbation $\delta\epsilon(\vec{r},t)$ can be expressed as
\begin{equation}
\frac{\delta\omega(t)}{\omega_0} = -\frac{1}{2}\frac{\int_{\delta\epsilon} \delta\epsilon(\vec{r},t) |\vec{E}(\vec{r})|^2 \mathrm{d}^3\vec{r}}{\int \epsilon |\vec{E}(\vec{r})|^2 \mathrm{d}^3\vec{r}}\approx -\frac{\int_{\delta n} n\, \delta n(\vec{r},t) |\vec{E}(\vec{r})|^2 \mathrm{d}^3\vec{r}}{\int \epsilon |\vec{E}(\vec{r})|^2 \mathrm{d}^3\vec{r}}= -\frac{1}{V_\text{eff}}\frac{\int_{\delta n} n\, \delta n(\vec{r},t) |\vec{E}(\vec{r})|^2 \mathrm{d}^3\vec{r}}{\text{max}\lbrace \epsilon|\vec{E}(\vec{r})|^2\rbrace}
\label{intForm}
\end{equation}
where we have made the approximation $\delta\epsilon\approx 2n\delta n$ and introduced the standard mode volume
\begin{equation}
V_\text{eff} = \frac{\int \epsilon(\vec{r})|\vec{E}(\vec{r})|^2 \mathrm{d}^3\vec{r}}{\text{max}\lbrace \epsilon|\vec{E}(\vec{r})|^2\rbrace}.
\end{equation}
The change in refractive index $\delta n(\vec{r},t)$ is directly proportional to temperature change $\delta {T}(\vec{r},t)$, with the thermo-optic coefficient $\alpha_{TO} = dn/dT$ serving as the constant of proportionality. We therefore find
\begin{equation}
\frac{\delta\omega(t)}{\omega_0} = -\frac{1}{V_\text{eff}}\frac{\int_{\delta n} \alpha_\text{TO}\delta T(\vec{r},t) n|\vec{E}(\vec{r})|^2 \mathrm{d}^3\vec{r}}{\text{max}\lbrace \epsilon|\vec{E}(\vec{r})|^2\rbrace}.
\label{intResult}
\end{equation}
Alternatively, Eqn. \ref{intForm} can be evaluated for a uniform, mode averaged temperature change $\delta \bar{T}(t)$ assuming complete confinement of the mode within a homogeneous medium. This approach yields
\begin{equation}
\frac{\delta\omega(t)}{\omega_0} = -\frac{1}{n}\alpha_\text{TO}\delta\bar{T}(t).
\label{avgForm}
\end{equation}
Comparing Eqns. \ref{intResult} and \ref{avgForm}, we find
\begin{equation}
\delta \bar{T}(t) = \frac{1}{V_\text{eff}} \frac{\int \delta T(\vec{r},t) \epsilon(\vec{r})|\vec{E}(\vec{r})|^2\mathrm{d}^3\vec{r}}{\text{max}\lbrace \epsilon|\vec{E}(\vec{r})|^2\rbrace}.
\label{modeavgT}
\end{equation}
Eqn. \ref{avgForm} can now be solved using the mode averaged temperature change, whose evolution is derived from the heat equation (given a thermal diffusivity $D_T$)
\begin{equation}
\frac{\partial \delta T(\vec{r},t)}{\partial t} - D_T\nabla^2\delta T = F_T(\vec{r},t)
\label{heatEqn}
\end{equation}
driven by a thermal Langevin source $F_T(\vec{r},t)$ with the statistics \cite{Braginsky1999}
\begin{equation}
\langle F_T(\vec{r}_1,t_1)F_T^*(\vec{r}_2,t_2)\rangle = \frac{2D_Tk_BT_0^2}{c_V}\delta(t_1-t_2)\vec{\nabla}_{\vec{r}_1}\cdot \vec{\nabla}_{\vec{r}_2}[\delta(\vec{r}_1-\vec{r}_2)]
\label{eqn:forceStat}
\end{equation}
that satisfy the fluctuation-dissipation theorem. Averaging over the mode, we find the approximation
\begin{equation}
\frac{\mathrm{d} (\delta\bar{T}(t))}{\mathrm{d}t} + \Gamma_T \delta \bar{T}(t) = \bar{F}_T(t),
\label{heatEqn-avg}
\end{equation}
where $\Gamma_T$ is introduced as a phenomenological thermal decay rate whose form will be chosen later to ensure the form of $\langle \delta\bar{T}^2\rangle$ matches the canonical result from statistical mechanics (Eqn.~\ref{thermoBasic}). In analog with $\delta\bar{T}(t)$, the mode averaged thermal force $\bar{F}_T(t)$ is
\begin{equation}
\bar{F}_T(t) = \frac{1}{V_\text{eff}} \frac{\int F_T(\vec{r},t) \epsilon(\vec{r})|\vec{E}(\vec{r})|^2\mathrm{d}^3\vec{r}}{\text{max}\lbrace \epsilon|\vec{E}(\vec{r})|^2\rbrace}.
\label{eqn:avgForce}
\end{equation}
The steady-state solution of Eqn.~\ref{heatEqn-avg},
\begin{equation}
\delta\bar{T}(t) = \int_{-\infty}^t  \bar{F}(t')e^{-\Gamma_T(t-t')} \mathrm{d}t',
\end{equation}
can then be used to find the corresponding statistics of the temperature fluctuation at equilibrium (i.e. long $t$):
\begin{equation}
\langle\delta\bar{T}(t)\delta\bar{T}(t+\tau)\rangle = \int_{-\infty}^t \mathrm{d}t' \int_{-\infty}^{t+\tau} \mathrm{d}t'' \langle \bar{F}_T(t')\bar{F}_T(t'')\rangle e^{-\Gamma_T(t-t')}e^{-\Gamma_T(t+\tau-t'')}.
\label{needFauto}
\end{equation}
The result requires the autocorrelation of $\bar{F}_T(t)$, which is readily evaluated using the mode averaged form of Eqn.~\ref{eqn:avgForce}:
\begin{equation}
\langle \bar{F}_T(t)\bar{F}^*_T(t+\tau)\rangle = \underbrace{\frac{2D_Tk_BT_0^2}{c_V V_\text{eff}^2}\frac{\int \mathrm{d}^3\vec{r} \left[ \vec{\nabla} \left( \epsilon(\vec{r})|\vec{E}(\vec{r})|^2\right)\right]^2}{(\text{max}\lbrace \epsilon|\vec{E}(\vec{r})|^2\rbrace)^2}}_{\mathcal{R}_{\bar{F}_T\bar{F}_T}(0)} \delta(\tau) = \mathcal{R}_{\bar{F}_T\bar{F}_T}(0) \delta(\tau).
\label{compareDef1}
\end{equation}
Inserting this result into Eqn. \ref{needFauto} along with the change of variables $t''\rightarrow t'+\tau'$ yields
\begin{align}
\langle\delta\bar{T}(t)\delta\bar{T}(t+\tau)\rangle &\approx \int_{-\infty}^t \mathrm{d}t' \int_{-t'}^{t+|\tau|-t'} {\rm d}\tau' \langle \bar{F}_T(t')\bar{F}^*_T(t'+\tau')\rangle e^{-2\Gamma_T(t-t')} e^{-\Gamma_T(|\tau|-\tau')}\nonumber \\
&\approx \int_{-\infty}^t \mathrm{d}t' e^{-2\Gamma_T(t-t')} \int_{-t'}^{t+|\tau|-t'} {\rm d}\tau' \mathcal{R}_{\bar{F}_T\bar{F}_T}(0)\delta(\tau') e^{-\Gamma_T(|\tau|-\tau')} \nonumber \\
\langle\delta\bar{T}(t)\delta\bar{T}(t+\tau)\rangle &\approx \frac{\mathcal{R}_{\bar{F}_T\bar{F}_T}(0)}{2\Gamma_T}  e^{-\Gamma_T|\tau|}.
\label{eqn:tempAutocorrelation}
\end{align}
Correspondence with Eqn.~\ref{thermoBasic} therefore requires
\begin{equation}
\frac{\mathcal{R}_{\bar{F}_T\bar{F}_T}(0)}{2\Gamma_T} = \frac{k_BT_0^2}{c_V V} \Rightarrow \mathcal{R}_{\bar{F}_T\bar{F}_T}(0) = \frac{2k_BT_0^2\Gamma_T}{c_V V}.
\label{compareDef2}
\end{equation}
The same correspondence for a general frequency-domain solution to Eqn.~\ref{heatEqn} in an infinite homogeneous medium (where the phenomenological parameter $\Gamma_T$ is not introduced; see Section~\ref{sec:Vtderivation}) mandates \footnote{The implications of this definition are especially important for the ultra-small move volume cavities proposed in recent studies \cite{Robinson2005a, Choi, Hu2016}. In these cavities, we found that $V_\text{eff}^{(2)} < V_\text{eff}$ \cite{Panuski2019}, implying that thermodynamic fluctuations in these cavities are significantly suppressed.}
\begin{equation}
V = V_T = \frac{V_\text{eff}^2}{V_\text{eff}^{(2)}} = \left[ \frac{\int \epsilon|\vec{E}(\vec{r})|^2 \mathrm{d}^3\vec{r}}{\text{max}\lbrace \epsilon|\vec{E}(\vec{r})|^2\rbrace}\right]^2\bigg/\underbrace{\frac{\int \epsilon^2|\vec{E}(\vec{r})|^4 \mathrm{d}^3\vec{r}}{\text{max}\lbrace \epsilon^2|\vec{E}(\vec{r})|^4\rbrace}}_{V_\text{eff}^{(2)}}
\label{eqn:Vt}
\end{equation}
Comparing Eqn.~\ref{compareDef2} to Eqn.~\ref{compareDef1} then lends a calculable form of the decay rate $\Gamma_T$:
\begin{equation}
\Gamma_T = D_T\frac{\int \mathrm{d}^3\vec{r} \left[ \vec{\nabla} \left( \epsilon(\vec{r})|\vec{E}(\vec{r})|^2\right)\right]^2}{\int \mathrm{d}^3\vec{r} \epsilon(\vec{r})^2|\vec{E}(\vec{r})|^4}.
\label{gammaT}
\end{equation}

The autocorrelation of the cavity resonance frequency is directly proportional to this result given $\delta\bar{\omega}_0(t)\approx -(\omega_0/n)\alpha_\text{TO}\delta\bar{T}(t)$ (as dictated by first-order perturbation theory):
\begin{equation}
\boxed{\langle\delta\omega(t)\delta\omega(t+\tau)\rangle \approx \left(\frac{\omega_0}{n}\alpha_\text{TO} \right)^2 \frac{\mathcal{R}_{\bar{F}_T\bar{F}_T}(0)}{2\Gamma_T}  e^{-\Gamma_T|\tau|}= \left(\frac{\omega_0}{n}\alpha_\text{TO} \right)^2 \frac{k_BT_0^2}{c_V V_T}  e^{-\Gamma_T|\tau|} = \delta\omega_\text{rms}^2e^{-\Gamma_T|\tau|}.}
\label{eqn:freqAutocorrelation}
\end{equation}

\subsection{Derivation of the Thermal Mode Volume}
\label{sec:Vtderivation}
For completeness, we recapitulate the derivation of the thermal mode volume $V_T$ (introduced in Section~\ref{sec:TRNstatistics}) in an infinite homogeneous medium \cite{Gorodetsky2004}, where Fourier modes can be used to solve solve Eqn.~\ref{heatEqn}:
\begin{equation}
\frac{\partial \delta T(\vec{r},t)}{\partial t} - D_T\nabla^2\delta T = F_T(\vec{r},t) \Rightarrow \delta T(\omega,\vec{k}) = \frac{F(\omega,\vec{k})}{i\omega+D_T|\vec{k}|^2}.
\end{equation}
Taking the temperature mode average in Eqn.~\ref{modeavgT} and inverse Fourier transforming yields
\begin{align}
\delta\bar{T}(t) = \frac{1}{(2\pi)^4V_\text{eff}}\int \mathrm{d}^3\vec{r} \frac{\epsilon |\vec{E}(\vec{r})|^2}{\text{max}\lbrace \epsilon|\vec{E}(\vec{r})|^2 \rbrace}\int \mathrm{d}\omega~ e^{-i\omega t} \int \mathrm{d}^3\vec{k}~ e^{i\vec{k}\cdot \vec{r}}\frac{F(\omega, \vec{k})}{i\omega+D_T|\vec{k}|^2}.
\end{align}
We can then solve for the autocorrelation of $\delta\bar{T}$:
\begin{align}
\langle \delta\bar{T}(t)\delta\bar{T}^*(t+\tau)\rangle &= \frac{(\text{max}\lbrace \epsilon|\vec{E}(\vec{r})|^2 \rbrace)^{-2}}{(2\pi)^8V_\text{eff}^2}\int \mathrm{d}^3\vec{r}_1~ \epsilon(\vec{r}_1) |\vec{E}(\vec{r}_1)|^2 \int \mathrm{d}^3\vec{r}_2~ \epsilon(\vec{r}_2) |\vec{E}(\vec{r}_2)|^2 \nonumber \\
& \qquad \times \int \mathrm{d}\omega_1~e^{-i\omega_1 t}\int \mathrm{d}\omega_2 ~e^{i\omega_2(t+\tau)} \nonumber\\
& \qquad \times \int~\mathrm{d}^3\vec{k}_1~ e^{i\vec{k}_1\cdot\vec{r}_1} \int~\mathrm{d}^3\vec{k}_2~e^{-i\vec{k}_2\cdot\vec{r}_2}\frac{\langle F(\omega_1,\vec{k}_1)F^*(\omega_2,\vec{k}_2)\rangle}{(i\omega_1+D_T|\vec{k}_1|^2)(-i\omega_2+D_T|\vec{k}_2|^2)}.
\label{eqn:3dFourierTempAutocorrelation}
\end{align}
Inputting the frequency-space autocorrelation of the Langevin driving force (compare Eqn.~\ref{compareDef1}) \cite{Braginsky2000},
\begin{equation}
\langle F(\omega_1,\vec{k}_1)F^*(\omega_2, \vec{k}_2)\rangle = (2\pi)^4\frac{2k_BT_0^2D_T}{c_V}|\vec{k}|^2\delta(\vec{k}_1-\vec{k}_2)\delta(\omega_1-\omega_2),
\end{equation}
and computing the remaining $\omega_1$ integral, we can simplify to
\begin{equation}
\langle \delta\bar{T}(t)\delta\bar{T}^*(t+\tau)\rangle = \frac{(\text{max}\lbrace \epsilon|\vec{E}(\vec{r})|^2 \rbrace)^{-2}}{(2\pi)^3V_\text{eff}^2}\frac{k_BT_0^2}{c_V} \int~\mathrm{d}^3\vec{k}~e^{-D_T|\vec{k}|^2\tau} \bigg|\int \mathrm{d}^3\vec{r}~ \epsilon(\vec{r}) |\vec{E}(\vec{r})|^2 e^{i\vec{k}\cdot \vec{r}}\bigg|^2.
\label{gettingThere}
\end{equation}
Eqn.~\ref{gettingThere} must equal Eqn.~\ref{thermoBasic} for $\tau=0$, which reveals the final solution for the thermal mode volume $V_T$:
\begin{align}
V_T &= \frac{(2\pi)^3 V_\text{eff}^2 (\text{max}\lbrace \epsilon|\vec{E}(\vec{r})|^2 \rbrace)^2}{\int~\mathrm{d}^3\vec{k}~ \big|\int \mathrm{d}^3\vec{r}~ \epsilon(\vec{r}) |\vec{E}(\vec{r})|^2 e^{i\vec{k}\cdot \vec{r}}\big|^2} = \frac{V_\text{eff}^2 (\text{max}\lbrace \epsilon|\vec{E}(\vec{r})|^2 \rbrace)^2}{\int \mathrm{d}^3\vec{r}~ \epsilon(\vec{r})^2 |\vec{E}(\vec{r})|^4} \nonumber \\
\therefore V_T &= \frac{V_\text{eff}^2}{\frac{\int \mathrm{d}^3\vec{r}~ \epsilon(\vec{r})^2 |\vec{E}(\vec{r})|^4}{\text{max}\lbrace \epsilon^2|\vec{E}|^4 \rbrace}} = \frac{V_\text{eff}^2}{V_\text{eff}^{(2)}}
\end{align}
where
\begin{equation}
V_\text{eff}^{(2)} = \frac{\int \mathrm{d}^3\vec{r}~ \epsilon(\vec{r})^2 |\vec{E}(\vec{r})|^4}{\text{max}\lbrace \epsilon^2|\vec{E}|^4 \rbrace}.
\label{eqn:Veff_2}
\end{equation}
This result matches Gorodetsky's original result \cite{Gorodetsky2004} with the exception of different normalization conditions. 

\subsection{Comparison to Multi-mode Thermal Decay in a 2D PhC Slab}
\label{sec:MMtheory}
Under the single-mode approximation derived in the preceding sections, Eqn.~\ref{eqn:freqAutocorrelation} implies a Lorentzian TRN spectrum 
\begin{equation}
S_{\omega\omega}(\omega) = \left(\frac{\omega_0}{n}\alpha_\text{TO} \right)^2 \frac{k_BT_0^2}{c_V V_T}\frac{2\Gamma_T}{\Gamma_T^2+\omega^2}.
\label{eqn:Sww}
\end{equation}
As noted in the main text, this approximate spectrum can be evaluated for any optical microcavity (photonic crystals, microtoroids, microbottles, ring resonators, micropillars, microdisks, and so on) independent of its exact confining geometry. This allows us to derive general noise limits as illustrated in the main text and derived in Section~\ref{sec:cavityDynamics}. If a particular experimental system is of interest, we can verify the accuracy of this approximation by solving the stochastic heat equation (Eqn.~\ref{heatEqn}) for that particular cavity geometry. Here, since we measure TRN in high-$Q/V_\text{eff}$ 2D slab photonic crystal cavities (see main text and supplement Section~\ref{sec:experiments}), we demonstrate this evaluation for a Gaussian mode confined within an infinite two dimensional slab. The heat equation in this case lends logarithmically --- as opposed to exponentially --- decaying temperature fluctuations in time.

For a slab of thickness $w$ lying atop the $xy$ plane, the local temperature change $\delta T(\vec{r},t)=\sum_n T_n(\vec{r}_\parallel,t)\phi_n(z)$ can be expanded in terms of the out-of-plane eigenfunctions, $\phi_n(z) = \cos(n\pi z/w)$ assuming insulating boundary conditions on the top and bottom of the slab. The stochastic heat equation then simplifies to the form
\begin{equation}
\frac{\partial T_n(\vec{r}_\parallel,t)}{\partial t} = D_T\left[ \nabla^2-(n\pi/w)^2\right]T_n(\vec{r}_\parallel,t)+\frac{1}{w}\int \phi^*_n(z)F_T(\vec{r},t) \mathrm{d}z.
\label{eqn:heatEqnExpansion}
\end{equation}
If assume a two-dimensional Gaussian mode profile
\begin{equation}
\epsilon(\vec{r})|\vec{E}(\vec{r})|^2 = \begin{cases}
\frac{1}{2\pi\sigma^2}e^{-|\vec{r}_\parallel|^2/2\sigma^2} & 0<z<w \\
0 & \text{else}
\end{cases},
\label{eqn:gaussianProfile}
\end{equation}
all $n\neq 0$ terms in the temperature expansion have zero contribution to the mode-averaged temperature fluctuation (Eqn.~\ref{modeavgT}) of interest, which involves the integral $\int\phi_n(z)dz$. Eqn.~\ref{eqn:heatEqnExpansion} then simplifies to the two-dimensional form 
\begin{equation}
\frac{\partial T(\vec{r}_\parallel,t)}{\partial t} = D_T \nabla^2T(\vec{r}_\parallel,t)+ F_T^\parallel(\vec{r}_\parallel,t),
\label{eqn:heatEqn2D}
\end{equation}
where we have dropped the $n=0$ subscript and introduced a modified fluctuation $F_T^\parallel(\vec{r}_\parallel,t)$ with statistics
\begin{equation}
\langle F_T^{\parallel *}(\vec{r}_\parallel,t)F_T^\parallel(\vec{r}_\parallel ',t')\rangle = \frac{1}{w^2}\int \mathrm{d}z \int \mathrm{d}z' \phi^*_0(z)\phi_0(z')\langle F_T^*(\vec{r},t) F_T(\vec{r}',t')\rangle = \frac{1}{w}\frac{2D_Tk_BT_0^2}{c_V}\delta(t-t')\vec{\nabla}_{\vec{r}_\parallel}\cdot  \vec{\nabla}_{\vec{r}_\parallel '}\delta(\vec{r}_\parallel-\vec{r}_\parallel ').
\label{eqn:2dautocorrelation}
\end{equation}
Comparing Eqns.~\ref{eqn:heatEqn2D},~\ref{eqn:2dautocorrelation} to their three-dimensional analogs (Eqns.~\ref{heatEqn},~\ref{eqn:forceStat}), we see that projecting onto the $n=0$ subspace reduces the finite-thickness slab to an infinite two-dimensional problem where $F_T$ is scaled by $w^{-1/2}$. We can then apply the techniques of Section~\ref{sec:Vtderivation} (expansion in Fourier normal modes) to solve for the spectrum of temperature (and therefore resonant frequency) fluctuations. Without inverse Fourier transforming frequency, Eqn.~\ref{eqn:3dFourierTempAutocorrelation} gives
\begin{align}
S_{\omega\omega}^\text{mm}(\omega) = \left( \frac{\omega_0}{n}\alpha_\text{TO}\right)^2 \frac{\langle \delta\bar{T}(\omega)\delta\bar{T}(\omega')\rangle}{\delta(\omega-\omega')} &= 2\left( \frac{\omega_0}{n}\alpha_\text{TO}\right)^2 \frac{k_BT_0^2}{wc_V} \int \frac{D_T k_\parallel^2}{(D_T k_\parallel^2)^2+\omega^2} |\epsilon E^2(k_\parallel)|^2 \mathrm{d}^2 k_\parallel \nonumber \\
S_{\omega\omega}^\text{mm}(\omega) &= \left( \frac{\omega_0}{n}\alpha_\text{TO}\right)^2 \frac{k_BT_0^2}{2\pi wc_vD_T}\underbrace{\int_0^\infty \frac{x}{x^2+(\omega\sigma^2/D_T)^2}e^{-x} \mathrm{d}x}_{I_\text{mm}(\omega\sigma^2/D_T)}
\label{eqn:Sww-mm}
\end{align}
with the change of variables $(k_\parallel\sigma)^2\rightarrow x$. Note that we treat the effect of patterned holes in our experimental structures through a reduced thermal conductivity, and therefore thermal diffusivity, as a function of the slab porosity (see Section \ref{sec:experiments} for further detail). We can compare this result with the single-mode approximation, which (by evaluating Eqns.~\ref{eqn:Vt},~\ref{gammaT} for the Gaussian mode profile in Eqn.~\ref{eqn:gaussianProfile}) gives
\begin{equation}
V_T = 4\pi w\sigma^2 ~~~ \Gamma_T = \frac{D_T}{\sigma^2}, ~~~ \Rightarrow ~~~ S_{\omega\omega}^\text{sm}(\omega) = \left( \frac{\omega_0}{n}\alpha_\text{TO}\right)^2 \frac{k_BT_0^2}{2\pi wc_vD_T}\underbrace{\frac{1}{1+(\omega\sigma^2/D_T)^2}}_{I_\text{sm}(\omega\sigma^2/D_T)}.
\label{eqn:Sww-sm}
\end{equation}

\begin{figure}
\includegraphics[width=0.9\textwidth]{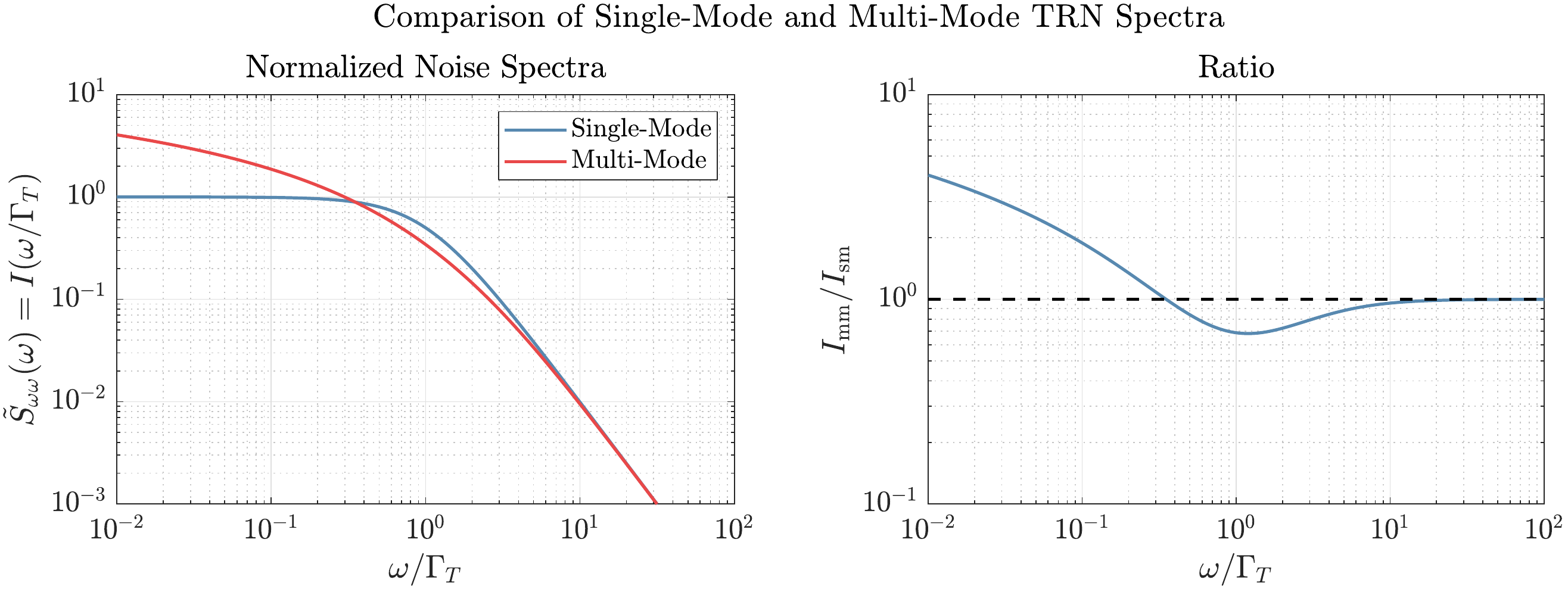}
\caption{Normalized noise spectra $\tilde{S}_{\omega\omega}(\omega) =I(\omega/\Gamma_T)$ for single-mode (Eqn.~\ref{eqn:Sww-sm}) and multi-mode (Eqn.~\ref{eqn:Sww-mm}) TRN in an infinite slab of finite thickness.}
\label{fig:SM-MMcompare}
\end{figure}

As expected, the integral $\int_{-\infty}^\infty S_{\omega\omega}\mathrm{d}\omega/2\pi$ of either spectra yields $\langle \delta\omega^2\rangle = (\omega_0\alpha_\text{TO}/n)^2\langle \delta T^2\rangle = (\omega_0\alpha_\text{TO}/n)^2 k_BT_0^2/c_VV_T$ in correspondence with Eqn.~\ref{thermoBasic}. Fig.~\ref{fig:SM-MMcompare} plots each normalized spectrum for comparison along with the ratio $I_\text{mm}/I_\text{sm}$. These results substantiate the claims in the main text: the single-mode approximation undershoots at low frequency $\omega\ll \Gamma_T$, slightly overshoots at intermediate frequencies $\omega\sim \Gamma_T$, and converges to the multi-mode spectrum at high frequencies $\omega\gg \Gamma_T$. We further note that the error of the multi-mode spectrum increases at low frequencies for any finite volume system: in our experiment, for example, the multi-mode estimate does not account for low frequency heat transfer through the underlying oxide around the released membrane. Thus, in the range of frequencies of interest (i.e. near the thermal cutoff frequency $\Gamma_T = D_T/\sigma^2$), single-mode thermal decay is an appropriate simplifying assumption that allows the thermal noise spectrum to be well-approximated irrespective of the cavity's exact geometry.

\subsection{Derivation of Driven Cavity Dynamics}
\label{sec:cavityDynamics}
To determine the practical impact of thermo-refractive noise on microcavity dynamics, we now consider the case of a cavity driven by a monochromatic laser with frequency $\omega_L$. Intuitively, we would expect that the large (relative to the cavity linewidth), fast (relative to the cavity decay time) stochastic deviations of the resonance frequency in the high-$Q/V_\text{eff}$ limit would restrict the maximum intensity in the cavity, as a narrow linewidth laser would no longer always be on resonance with the fluctuating cavity resonance. A mode volume-dependent maximum ``effective" quality factor describing the stored energy should result. A similar effective quality factor could be derived if the coherence of the intracavity field --- rather than the stored energy alone --- is also of interest. 

To prove these suppositions, we solve the driven temporal coupled mode theory relation
\begin{equation}
\frac{\mathrm{d}a(t)}{\mathrm{d}t}= \left[ i\omega_0(t)-\Gamma_l\right] a(t) +i\sqrt{2\Gamma_c}s_\text{in}(t),
\label{eqn:TCMT}
\end{equation}
for the cavity field $a(t)$, where $\omega_0(t)=\omega_0+\delta\omega(t)$ is the instantaneous resonant frequency, $\Gamma_l = \langle \omega_0(t)\rangle /2Q_l$ is the amplitude decay rate of $a$ corresponding to a loaded quality factor $Q_l$, and $\Gamma_c$ is the amplitude coupling rate of the drive field $s_\text{in}(t)=\tilde{s}_\text{in}e^{i\omega_L t} + \text{ c.c}$, detuned from resonance by $\Delta=\omega_L-\omega_0$. In the presence of TRN, $\delta\omega(t)$ is non-Markovian, zero-mean Gaussian noise with the autocorrelation given by Eqn.~\ref{eqn:freqAutocorrelation}. Solving with an integrating factor and introducing the slowly varying cavity amplitude $\tilde{a}(t)=a(t)e^{-i\omega_0 t}$ in a reference frame co-rotating with the static cavity resonance, we find
\begin{equation}
\tilde{a}(t) = i\sqrt{2\Gamma_c}\tilde{s}_\text{in} \int_{-\infty}^t \mathrm{d}t' e^{-(i\Delta+\Gamma_l)(t-t')} e^{-i\int_{t'}^{t} \mathrm{d}t'' \delta\omega(t'')}.
\label{steadyState}
\end{equation}
Since the steady state solution is desired, we assume that the integration starts at $t=-\infty$ such that the system has no ``memory" of the initial conditions. Using Eqn.~\ref{steadyState}, we can compute $\langle \tilde{a}(t)\rangle$ and $\langle |\tilde{a}(t)|^2\rangle$, the mean cavity field amplitude and stored energy, respectively. In certain limiting cases, the noise spectrum $S_{aa}(\omega)$ of the intra-cavity field can also be derived.

\subsubsection{Cavity Spectrum in the Perturbative Limit}
One of these limiting cases is the perturbative regime commonly studied in the literature for low-$Q/V_\text{eff}$ cavities, wherein $\delta\omega_\text{rms}\ll \Gamma_l$. In this case, $\tilde{a}(t)$ and $\delta\omega(t)$ --- described by (from Eqns.~\ref{heatEqn-avg},~\ref{eqn:tempAutocorrelation})
\begin{equation}
\frac{\mathrm{d}\,\delta\omega(t)}{\mathrm{d}t} = -\Gamma_T\delta\omega(t)+\delta\omega_\text{rms}\sqrt{2\Gamma_T}W(t)
\label{eqn:resFreqDiffEq}
\end{equation}
for a Wiener process $W(t)$ with $\langle W(t)W(t')\rangle=\delta(t-t')$ --- can be expanded in orders of $\delta\omega_\text{rms}\sqrt{2\Gamma_T}$. The zeroth- and first-order evolution equations (with subscripts 0 and 1, respectively) are
\begin{align}
\frac{\mathrm{d}\tilde{a}_0(t)}{\mathrm{d}t} = [i(\delta\omega_0(t)+\Delta)-\Gamma_l]\tilde{a}_0(t)+\sqrt{2\Gamma_l}\tilde{s}_\text{in} & && \frac{\mathrm{d}\delta\omega_0(t)}{\mathrm{d}t} = -\Gamma_T\delta\omega_0(t) \\
\frac{\mathrm{d}\tilde{a}_1(t)}{\mathrm{d}t} = [i(\delta\omega_0(t)+\Delta)-\Gamma_l]a_1(t)+i\delta\omega_1(t)\tilde{a}_0(t) & && \frac{\mathrm{d}\delta\omega_1(t)}{\mathrm{d}t} = -\Gamma_T\delta\omega_1(t)+\delta\omega_\text{rms}\sqrt{2\Gamma_T}W(t).
\end{align}
Solving in the frequency domain yields
\begin{equation}
\tilde{a}_1(\omega) = \frac{\sqrt{2\Gamma_l}\tilde{s}_\text{in}}{\Gamma_l-i\Delta}\frac{i\delta\omega_\text{rms}\sqrt{2\Gamma_T}W(\omega)}{(\Gamma_T-i\omega)[\Gamma_l+i(\omega-\Delta)]},
\end{equation}
corresponding to the frequency spectrum
\begin{equation}
\boxed{S_{aa}(\omega) = \langle \tilde{a}_1(\omega)^*\tilde{a}_1(\omega)\rangle = \frac{2\Gamma_l|\tilde{s}_\text{in}|^2}{\Gamma_l^2+\Delta^2}\frac{2\Gamma_T\delta\omega_\text{rms}^2}{(\Gamma_T^2+\omega^2)(\Gamma_l^2+(\omega-\Delta)^2)}.}
\label{eqn:perturbativeSpectrum}
\end{equation}
The intra-cavity noise spectrum can therefore be approximated as the product of two Lorentzians with spectral widths $2\Gamma_T$ and $2\Gamma_l$. When $\Gamma_T\ll\Gamma_l$, which often coincides with the perturbative limit $\delta\omega_\text{rms}\ll\Gamma_l$ for large mode volumes ($\delta\omega_\text{rms}\propto V_T^{-1/2}$ and $\Gamma_T\propto V_T^{-2/3}$ for a three-dimensional Gaussian mode), the resonant frequency fluctuations are small and occur over timescales much longer than that of intra-cavity photon decay. TRN thus leads to a weak inhomogeneous broadening of the resonant mode that can often be neglected for common applications of low-$Q/V_\text{eff}$ optical cavities. Gravitational wave interferometry \cite{Braginsky2000,Martynov2016} and ultra-stable optical frequency references \cite{Lim,Zhang2019} are two notable exceptions that have led to significant interest in perturbative TRN.

\subsubsection{General Derivation for $\langle\tilde{a}(t)\rangle$}
Our work focuses on the transition to non-perturbative TRN in high-$Q/V_\text{eff}$ microcavities, where we are interested in general solutions for the statistical moments of Eqn.~\ref{steadyState} in the presence of TRN. Specifically, $\langle \tilde{a}(t) \rangle$ provides insight into thermal noise-induced dephasing while $\langle \tilde{a}^2(t) \rangle$ lends a bound on the maximum allowable stored energy.

The expected intra-cavity field amplitude
\begin{equation}
\langle \tilde{a}(t)\rangle = i\sqrt{2\Gamma_c}\tilde{s}_\text{in} \int_{-\infty}^t \mathrm{d}t' e^{-(i\Delta+\Gamma_l)(t-t')} \langle e^{-\int_{t'}^{t} \mathrm{d}t'' i\delta\omega(t'')}\rangle
\label{eqn:cavAmp}
\end{equation}
follows directly from Eqn.~\ref{steadyState}, where the average on the right-hand side has a similar form to the characteristic functional \cite{Feynman1965}
\begin{equation}
\Phi[k(t)] = \langle e^{i\int k(t)f(t)\mathrm{d}t}\rangle = \frac{\int e^{i\int k(t)f(t)\mathrm{d}t} P[f(t)]\mathcal{D}f(t)}{\int P[f(t)]\mathcal{D}f(t)},
\label{NCF}
\end{equation}
a normalized average of $e^{i\int k(t)f(t)\mathrm{d}t}$ along the paths $f(t)$ with respective probabilities $P[f(t)]$. For the special case of Gaussian noise, the moment-generating properties of the characteristic functional allow Eqn.~\ref{NCF} to be simplified to
\begin{equation}
\Phi[k(t)] = e^{i\int k(t)M(t)\mathrm{d}t}e^{-1/2\int \mathrm{d}t \int \mathrm{d}t' k(t)k(t')\langle f(t)f(t')\rangle},
\end{equation}
which is characterized by two parameters only: 1) the mean path $M(t)$, and the autocorrelation of the noise $f(t)$, $\langle f(t)f(t+t')\rangle$. Comparing Eqn.~\ref{eqn:cavAmp} to Eqn.~\ref{NCF}, we find 
$k(t'')= \begin{cases} 1 & t'<t''<t \\ 0 & \text{else} \end{cases}$ 
and $f(t) = \delta\omega(t'')$. Since $\langle \omega_0(t) \rangle=0$,
\begin{align}
\langle e^{-\int_{t'}^{t} \mathrm{d}t'' i\delta\omega(t'')}\rangle &= \exp\left[-\frac{1}{2}\int_{t'}^t \mathrm{d}t_2' \int_{t'}^t \mathrm{d}t_2'' \delta\omega_\text{rms}^2e^{-\Gamma_T|t_2''-t_2'|}\right] \nonumber \\
&= \exp\left[-\int_{0}^{t-t'} \mathrm{d}\tau (t-t'-\tau) \delta\omega_\text{rms}^2e^{-\Gamma_T|\tau|}\right] \nonumber \\
\langle e^{-\int_{t'}^{t} \mathrm{d}t'' i\delta\omega(t'')}\rangle &= \exp\left[\frac{\delta\omega_\text{rms}^2}{\Gamma_T^2}\left(1-\Gamma_T(t-t')- e^{-\Gamma_T(t-t')}\right)\right].
\label{F'(t)}
\end{align}
Combined with the substitution $\tilde{\tau} = \left(\delta\omega_\text{rms}/\Gamma_T\right)^2e^{-\Gamma_T(t-t')}$, Eqn.~\ref{F'(t)} simplifies to
\begin{equation}
\langle \tilde{a}(t)\rangle =  i\frac{\sqrt{2\Gamma_c}\tilde{s}_\text{in}}{\Gamma_T}e^{(\delta\omega_\text{rms}/\Gamma_T)^2}\left(\frac{\delta\omega_\text{rms}}{\Gamma_T}\right)^{2\left[\frac{-i\Delta-\Gamma_l-\delta\omega_\text{rms}^2/\Gamma_T}{\Gamma_T}\right]}\int_0^{(\delta\omega_\text{rms}/\Gamma_T)^2} \tilde{\tau}^{\frac{i\Delta+\Gamma_l+\delta\omega_\text{rms}^2/\Gamma_T}{\Gamma_T}-1}e^{-\tilde{\tau}}\mathrm{d}\tilde{\tau},
\end{equation}
which is in the form of the lower incomplete Gamma function
\begin{equation}
\gamma_l(s,x) = \int_0^x \tilde{\tau}^{s-1}e^{-\tilde{\tau}}\mathrm{d}\tilde{\tau}.
\end{equation}
The final closed-form solution is therefore
\begin{empheq}[box=\widefbox]{align}
  \langle \tilde{a}(t)\rangle &=  i\frac{\sqrt{2\Gamma_c}\tilde{s}_\text{in}}{\Gamma_T}e^x x^{-s}\gamma_l(s,x) \label{aSoln}\\
  x &\equiv \left(\frac{\delta\omega_\text{rms}}{\Gamma_T}\right)^2 \label{xDef}\\
  s &\equiv \frac{\Gamma_l+i\Delta}{\Gamma_T}+x. \label{sDef}
\end{empheq}
To confirm this solution, we can evaluate the limiting case of $\delta\omega_\text{rms}\rightarrow 0$ ($x\rightarrow 0$), corresponding to a noiseless thermal reservoir when $T\rightarrow 0$. Using the series expansion of $\gamma_l(s,x)$ in terms of $s$, $x$ and the standard Gamma function $\gamma_f(z)$, we find
\begin{align}
\langle |\tilde{a}(t)| \rangle_{T=0}&=\lim_{x \rightarrow 0} i\frac{\sqrt{2\Gamma_c}\tilde{s}_\text{in}}{\Gamma_T} e^x x^{-s} \left[x^s e^{-x} \gamma_f(s) \sum_{k=0}^\infty \frac{x^k}{\gamma_f(s+k+1)} \right] \nonumber \\
&= i\frac{\sqrt{2\Gamma_c}\tilde{s}_\text{in}}{\Gamma_T}\frac{\gamma_f(s)}{\gamma_f(s+1)} \nonumber \\
\langle |\tilde{a}(t)|\rangle_{T=0}  &= i\frac{\sqrt{2\Gamma_c}\tilde{s}_\text{in}}{\Gamma_T}\frac{1}{s}\bigg|_{x=0} = i\frac{\sqrt{2\Gamma_c}\tilde{s}_\text{in}}{\Gamma_l+i\Delta},
\label{ampSS}
\end{align}
as expected from standard (noiseless) temporal coupled mode theory. 

Assuming critical coupling ($\Gamma_c = \Gamma_l/2$) and resonant excitation ($\delta=0$), we find the ``effective" quality factor 
\begin{equation}
Q_\text{eff} = \frac{\omega_0|\langle\tilde{a}(t)\rangle|^2}{2|\tilde{s}_\text{in}|^2} = Q_l \left(\frac{\Gamma_l}{\Gamma_T}\right)^2e^{2x}x^{-2s}\gamma_l^2(s,x)
\end{equation}
by analogy to the noiseless result where $Q=\omega_0\langle |\tilde{a}(t)| \rangle_{T=0}^2/2|\tilde{s}_\text{in}|^2$.

This result is used in the main text to describe dephasing in the qubit limit of cavity nonlinear optics. For a given mode volume, the optimum loaded quality factor $Q_l^\text{opt} \approx \omega_0\Gamma_T/2\delta\omega_\text{rms}^2$ (assuming $\delta\omega_\text{rms}\ll \Gamma_T$, which is valid for the range of mode volumes plotted in Fig.~4 of the main text) maximizes the resonant cavity amplitude: lower quality factors incur excess loss, whereas higher quality factors allow the qubit to ``explore" a larger region of the phase space, thereby decaying the integrated cavity amplitude. Intuitively, the resulting maximum amplitude $\langle |\tilde{a}(t)| \rangle$ increases with increasing mode volume due to the reduced magnitude of temperature fluctuations.

\subsubsection{General Derivation for $\langle\tilde{a}^2(t)\rangle$}
Solving for $\langle \tilde{a}(t)^2\rangle$ generally follows the same procedure, and reveals a limit on the allowable intra-cavity optical energy in the presence of TRN. Starting from Eqn.~\ref{steadyState}, the autocorrelation of $\tilde{a}$ takes the form
\begin{equation}
\langle \tilde{a}(t)\tilde{a}^*(0) \rangle = 2\Gamma_c|\tilde{s}_\text{in}|^2\int_{-\infty}^t \mathrm{d}t' e^{-(i\Delta+\Gamma_l)(t-t')}\int_{-\infty}^0 \mathrm{d}t'' e^{-(i\Delta-\Gamma_l)t''}\langle e^{\int_{t'}^t i\delta\omega(t_2) \mathrm{d}t_2-\int_{t''}^0 i\delta\omega(t_2) \mathrm{d}t_2} \rangle.
\end{equation}
\begin{figure}
\includegraphics[width=0.5\textwidth]{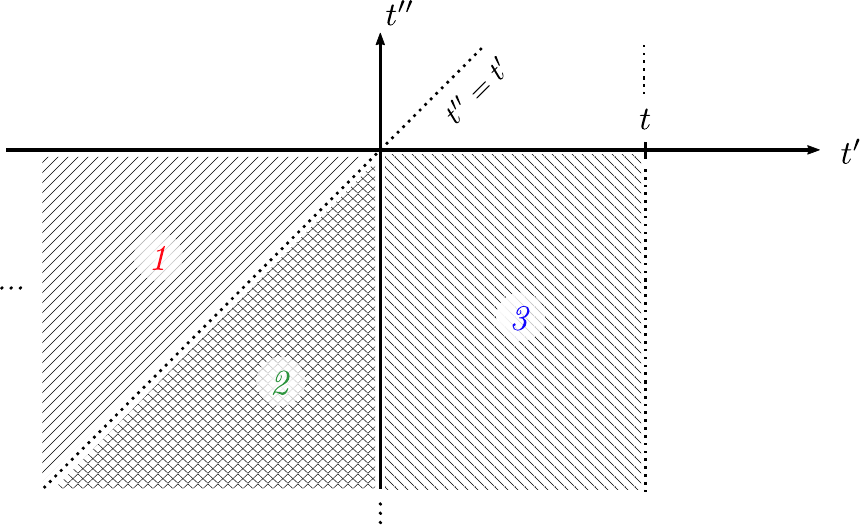}
\caption{The complete region of integration can be divided into three sub-spaces which yield different conditions for $k(t)$.}
\label{fig:intRegions}
\end{figure}
Following the method of \cite{Zhang2014a}, the average can be expressed in the form of Eqn.~\ref{NCF},
\begin{equation}
\langle e^{\int_{t'}^t i\delta\omega(t_2) \mathrm{d}t_2-\int_{t''}^0 i\delta\omega(t_2) \mathrm{d}t_2} \rangle = \langle e^{i\int_{-\infty}^\infty k(t_2) \delta\omega(t_2) \mathrm{d}t_2} \rangle = e^{-1/2\int_{-\infty}^\infty \mathrm{d}t_2' \int_{-\infty}^\infty \mathrm{d}t_2'' k(t_2')k(t_2'')\langle \delta\omega(t_2')\delta\omega(t_2'')\rangle},
\end{equation}
by appropriately defining $k(t)$. As illustrated in Fig.~\ref{fig:intRegions}, the dependence of $k(t_2)$ upon $t'$, $t''$, and $t$ differs in three sectors of the region of integration. Simple diagrams in each of the three scenarios can be used to show 
\begin{align}
k_1(t_2) =   \begin{cases}
    1, & t' \leq t_2 \leq t'' \\
    1, & 0 \leq t_2 \leq t \\
    0, & \text{else}
  \end{cases}
&&
k_2(t_2) =   \begin{cases}
    -1, & t'' \leq t_2 \leq t' \\
    1, & 0 \leq t_2 \leq t \\
    0, & \text{else}
  \end{cases}
&&
k_3(t_2) =   \begin{cases}
    -1, & t'' \leq t_2 \leq 0 \\
    1, & t' \leq t_2 \leq t \\
    0, & \text{else}
  \end{cases}
\end{align}
Combining these conditions with those of each region, we find a closed form for $k(t)$:
\begin{equation}
k(t_2) =   \begin{cases}
    \text{sign}(t''-t'), & t'<0 ~\&~ \text{min}(t',t'') \leq t_2 \leq \text{max}(t',t'') \\
    1, & (t'<0 ~\&~ 0 \leq t_2 \leq t)~||~(t'>0 ~\&~ t' \leq t_2 \leq t) \\
    -1, & t'>0 ~\&~ t''\leq t_2\leq 0\\
    0, & \text{else}.
  \end{cases}
  \label{fullK(t)}
\end{equation}
This definition allows us to rewrite the autocorrelation as 
\begin{multline}
\mathcal{R}_{aa}(t) = \langle \tilde{a}(t)\tilde{a}^*(t+\tau) \rangle = 2\Gamma_c|\tilde{s}_\text{in}|^2\int_{-\infty}^t \mathrm{d}t' e^{-(i\Delta+\Gamma_l)(t-t')}\int_{-\infty}^0 \mathrm{d}t'' e^{-(i\Delta-\Gamma_l)t''}\\ \times\exp\left[-\frac{1}{2}\int_{-\infty}^\infty \mathrm{d}t_2' \int_{-\infty}^\infty \mathrm{d}t_2'' k(t_2')k(t_2'') \delta\omega_\text{rms}^2 e^{-\Gamma_T|t_2'-t_2''|} \right].
\label{fullAutocorrelation}
\end{multline}
While a general solution to the full autocorrelation in Eqn.~\ref{fullAutocorrelation} appears intractable, we can find $\langle |\tilde{a}|^2\rangle=\langle |\tilde{a}(0)|^2\rangle$ by evaluating Eqn.~\ref{fullAutocorrelation} at $t=0$ (thereby eliminating integration region \#3 in Fig.~\ref{fig:intRegions}), which yields
\begin{multline}
\langle |\tilde{a}|^2\rangle = 2\Gamma_c|\tilde{s}_\text{in}|^2\int_{-\infty}^0 \mathrm{d}t' e^{(i\Delta+\Gamma_l)t'}\int_{-\infty}^0 \mathrm{d}t'' e^{-(i\Delta-\Gamma_l)t''}\\ \times 
\begin{cases}
\exp\left[\left(\frac{\delta\omega_\text{rms}}{\Gamma_T}\right)^2\left(1-\Gamma_T(t''-t')- e^{-\Gamma_T(t''-t')}\right)\right], & \color{Red} \text{Region 1 } \color{Black} (t'<0 ~\&~ t''>t') \\
\exp\left[\left(\frac{\delta\omega_\text{rms}}{\Gamma_T}\right)^2\left(1-\Gamma_T(t'-t'')- e^{-\Gamma_T(t'-t'')}\right)\right], & \color{ForestGreen} \text{Region 2 } \color{Black} (t'<0 ~\&~ t'>t'')
\end{cases} 
\label{t=0Form}
\end{multline}
Note that the result in either region is nearly the same -- exchanging $t'$ and $t''$ in either region returns the integral for the other region, but conjugates $i\Delta$. Therefore, we focus on evaluating Eqn.~\ref{t=0Form} in Region 1, and then generalize this result to the other region by taking the complex conjugate. In Region 1, the substitution $\tilde{\tau} = (\delta\omega_\text{rms}/\Gamma_T)^2 \exp\left[-\Gamma_T(t''-t')\right]$ yields
\begin{align}
\color{Red}\langle |\tilde{a}(0)|^2\rangle_{1} &= 2\Gamma_c|\tilde{s}_\text{in}|^2\int_{-\infty}^0 \mathrm{d}t' e^{2\Gamma_l t'}\int_{t'}^0 \mathrm{d}t'' e^{(-i\Delta+\Gamma_l)(t''-t')}\exp\left[\left(\frac{\delta\omega_\text{rms}}{\Gamma_T}\right)^2\left(1-\Gamma_T(t''-t')- e^{-\Gamma_T(t''-t')}\right)\right] \nonumber\\
&= 2\Gamma_c|\tilde{s}_\text{in}|^2\int_{-\infty}^0 \mathrm{d}t' e^{2\Gamma_l t'}\frac{e^x}{\Gamma_T}x^{-s'}\int_{xe^{\Gamma_T t'}}^{x}\tilde{\tau}^{s'-1}e^{-\tilde{\tau}} d\tilde{\tau} \nonumber \\
\color{Red}\langle |\tilde{a}(0)|^2\rangle_{1} &= 2\Gamma_c|\tilde{s}_\text{in}|^2\frac{e^x}{\Gamma_T}x^{-s'}\int_{-\infty}^0 \mathrm{d}t' e^{2\Gamma_l t'} \left[\gamma_l(s',x)-\gamma_l(s',xe^{\Gamma_T t'}) \right]
\end{align}
where $x=(\delta\omega_\text{rms}/\Gamma_T)^2$ and $s'=(\delta\omega_\text{rms}^2/\Gamma_T-\Gamma_l+i\Delta)/\Gamma_T$. The first term can be directly evaluated, while the second can be simplified with integration by parts using the relationship
\begin{equation}
\frac{\partial\gamma_l(s',x)}{\partial x} = x^{s'-1}e^{-x}.
\end{equation}
With a second substitution $\tilde{\tau}_2= xe^{\Gamma_Tt'}$ we find
\begin{align}
\color{Red}\langle |\tilde{a}(0)|^2\rangle_{1} &= 2\Gamma_c|\tilde{s}_\text{in}|^2\frac{e^x}{\Gamma_T}x^{-s'} \left[\frac{\gamma_l(s',x)}{2\Gamma_l}-\frac{x^{-\frac{2\Gamma_l}{\Gamma_T}}}{\Gamma_T}\int_0^x\tilde{\tau}_2^{\frac{2\Gamma_l}{\Gamma_T}-1}\gamma_l(s',\tilde{\tau}_2) \mathrm{d}\tilde{\tau}_2 \right] \nonumber \\ 
&= 2\Gamma_c|\tilde{s}_\text{in}|^2\frac{e^x}{\Gamma_T}x^{-s'} \left\lbrace\frac{\gamma_l(s',x)}{2\Gamma_l}-\frac{x^{-\frac{2\Gamma_l}{\Gamma_T}}}{\Gamma_T}\left(\frac{\Gamma_T}{2\Gamma_l}\right)\left[\tilde{\tau}_2^{\frac{2\Gamma_l}{\Gamma_T}}\gamma_l(s',\tilde{\tau}_2)+\gamma_u(s'+\frac{2\Gamma_l}{\Gamma_T},\tilde{\tau}_2) \right]_{\tilde{\tau}_2=0}^{\tilde{\tau}_2=x}\right\rbrace
\end{align}
where $\gamma_u(s',x)$ is the upper incomplete Gamma function defined by 
\begin{equation}
\gamma_u(s',x) = \int_x^\infty \tau^{s'-1}e^{-\tau} d\tau.
\end{equation}
Evaluating the final terms, the result simplifies nicely to
\begin{align}
\color{Red}\langle |\tilde{a}(0)|^2\rangle_{1} &= \Gamma_c|\tilde{s}_\text{in}|^2\frac{e^xx^{-s'}}{\Gamma_T\Gamma_l} \left\lbrace \gamma_l(s',x)-\left[\gamma_l(s,x)-x^{\frac{2\Gamma_l}{\Gamma_T}}\gamma_l(s'+\frac{2\Gamma_l}{\Gamma_T},x) \right]\right\rbrace \nonumber \\
&= \frac{|\tilde{s}_\text{in}|^2}{\Gamma_T}\left(\frac{\Gamma_c}{\Gamma_l}\right)e^xx^{-(s'+\frac{2\Gamma_l}{\Gamma_T})}\gamma_l(s'+\frac{2\Gamma_l}{\Gamma_T},x)
\end{align}
To find the complete result, we simply add the second term in Eqn.~\ref{t=0Form} to find
\begin{empheq}[box=\widefbox]{align}
  \langle |\tilde{a}|^2\rangle &= \frac{|\tilde{s}_\text{in}|^2}{\Gamma_T}\left(\frac{\Gamma_c}{\Gamma_l}\right)e^x\left[x^{-s}\gamma_l(s,x)+x^{-s^*}\gamma_l(s^*,x)\right] \label{a2Soln}\\
  x &\equiv \left(\frac{\delta\omega_\text{rms}}{\Gamma_T}\right)^2 \label{xDef2} \\
  s &\equiv \frac{\Gamma_l+i\Delta}{\Gamma_T}+x. \label{sDef2}
\end{empheq}
Note the similarity to Eqns.~\ref{aSoln}-\ref{sDef}. Once again, we must ensure that our solution corresponds to the noiseless result expected when $\delta\omega_\text{rms}\rightarrow 0$. Using the series expansion of $\gamma_l(s,x)$, we find 
\begin{align}
\langle |\tilde{a}|^2\rangle_{T=0} &= \lim_{x\rightarrow 0} \frac{|\tilde{s}_\text{in}|^2}{\Gamma_T}\left(\frac{\Gamma_c}{\Gamma_l}\right)e^x\left[x^{-s}\gamma_l(s,x)+x^{-s^*}\gamma_l(s^*,x)\right] \nonumber \\
&= \frac{|\tilde{s}_\text{in}|^2}{\Gamma_T}\left(\frac{\Gamma_c}{\Gamma_l}\right)\left[\frac{\gamma_f(s|_{x=0})}{\gamma_f(s|_{x=0}+1)}+\frac{\gamma_f(s^*|_{x=0})}{\gamma_f(s^*|_{x=0}+1)}\right] \nonumber \\
&= \frac{|\tilde{s}_\text{in}|^2}{\Gamma_T}\left(\frac{\Gamma_c}{\Gamma_l}\right)\left[ \frac{\Gamma_T}{\Gamma_l+i\Delta}+\frac{\Gamma_T}{\Gamma_l-i\Delta}\right]\nonumber \\
\langle |\tilde{a}|^2\rangle_{T=0} &= \frac{2\Gamma_c|\tilde{s}_\text{in}|^2}{\Delta^2+\Gamma_l^2}
\label{energySS}
\end{align}
as expected. Similar to the solution for $\langle\tilde{a}(t)\rangle$, we define the effective quality factor  
\begin{equation}
Q_\text{eff} = \frac{\omega_0|\langle\tilde{a}(t)\rangle|^2}{2|\tilde{s}_\text{in}|^2} = \frac{\omega_0}{2\Gamma_T}e^{x}x^{-s}\gamma_l(s,x)
\end{equation}
for resonant excitation ($\Delta=0$) and critical coupling ($\Gamma_c=\Gamma_l/2$). 

As opposed to the non-monotonic scaling of the mean field amplitude $\langle |\tilde{a}(t)| \rangle$ with $Q_l$, the stored energy $\langle|\tilde{a}(t)|^2\rangle$ increases monotonically with increasing $Q_l$. This is intuitively described in the main text: continuing to increase $Q_l$ decreases the cavity linewidth until $Q_\text{eff}$ is saturated by mode volume-dependent thermal noise in the high-$Q_l/V_\text{eff}$ regime. Finally, we note that the maximum energy storage (although not necessarily the maximum intensity, which also depends on the mode volume) is a achieved with large mode volumes due to reduced thermo-optic noise.

\subsubsection{Cavity Spectrum in the White Noise Limit}
The complete field autocorrelation $\mathcal{R}_{aa}(t)$ in Eqn.~\ref{fullAutocorrelation} simplifies considerably in the high-$Q_l$ limit where $\Gamma_T\gg\Gamma_l$, as the cavity resonant frequency $\omega_0(t)$ can be assumed to directly track the temperature noise over the relevant timescales. The frequency noise  is then effectively delta-correlated in time, and the aforementioned --- albeit tedious --- ``integration by regions" technique can then be similarly applied to solve for the field noise spectrum $S_{aa}(\omega)$. A more intuitive approach to this solution is though adiabatic elimination of $\omega_0(t)$'s dynamics following the procedure in \cite{Gardiner1984}. Converting the optical field and resonant frequency evolution equations (Eqns.~\ref{eqn:TCMT}, \ref{eqn:resFreqDiffEq}) into stochastic differential equations yields
\begin{align}
\mathrm{d}\tilde{a}(t) = \left\lbrace \left[ i (\delta\omega(t)+\Delta)-\Gamma_l\right]\tilde{a}(t)+\sqrt{2\Gamma_c}\tilde{s}_\text{in}\right\rbrace \mathrm{d}t && \mathrm{d}\delta\omega(t) = -\Gamma_T\delta\omega(t)\mathrm{d}t+\delta\omega_\text{rms}\sqrt{2\Gamma_T}\mathrm{d}W(t)
\label{eqn:SDEs}
\end{align}
for both It\^{o} and Stratonovich forms since the frequency noise is additive ($\delta\omega_\text{rms}\sqrt{2\Gamma_T}$ is constant). In the limit $\Gamma_T\rightarrow\infty$, we can adiabatically eliminate the resonant frequency dynamics, yielding a steady state value $\delta\omega(t) = \sqrt{2/\Gamma_T}\delta\omega_\text{rms}\mathrm{d}W(t)/\mathrm{d}t$. The cavity evolution can then be simplified to
\begin{align}
\mathrm{d}\tilde{a}_S(t) &= \left[ (i\Delta-\Gamma_l)\tilde{a}_S(t)+\sqrt{2\Gamma_c}\tilde{s}_\text{in}\right] \mathrm{d}t + \sqrt{\frac{2}{\Gamma_T}}\delta\omega_\text{rms}\tilde{a}_S(t)\mathrm{d}W(t)\\
\mathrm{d}\tilde{a}_I(t) &= \left\lbrace \left[ i\Delta-\left(\Gamma_l+\frac{\delta\omega_\text{rms}^2}{\Gamma_T}\right)\right]\tilde{a}_I(t)+\sqrt{2\Gamma_c}\tilde{s}_\text{in}\right\rbrace \mathrm{d}t + \sqrt{\frac{2}{\Gamma_T}}\delta\omega_\text{rms}\tilde{a}_I(t)\mathrm{d}W(t)
\end{align}
in Stratonovich and It\^{o} forms, respectively. Applying the It\^{o} rule $(\mathrm{d}W(t))^2=\mathrm{d}t$ to the latter, we can solve for the steady-state moments 
\begin{align}
\langle \tilde{a}(t)\rangle = \frac{\sqrt{2\Gamma_c}\tilde{s}_\text{in}}{\left[\Gamma_l+\delta\omega_\text{rms}^2/\Gamma_T \right] -i\Delta} && \langle |\tilde{a}(t)|^2\rangle = \frac{2\left(\Gamma_c+\delta\omega_\text{rms}^2/\Gamma_T \right) \tilde{s}_\text{in}}{\left[\Gamma_l+\delta\omega_\text{rms}^2/\Gamma_T \right]^2 +\Delta^2}
\label{eqn:SDEmoments}
\end{align}
which by comparison to Eqn.~\ref{ampSS} and Eqn.~\ref{energySS} immediately reveals a thermal broadening $2\Gamma_l\rightarrow 2\Gamma_l+2\delta\omega_\text{rms}^2/\Gamma_T$ of the microcavity linewidth. We can also derive an equation of motion for the autocorrelation, 
\begin{equation}
\frac{\mathrm{d}}{\mathrm{d}\tau} \langle \tilde{a}^*(t+\tau)\tilde{a}(t)\rangle = \left[ i\Delta-\left(\Gamma_l+\frac{\delta\omega_\text{rms}^2}{\Gamma_T}\right)\right]\langle \tilde{a}^*(t+\tau)\tilde{a}(t)\rangle + \sqrt{2\Gamma_c}\tilde{s}_\text{in}\langle \tilde{a}(t)\rangle.
\label{eqn:non-perturbativeAutocorrelation}
\end{equation}
Solving Eqn.~\ref{eqn:non-perturbativeAutocorrelation} subject to the $\tau=0$ conditions of Eqn.~\ref{eqn:SDEmoments}, we find
\begin{equation}
R_{aa}(t) = \frac{2\delta\omega_\text{rms}^2|s_\text{in}|^2/\Gamma_T}{\left[\Gamma_l+\delta\omega_\text{rms}^2/\Gamma_T \right]^2 +\Delta^2}e^{-\left(i\Delta+\Gamma_l +\delta\omega_\text{rms}^2/\Gamma_T \right) t},
\end{equation}
corresponding to the optical noise spectrum
\begin{equation}
S_{aa}(\omega) = \frac{2\delta\omega_\text{rms}^2|s_\text{in}|^2/\Gamma_T}{\left[\Gamma_l+\delta\omega_\text{rms}^2/\Gamma_T \right]^2 +\Delta^2}\frac{2(\Gamma_l +\delta\omega_\text{rms}^2/\Gamma_T)}{(\Gamma_l +\delta\omega_\text{rms}^2/\Gamma_T)^2+(\omega-\Delta)^2}.
\label{eqn:Saa-white}
\end{equation}
Eqn.~\ref{eqn:Saa-white} evaluated in the perturbative limit $\delta\omega_\text{rms}\ll\Gamma_l$ coincides with the low-frequency ($\omega\ll\Gamma_T$) limit of the previous perturbative spectrum (Eqn.~\ref{eqn:perturbativeSpectrum}).

\section{Experimental TRN in Photonic Crystal Cavities}
\label{sec:experiments}
The single-mode thermal decay approximation made in Eqn.~\ref{heatEqn-avg} implies the decay rate of Eqn.~\ref{gammaT} and the spectral density of cavity resonant frequency in Eqn.~\ref{eqn:Sww}. This result is commonly used as a simplifying assumption for temperature fluctuations \cite{Chui1992,Sun2017}; however, it is not immediately clear that the single-mode approximation holds in the case of small mode volume optical microcavities, where the characteristic length scales of the near diffraction-limited optical mode ($\resim \lambda/n$) can approach the phonon mean free paths \cite{Cuffe2015}. In absence of any experimental data in the literature to verify the assumption, we constructed an experiment to measure thermo-refractive noise in high-$Q_l/V_\text{eff}$ silicon photonic crystal cavities. The experiment also allows us to compare measured TRN with the spectra derived from our multi-mode theory (Section~\ref{sec:MMtheory}) or computed through finite-element Fluctuation-Dissipation simulations (Section~\ref{sec:FDTsim}).

\subsection{Photonic Crystal Cavity Sample Details}
\label{sec:samples}
The L3 and L4/3 photonic crystal cavities were fabricated by Applied Nanotools foundry via electron-beam patterning and dry-etching of 220 nm-thick undoped silicon-on-insulator wafers with a 2 $\upmu$m-thick buried oxide layer. To suspend the devices, the buried oxide was subsequently released via a 60 second timed wet etch in 49\% hydrofluoric (HF) acid. The designs were adapted from Refs.~ \cite{Minkov2014,Minkov2017}. As shown in Fig.~2 of the main text, superimposed gratings were added to improve vertical coupling efficiency. The gratings are formed via periodic hole radii perturbations ranging from $\Delta r = 0\rightarrow 0.05 r$ at a period equal to twice the lattice constant $a$. Although devices with quality factors as large as 600,000 were measured for small values of $\Delta r$, the results presented in the main text use $\Delta r=0.05r$, which significantly improves collection efficiency into our fiber-coupled detector.

\subsection{Experimental Setup}
\begin{figure}[h!]
\includegraphics[width=0.8\textwidth]{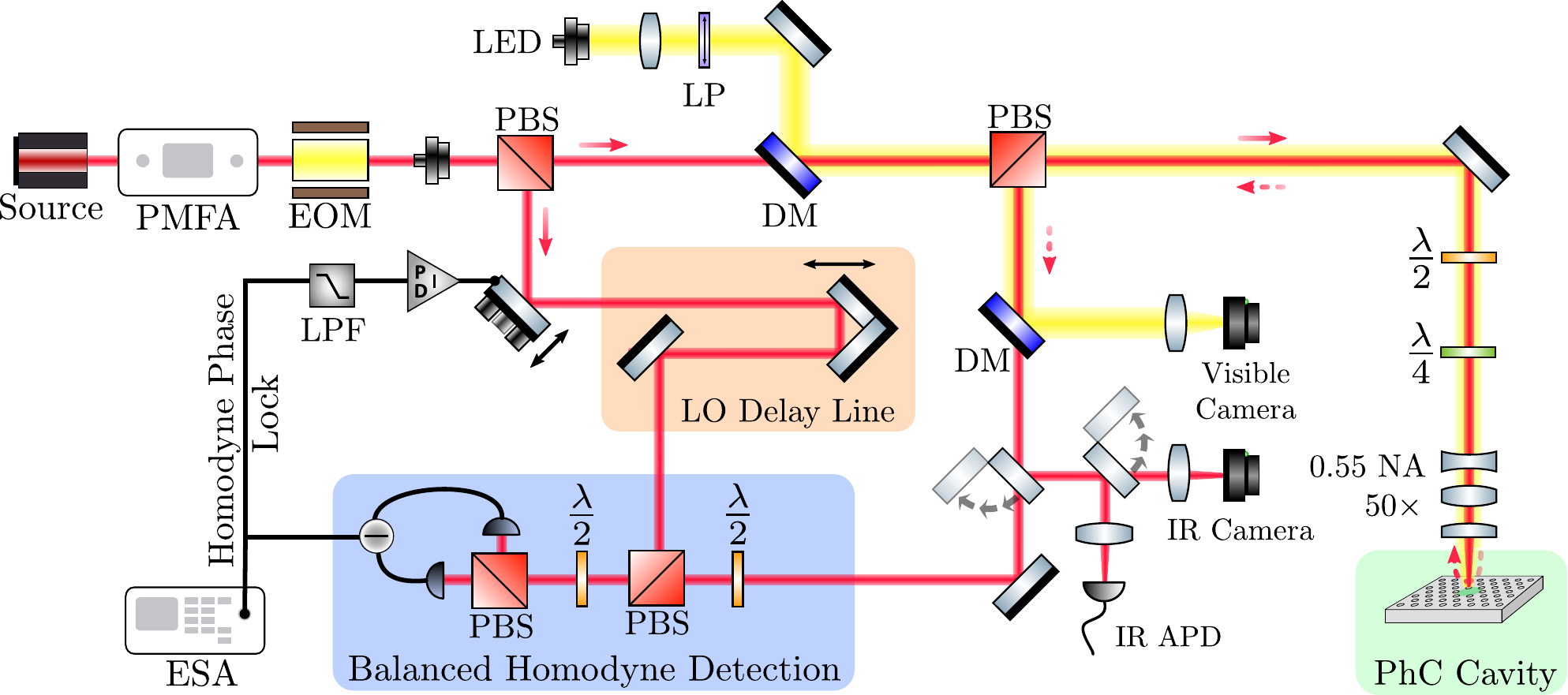}
\caption{Schematic of the setup built to measure TRN in photonic crystal cavities. An amplified (PriTel PMFA) continuous wave laser (Santec TSL-710) is separated into a local oscillator and cavity signal by a polarizing beamsplitter (PBS). The LO line is passively path-length matched to the cavity signal using a tunable retroreflector delay line. The cavity signal is combined with a linearly polarized (LP) white light source (LED) using a dichroic mirror (DM) and is reflected from a PhC cavity rotated $45^\circ$ from the incident polarization (adjustable with a half-wave plate, $\lambda/2$), which allows the cavity signal to be isolated from the specular reflection using a PBS. A quarter-wave plate ($\lambda/4$) allows the specular reflection to be extracted for comparison to the cavity-only reflection. The reflected illumination light is separated and imaged onto a silicon CCD. The cavity signal can be directed with flip-mirrors towards an IR camera for imaging, an IR avalanche photodetector (ThorLabs PDB410C 10 MHz InGaAs APD) to collect low-noise reflection spectra, or towards the balanced homodyne detector. For the latter, a balanced photodetector (ThorLabs PDB480C-AC 1.6 GHz InGaAs p-i-n Photodetector) measures the homodyne signal from the recombined cavity reflection and local oscillator, and the result is recorded on an electronic spectrum analyzer (ESA; Agilent N9010A EXA Signal Analyzer). The DC signal extracted from a low-pass filter (LPF) is used as the feedback signal for a digital PID controller which stabilizes the signal-LO phase difference by actuating a piezo-actuated mirror. An electro-optic modulator (EOM) provides a known phase noise which can be used to calibrate the frequency noise of the PhC cavity. The sample stage is temperature-stabilized to $\Delta T<0.01$ K using a peltier plate and a feedback temperature controller. }
\label{fig:setup}
\end{figure}

A more detailed version of the experimental setup depicted in the main text is provided in Fig.~\ref{fig:setup}. The setup consists of a typical polarized light microscope, where the signal reflected from a PhC cavity is measured with balanced homodyne detection. 
The homodyne detector is balanced to quadrature by zeroing the DC component of the homodyne signal with a digital PID feedback controller connected to a piezo-actuated mirror. In this configuration, the homodyne voltage signal 
\begin{equation}
v_h \sim |\tilde{a}_\text{LO}||\tilde{a}_\text{cavity}|\sin(\delta\phi_\text{cavity}(t)+\cancelto{0}{\phi_\text{cavity}^0-\phi_\text{LO}}) = |\tilde{a}_\text{LO}||\tilde{a}_\text{cavity}|\delta\phi_\text{cavity}(t)
\end{equation}
for a local oscillator signal $\tilde{a}_\text{LO}e^{i\phi_\text{LO}}$ is directly proportional to the cavity amplitude $|\tilde{a}_\text{cavity}|$ and phase fluctuations $\delta\phi_\text{cavity}(t)$ resulting from the stochastic resonant frequency. An electronic spectrum analyzer is used to measure the power spectral density $S_{vv}$ of this homodyne voltage signal.

\subsection{Phase Noise Calibration}
The resonant frequency noise spectral density $S_{\omega\omega}$ can then be determined from $S_{vv}$ using the absolute calibration technique discussed in Refs.~\cite{Schliesser2008, Gorodetsky2010}. For example, consider a Mach-Zehnder interferometer with input power $P_\text{in}$ and splitting ratio $\eta_h$, which creates the in-phase local oscillator and cavity input signals 
\begin{align}
\tilde{a}_\text{LO} = \sqrt{\eta_HP_\text{in}} && \tilde{a}_\text{in} = \sqrt{(1-\eta_H)P_\text{in}}.
\end{align}
Assuming resonant drive ($\Delta = \omega_L-\omega_0=0$) for a cavity with input/output power coupling rate $\Gamma_c$, total loss rate $\Gamma_l$, and a perturbative resonant frequency noise $\delta\omega(t)$, the output cavity signal is then
\begin{align}
\tilde{a}_\text{out} = \sqrt{(1-\eta_H)P_\text{in}}\frac{\Gamma_c}{\Gamma_l+i\delta\omega(t)},
\end{align}
yielding a homodyne detection voltage
\begin{align}
v_h(t) &\approx 2G_c|\tilde{a}_\text{LO}||\tilde{a}_\text{out}|\delta\phi_\text{cavity}(t)\nonumber \\
v_h(t) &\approx 2G_c \sqrt{\eta_H(1-\eta_H)}P_\text{in}\frac{\Gamma_c}{\Gamma_l^2}\delta\omega(t)
\end{align}
for a detector conversion gain $G_c$. The final frequency noise spectral density
\begin{equation}
S_{vv}^{\delta\omega} \approx \underbrace{4G_c^2 \eta_H(1-\eta_H)\left(\frac{\Gamma_c}{\Gamma_l^2}\right)^2 P_\text{in}^2}_{K_\text{expt}}S_{\omega\omega} = K_\text{expt}S_{\omega\omega}
\end{equation}
is therefore a function of various experimental constants and cavity coupling parameters. 

However, the value of $K_\text{expt}$ can be exactly determined by injecting a known phase noise $\delta\phi(t) = \phi_m(V_p) cos(\omega_m t)$ into the interferometer with an electro-optic modulator driven with an electrical tone with frequency $\omega_m$ and peak voltage $V_p$. Under the same experimental conditions, the local oscillator and cavity input signals are
\begin{align}
\tilde{a}_\text{LO} = \sqrt{\eta_HP_\text{in}}e^{i\phi_m(V_p)\cos(\omega_m t)} && \tilde{a}_\text{in} = \sqrt{(1-\eta_H)P_\text{in}}e^{i\phi_m(V_p)\cos(\omega_m t)}.
\end{align}
With a small enough modulation depth $\phi_m(V_p) = \pi V_p/V_\pi$ (and therefore a small enough drive voltage $V_p$ for a given half-wave voltage $V_\pi$), the local oscillator can be approximated to first order as
\begin{equation}
\tilde{a}_\text{LO} \approx \sqrt{\eta_HP_\text{in}}\left(1+i\phi_m(V_p)\cos(\omega_m t)\right).
\end{equation}
Similarly, assuming $\omega_m\ll \Gamma_l$ (as is the case in our experiment), the cavity response yields the output signal 
\begin{align}
\tilde{a}_\text{out} &= \sqrt{(1-\eta_H)P_\text{in}}\left[\frac{\Gamma_c}{\Gamma_l+i\Delta}+i\frac{\phi_m(V_p)}{2}\left(\frac{\Gamma_c}{\Gamma_l+i(\Delta+\omega_m)}e^{i\omega_mt}+\frac{\Gamma_c}{\Gamma_l+i(\Delta-\omega_m)}e^{-i\omega_mt}\right)\right] \nonumber \\
\tilde{a}_\text{out} &\approx \sqrt{(1-\eta_H)P_\text{in}}\frac{\Gamma_c}{\Gamma_l}\left[1+i\phi_m(V_p)\left(\cos(\omega_mt)+\frac{\omega_m}{\Gamma_l}\sin(\omega_mt)\right)\right].
\end{align}
The homodyne signal
\begin{align}
v_h(t) &\approx 2G_c \sqrt{\eta_H(1-\eta_H)}P_\text{in}\frac{\Gamma_c}{\Gamma_l^2}\omega_m\phi_m(V)\sin(\omega_m t)
\end{align}
corresponds to a power spectral density
\begin{equation}
S_{vv}^{\delta\phi_m} \approx \underbrace{4G_c^2 \eta_H(1-\eta_H)\left(\frac{\Gamma_c}{\Gamma_l^2}\right)^2 P_\text{in}^2}_{K_\text{expt}}\omega_m^2S_{\phi\phi}\big|_{\omega=\omega_m},
\end{equation}
which, similar to $S_{vv}^{\delta\omega}$, is directly proportional to $K_\text{expt}$. $K_\text{expt}$ can therefore be eliminated to yield an absolute calibration for the resonant frequency noise spectral density:
\begin{equation}
S_{\omega\omega} \approx \frac{S_{vv}^{\delta\omega}}{K_\text{expt}} \approx \frac{\omega_m^2 S_{\phi\phi}\big|_{\omega=\omega_m}}{S_{vv}^{\delta\phi_m}\big|_{\omega=\omega_m}} S_{vv}^{\delta\omega}.
\label{eqn:cal}
\end{equation}
This result can be simplified by evaluating the phase spectral density
\begin{equation}
S_{\phi\phi}(\omega) = \int_{-\infty}^{\infty} \phi_m^2(V_p) \langle \cos(\omega_m t) \cos(\omega_m (t+ \tau))\rangle e^{-i\omega\tau} \mathrm{d}\tau = \frac{\phi_m^2(V_p)}{2} \left[\frac{1}{2}\delta(\omega-\omega_m)  \frac{1}{2}\delta(\omega+\omega_m) \right].
\end{equation}
The spectrum analyzer convolves the $\delta$-function with the input filter function $F(\omega)$, which is normalized such that $F(0) = \frac{1}{\text{ENBW}}$ \cite{Gorodetsky2010}, where the effective noise bandwidth $\text{ENBW}=\eta_\text{F}\text{RBW}$ for a resolution bandwidth RBW and a filter shape-dependent $\eta_\text{F}\approx 1$. Therefore, the measured noise spectral density evaluated at the modulation frequency $\omega_m$ becomes
\begin{equation}
S_{\phi\phi}(\omega=\omega_m) = \frac{\phi_m^2(V_p)}{4} F(\omega)*\delta(\omega-\omega_m) = \frac{\phi_m^2(V_p)}{4\cdot \text{ENBW}}.
\end{equation}
Using this result, the calibration term in Eqn.~\ref{eqn:cal} can be simplified to a final form
\begin{equation}
\boxed{S_{\omega\omega}(\omega) = \frac{\omega_m^2 \phi_m^2(V_p)}{4\eta_F\cdot\text{RBW}} \frac{S_{vv}^{\delta\omega}(\omega)}{S_{vv}^{\delta\phi_m}(\omega_m)}}
\label{calPSD}
\end{equation}
that agrees with Eqn.~20 of Ref.~\cite{Gorodetsky2010}.

\subsubsection{Electo-optic Phase Modulator Calibration}
Eqn.~\ref{calPSD} demonstrates that the calibrated frequency noise can be readily obtained by comparing the recorded RF power spectral density $S_{vv}^{\delta\omega}(\omega)$ to the calibration PSD $S_{vv}^{\delta\phi_m}(\omega_m)$ (which corresponds to a known phase spectral density) for a given calibration frequency $\omega_m/2\pi$ and spectrum analyzer RBW. The ENBW correction factor $\eta_F$ is a function of various spectrum analyzer settings (see Ref.~\cite{KeysightTechnologiesA} for example), and is therefore measured by comparing the noise marker amplitude (dBm/$\sqrt{\text{Hz}}$) to the measured PSD divided by the RBW. This technique yields $\eta_F\approx 1.057$, which is approximately equal to the value given in Ref.~\cite{KeysightTechnologiesA} assuming typical spectrum analyzer settings. 

\begin{figure}
\includegraphics[width=\textwidth]{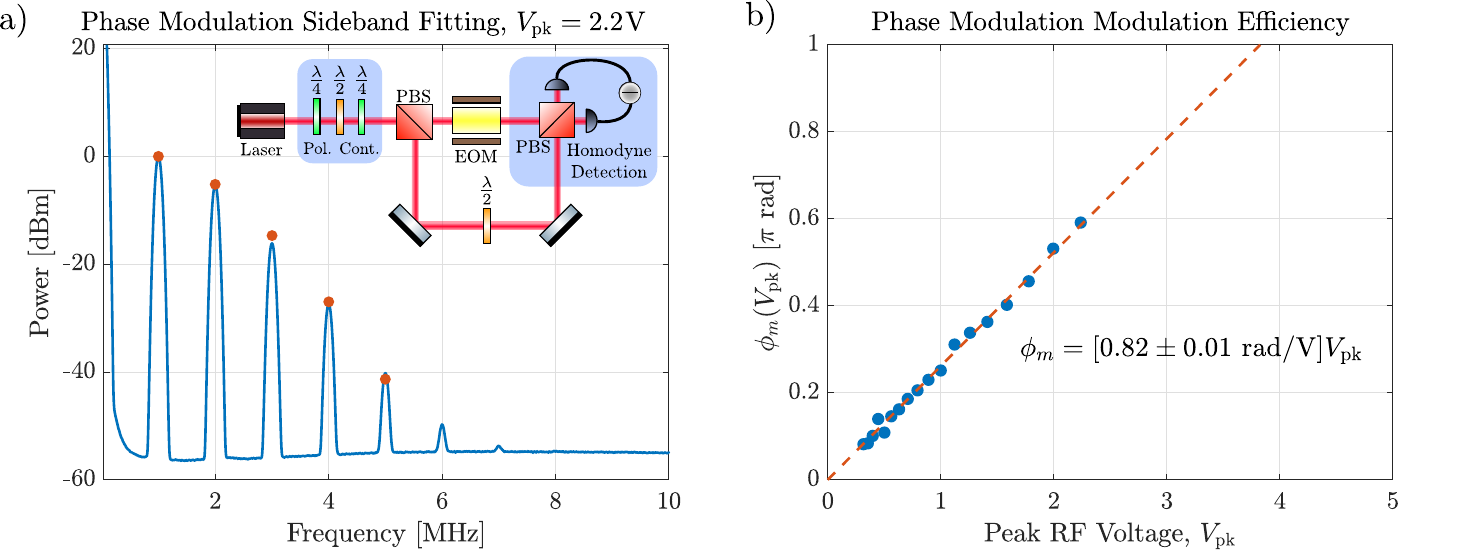}
\caption{Measurement of phase modulator modulation depth $\phi_m(V_p)$ at $\lambda=1550$ nm. A balanced homodyne measurement is performed on the output of a Mach-Zehnder interferometer with the electro-optic modulator (EOM) in one arm, yielding spectra similar to that of (a). The sideband amplitudes are fitted to find the modulation depth at each peak drive voltage $V_p$, and a linear fit is applied to find the modulation efficiency. The measured value $\phi_m/V_p = 0.82 \pm 0.01$ rad/V corresponds to a half-wave voltage $V_\pi= 3.83$ V.}
\label{fig:PMcal}
\end{figure}

The only remaining unknown parameter required for calibration is the peak-voltage-dependent modulation depth $\phi_m(V_p)$ of the phase modulator, which can be determined with a sideband fitting technique as shown in Fig.~\ref{fig:PMcal}. An electro-optic phase modulator (EOM) is embedded in one arm of an unbalanced Mach-Zehnder interferometer, yielding a homodyne signal 
\begin{equation}
v_h\propto \sum_n J_n(\phi_m)\cos(n\omega_mt)
\end{equation} 
for a modulation frequency $\omega_m$. The power spectrum observed on the spectrum analyzer therefore consists of a periodic sequence of $\delta$-like functions (spectrum analyzer filter functions $F(\omega-n\omega_m)$, to be explicit) at frequencies $\omega_n=n\omega_m$ with powers proportional to $J_n^2(\phi_m)$. Fitting the sideband powers (relative to the $n=1$ sideband, as the $n=0$ peak is inaccessible on the AC-coupled spectrum analyzer) via a least-squares regression yields $\phi_m(V_p)$ for any peak drive voltage $V_p$. Fig.~\ref{fig:PMcal}(a) illustrates the result for $\lambda=1550$ nm and $V_p=2.24$ V, where the Bessel functions evaluated at $\phi_m\approx 0.6$ rad (red points) are well fitted to the measured (blue curve) peak amplitudes. After repeating the experiment for multiple values of $V_p$, a linear fit (Fig.~\ref{fig:PMcal}(b)) gives the modulation efficiency
\begin{equation}
\boxed{\eta_\text{mod}=\frac{\phi_m}{V_p}\bigg|_{\lambda_0=\lambda_\text{cal}} = \frac{\lambda_\text{cal}}{\lambda_0}\frac{\phi_m}{V_p}\bigg|_{\lambda_0=\lambda_\text{cal}} = 0.82\pm 0.01 \text{ rad/V}}
\end{equation} 
corresponding to a half-wave voltage $V_\pi = 3.75$ V (roughly in line with the manufacturer quoted value of $3.17$ V) at the calibration wavelength.

Note that the DC phase of the fiber interferometer in this experiment was unstable, and was therefore allowed to drift while the measurement was averaged on a timescale much longer than the drift -- a standard technique \cite{Ludvigsen1998} which only affects the total power of the homodyne signal, not the quantity of interest (relative magnitude of the sidebands).

\subsubsection{Balanced Homodyne Detector Characterization}
Using the measured EOM modulation efficiency, the calibrated thermo-refractive noise measurements in Fig.~2 of the main text were obtained by measuring the cavity reflection with the stabilized homodyne detector in Fig.~\ref{fig:setup}. We confirmed that the balanced photodetection was shot noise limited (with 10 dB of shot noise clearance) for frequencies greater than $\resim$100 kHz and balanced the interferometer arms to well within 1 mm -- over an order of magnitude shorter than the expected cavity delay ($\resim$cm). This was achieved by tuning a  retroreflector-based delay line while observing pulse delays from a picosecond fiber laser on both interferometer paths. 


\subsection{Additional Measurements for Varying Probe Powers}
\begin{figure}
\includegraphics[width=0.6\textwidth]{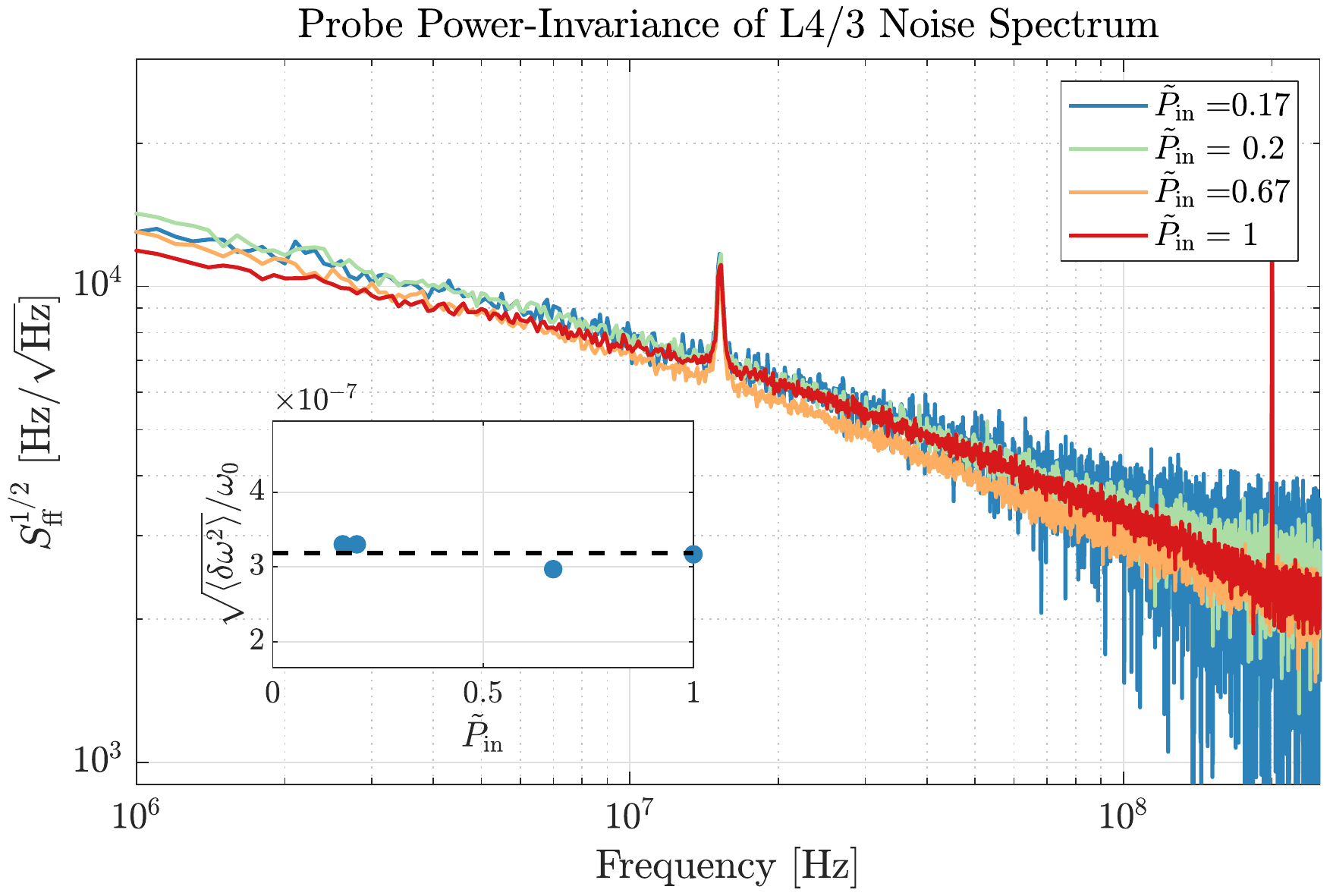}
\caption{L4/3 resonant frequency noise $S_\text{ff}$ measurements as a function of the input power $\tilde{P}_\text{in}$. $\tilde{P}_\text{in}$ is normalized to the maximum input probe power for the dataset. The rms fractional frequency fluctuation $\sqrt{\langle\delta \omega^2\rangle}/\omega_0$ is computed from the integrated noise over the plotted measurement bandwidth and plotted in the inset as a function of $\tilde{P}_\text{in}$. No significant power scaling or deviation from the mean value (black dashed line) is observed, indicating that noise contributions from nonlinear effects can be neglected.}
\label{fig:powerinvariance}
\end{figure}

In the main text, we note that the measured frequency noise spectra are independent of the cavity probe power for sufficiently low input powers. This claim is supported by Fig.~\ref{fig:powerinvariance}, which plots the calibrated, background-corrected L4/3 cavity frequency noise spectra for various normalized input powers $\tilde{P}_\text{in}$.
The inset depicts the nearly constant integrated noise as a function of $\tilde{P}_\text{in}$. The range of plotted input powers (spanning about an order of magnitude) is limited by the setup and characteristics of the PhC cavities: the homodyne measurement fails to properly lock if the input power is too low (further frustrated by the signal loss when coupling the non-Gaussian PhC emission into single mode fiber), whereas too high of an input power excites nonlinear intracavity effects, such as two-photon absorption (further details in Section~\ref{sec:noiseComparison}).

\subsection{Summary of Experimental Parameters}
Table~\ref{table:paramsExpt} summarizes the various experimental parameters used to generate the data and fit parameters shown in the main text. Note that, as described in the caption of Fig.~2 of the main text, the expected thermal diffusivity is based on thermal conductivity measurements in thin silicon films \cite{Cuffe2015} and the hole lattice porosity $\phi$ \cite{Jain2013}. The porosity $\phi=3\pi r^2/(3\sqrt{3} a^2/2-3\pi r^2)$ --- calculated as the ratio of hole area $3\pi r^2$ (assuming a hole radius $r$) to material area within a hexagonal unit cell of a lattice with lattice constant $a$ --- reduces the thin film diffusivity to $D_T=D(1-\phi)/(1+\phi)$ \cite{Jain2013}. This ``restricted" diffusivity is used to calculate the expected decay rates in Fig.~2 of the main text.

\renewcommand{\arraystretch}{1.2}
\begin{table}
\footnotesize
\centering
\begin{tabular}{@{} L{2in}C{0.5in}C{1.5in}C{2in} @{}} \toprule
\textbf{Parameter} & \textbf{Symbol} & \textbf{Value} & \textbf{Source} \\
\midrule
Temperature & $T$ & 295.68 K & Measured \\
Si Refractive Index & $n_\text{Si}$ & 3.48 & \cite{Komma2012} \\ 
Si Thermo-optic Coefficient & $\alpha_\text{TO}^\text{Si}$ & $1.8\times 10^{-4}$ K$^{-1}$ & \cite{Komma2012} \\ 
Si Specific Heat & $c_{V}^\text{Si}$ & 1.64 J/cm$^3\cdot$K & \cite{Cuffe2015} \\
Si Thermal Conductivity & $\kappa_\text{Si}$ & 70 W/m$\cdot$K & \cite{Cuffe2015} \\ 
Si Thermal Diffusivity (thin film) & $D^\text{Si}$ & 0.43 cm$^2$/s & $\kappa/c_V$ \\
Lattice Porosity & $\phi$ & $\lbrace 0.29, 0.26 \rbrace$ & Calculated \\ 
Patterned Thermal Diffusivity & $D_\text{T}^\text{Si}$ & $\lbrace 0.23, 0.25\rbrace$ cm$^2$/s & $D^\text{Si}(1-\phi)/(1+\phi)$ \cite{Jain2013} \\ 
Resonant Wavelength & $\lambda_0$ & $\lbrace 1559.3, 1551.5 \rbrace$ nm & Measured \\ 
Quality Factor & $Q_l$ & $\lbrace \text{168,000}, \text{163,000} \rbrace$ & Measured \\ 
Phase Modulator Efficiency & $\eta_\text{mod}$ & 0.821$\text{ rad/V}$ & Measured (1550 nm) \\
ESA Noise Correction Factor & $\eta_F$ & 1.057 & Measured \\
Mode Confinement Factor & $\gamma_\text{Si}$ & $\lbrace 0.96, 0.95 \rbrace$ & Simulated \\ 
Mode Volume & $\tilde{V}_\text{eff}$ & $\lbrace 0.95, 0.32 \rbrace$ & Simulated \\ 
Thermal Mode Volume & $\tilde{V}_T$ & $\lbrace 3.92, 1.51 \rbrace$ & Simulated (Eqn. \ref{eqn:Vt}) \\ 
\bottomrule
\end{tabular}
\caption{Parameters used for calibrating the noise spectrum and computing or fitting $\Gamma_T$ and $V_T$. Independent values $n$ for L3 and L4/3 microcavities are listed as $\lbrace n_\text{L3}, n_\text{L4/3}\rbrace$ for cavity-dependent parameters. The mode confinement factor $\gamma_\text{Si}\sim 1$ confirms the validity of Eqn.~\ref{calPSD}, which assumes complete confinement of the mode in silicon.}
\label{table:paramsExpt}
\end{table}


\subsection{Finite Element Fluctuation-Dissipation Simulations}
\label{sec:FDTsim}

\begin{figure}
\includegraphics[width=0.6\textwidth]{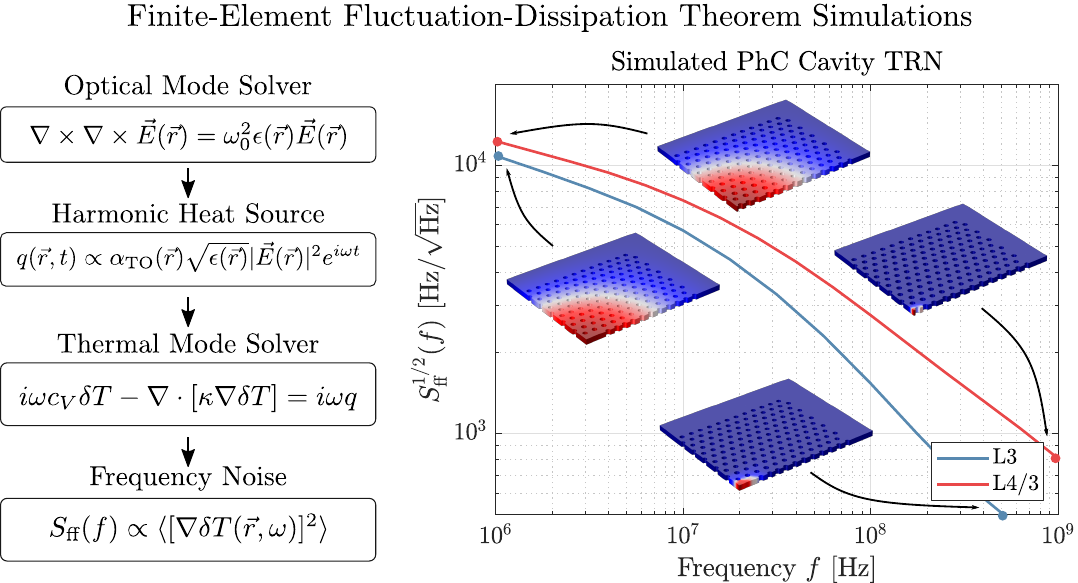}
\caption{Fluctuation-Dissipation simulations for thermo-refractive noise. A harmonic heat source with the same shape as the optical mode generates a harmonic temperature profile from which the dissipated power --- and therefore the fluctuation --- are calculated. The resulting noise spectra for the L3 and L4/3 cavities is shown (right) along with a few characteristic temperature profiles (where the color scale varies from blue to red for the range $\delta T \in [0,\delta T_\text{max}]$). At low frequencies, the thermal diffusion length is much longer than the characteristic dimensions of the optical mode. The resulting temperature profiles and noise spectral densities of either cavity converge. The opposite is true at high frequencies: the temperature profiles closely follow the shapes of the distinct optical modes, and $S_\text{ff}$ varies significantly between the two cavities.}
\label{fig:FDTsim}
\end{figure}

As schematically outlined in Fig.~\ref{fig:FDTsim}, the approximate spectrum of thermo-refractive noise in microcavities can be computed using Levin's modified form of the Fluctuation-Dissipation theorem \cite{Levin1997,Levin2008}. In this formulation, fluctuations of a readout variable $y = \int q(\vec{r}) x(\vec{r}) \mathrm{d}^3\vec{r}
$ --- the spatial average of the generalized coordinate $x(\vec{r})$ weighted by $q(\vec{r})$ --- are calculated by driving the momentum conjugate to $x$ with a harmonic force $F(\vec{r},t) = F_0 q(\vec{r}) \cos(\omega t)$. The resulting noise spectrum at the equilibrium temperature $T$,
\begin{equation}
S_{yy}(\omega) = \frac{2k_BT}{\pi\omega^2F_0^2}W_\text{diss},
\end{equation}
is then computed from the time-averaged dissipated power $W_\text{diss}$. In the case of thermo-refractive noise in a homogeneous medium (a suitable assumption for highly confined ``dielectric mode" PhC cavities),
\begin{equation}
y = |\delta\omega| = \int \underbrace{\frac{\omega_0}{n} \alpha_\text{TO}\frac{|\vec{E}(\vec{r})|^2}{\int |\vec{E}(\vec{r'})|^2\mathrm{d}^3\vec{r'}}}_{q(\vec{r})}\delta T(\vec{r}) \mathrm{d}^3\vec{r}
\end{equation}
The power dissipated from irreversible heat flow following the harmonic injection of the volumetric entropy density $F(\vec{r},t)$ (conjugate to $\delta T$) is \cite{Levin2008}
\begin{equation}
W_\text{diss} = \int \frac{\kappa}{T} \langle [\nabla\delta T(\vec{r},t)]^2\rangle \mathrm{d}^3\vec{r},
\end{equation}
where $\kappa$ is the material thermal conductivity and $\delta T(\vec{r},t)$ is the resulting harmonic temperature profile.

Fig.~\ref{fig:FDTsim} summarizes the computational steps to implement this procedure and the resulting noise spectra. We first use a finite element eigensolver to compute the optical mode $|\vec{E}(\vec{r})|$. A harmonic heat source with magnitude $Q=T F(\vec{r},t)$ is used as the source in a frequency-domain heat equation solver, which yields the harmonic temperature profile $\delta T(\vec{r})$ from which the noise spectral density $S_{\omega\omega}$ is computed at the drive frequency. Iteratively running the heat equation solver at each frequency of interest then yields the desired noise spectrum $S_{\omega\omega}(\omega)$. The material properties used in the simulations match those assumed for our experiment (and corresponding fits), and are therefore listed in Table~\ref{table:paramsExpt}.

\subsection{Comparison of Other Noise Sources}
\label{sec:noiseComparison}
Other stochastic processes can also produce resonant frequency noise. Here we consider two such sources: 1) multi-photon absorption leading to photothermal shot noise from free carrier recombination, and 2) self phase modulation. Both noise sources evaluated at their respective nonlinear thresholds --- as an estimate of the ``worst case" maximum noise levels --- are found to be more than a factor of two weaker than TRN. Since the cavity is measured well within the linear regime, we find that TRN dominates both other contributions combined, thus further confirming our experimental measurements.

\subsubsection{Multi-Photon Absorption}
\begin{figure}
\includegraphics[width=0.8\textwidth]{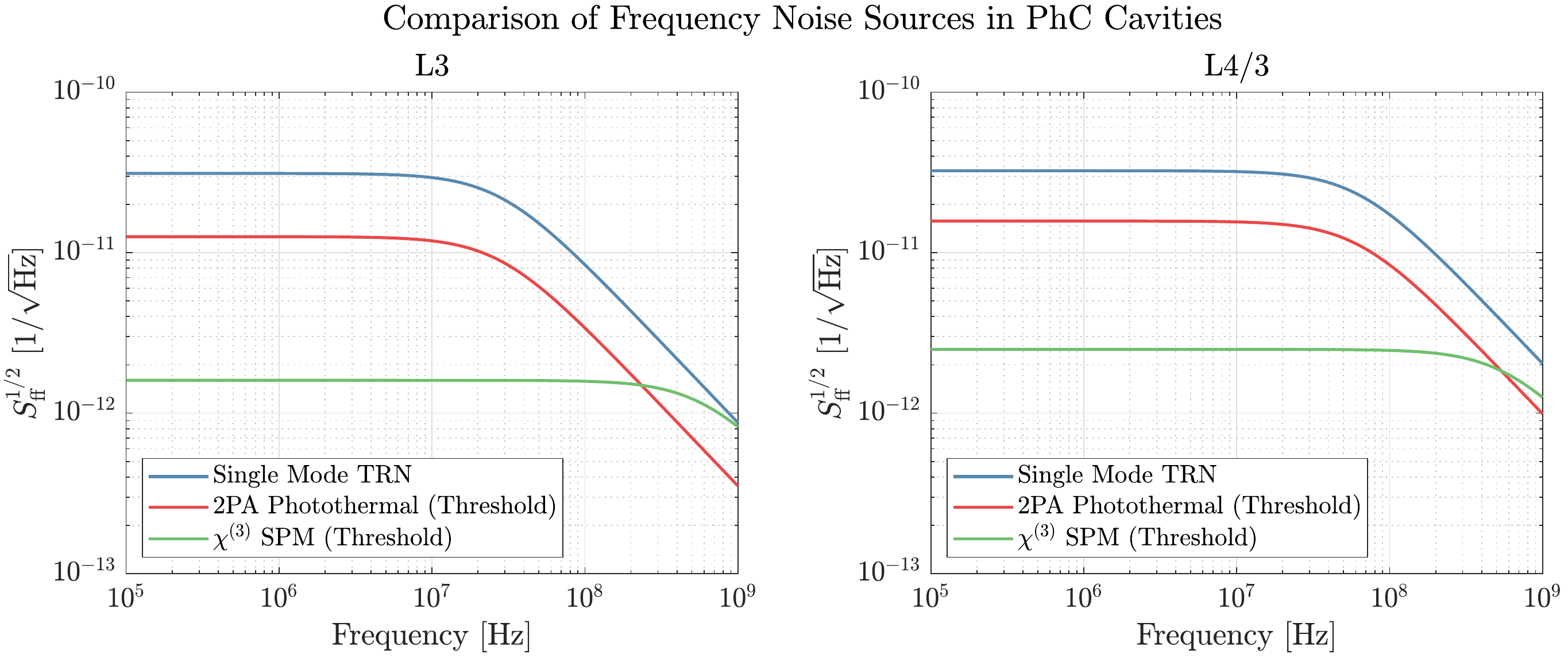}
\caption{Approximate spectrum of microcavity noise sources for the experimental parameters in Table~\ref{table:paramsExpt}. Note that $S_\text{ff}^{1/2}$ is plotted as a fractional stability (units 1/$\sqrt{\text{Hz}}$ to aid comparison with cavity stabilization literature. Noise from two-photon absorption (2PA) and self phase modulation (SPM) at their respective nonlinear threshold powers --- which approximates the maximum noise level --- is still smaller than TRN.}
\label{fig:NoiseComparison}
\end{figure}

Multi-photon absorption (MPA) within the resonator leads to a free carrier population that stochastically recombines, producing random local heating analogous to fundamental thermo-refractive noise. Considering this similarity, we can analyze the MPA photothermal shot noise by redefining the statistics of the mode averaged temperature driving force $\bar{F}_T(t)$ in Eqn.~\ref{heatEqn-avg}. The mean rate of intra-cavity $k$-photon absorption is \cite{Panuski2019}
\begin{equation}
\langle r_{k\text{PA}} \rangle = \frac{\beta_k}{k\hbar\omega_0}I_\text{pk}^kV_{k\text{PA}},
\label{eqn:kPArate}
\end{equation}
where $I_\text{pk}=c|\tilde{a}|^2/2n V_\text{eff}$ is the peak intensity of the stored energy $|\tilde{a}|^2$, $\beta_k$ is the $k$-photon absorption coefficient, $c$ is the speed of light, and $V_{k\text{PA}}=\int_\text{dielectric} |\vec{E}(\vec{r})|^{2k} \mathrm{d}^3 r/\text{max}\lbrace |\vec{E}(\vec{r})|^{2k} \rbrace$. Note that we assume that the heating produced by the photoexcited free carriers is local (i.e. no carrier diffusion). The variance of $\bar{F}_T(t)$ is then determined from the temperature change expected from the variance of MPA events within an infinitesimally small time (a Poisson process), yielding the autocorrelation
\begin{equation}
\langle \bar{F}_{T,k}^*(t) \bar{F}_{T,k}(t')\rangle = \frac{k\hbar\omega_0}{c_V^2}\beta_k I_\text{pk}^kV_{k\text{PA}}\delta(t-t').
\end{equation}
Following the method of Section~\ref{sec:TRNstatistics}, we arrive at the spectral density
\begin{equation}
S_{\omega\omega}^{k\text{PA}}(\omega) = \left(\frac{\omega_0}{n}\alpha_\text{TO} \right)^2 \frac{k\hbar\omega_0}{c_V^2} \beta_k I_\text{pk}^kV_{k\text{PA}}\frac{1}{\Gamma_T^2+\omega^2},
\end{equation}
which can be evaluated for any intra-cavity stored energy. Here, we consider $I_\text{pk}$ at the nonlinear threshold, i.e. the peak intensity for a linewidth resonance shift $|\langle \Delta\omega \rangle| /2\Gamma_l = \alpha_\text{TO}\langle \Delta T_{k\text{PA}}\rangle Q_l /n = 1$. The threshold intensity can therefore be derived from the steady state value of Eqns.~\ref{heatEqn-avg}, which lends the average temperature change
\begin{equation}
\langle \Delta T_{k\text{PA}} \rangle = \frac{k\hbar\omega_0\langle r_{k\text{PA}}\rangle}{c_V V_{k\text{PA}}\Gamma_T} = \frac{\beta_k I_\text{pk}^k}{c_V\Gamma_T}.
\end{equation}
Substituting this result into the spectral density equation assuming two-photon absorption as the dominant process (true for our silicon cavities driven at $\resim$1550 nm), we can simplify to the final result
\begin{equation}
\boxed{S_{\omega\omega}^{2\text{PA},\text{threshold}}(\omega) = \left(\frac{\omega_0^2}{n}\alpha_\text{TO} \right) \frac{\hbar\omega_0}{c_V Q_l V_\text{eff}^{(2)}} \frac{2\Gamma_T}{\Gamma_T^2+\omega^2}}
\end{equation}
for the nonlinear mode volume $V_\text{eff}^{(2)}=V_{2\text{PA}}$ (c.f. Eqn.~\ref{eqn:Veff_2}). This result is plotted in Fig.~\ref{fig:NoiseComparison} assuming the experimental parameters of our devices listed in Table~\ref{table:paramsExpt}. Comparing with Eqn.~\ref{eqn:Sww-sm}, we find $S_{\omega\omega}^{2\text{PA},\text{threshold}}/S_{\omega\omega}^\text{TRN}=(n V_T/\alpha_\text{TO}T)(\hbar\omega_0 V_\text{eff}^{(2)}/k_BT)$, which accounts for the factor of $\resim$2 weaker maximum photothermal shot noise in our devices as shown in Fig.~\ref{fig:NoiseComparison}. We operate with an input power much lower than the nonlinear threshold power (such that $\langle \Delta\omega_{2\text{PA}}\rangle\ll \Gamma_l$), so the experimental photothermal shot noise is substantially weaker than the maximum value calculated here.

\subsubsection{Kerr Self Phase Modulation}
When confined in a $\chi^{(3)}$ nonlinear material, Poissonian fluctuations of the mean intra-cavity photon number impart self phase modulational (SPM) noise on the resonant frequency. From first-order perturbation theory, the Kerr index change $\delta n(\vec{r}) = 3\chi^{(3)}\epsilon(\vec{r})|\vec{E}(\vec{r})|^2/8\epsilon_0n^3$ results in a resonant frequency shift
\begin{equation}
\left(\frac{\delta\omega(t)}{\omega_0}\right)_\text{Kerr} = -\frac{3\chi^{(3)}}{4\epsilon_0n^4V_\text{Kerr}}\delta|\tilde{a}(t)|^2
\end{equation}
where $\epsilon_0$ is the free space permittivity, $|\tilde{a}(t)|^2$ is the stored energy, and the Kerr mode volume $V_\text{Kerr}$ is equal to the thermal mode volume $V_T$ \cite{Choi}. When driven with a classical source, the intra-cavity energy autocorrelation 
\begin{equation}
\langle |\tilde{a}(t)|^2|\tilde{a}(t')|^2\rangle = \left( \frac{2\Gamma_c}{\Gamma_l^2+\omega^2}\right)^2\hbar\omega_0\langle |\tilde{s}_\text{in}|^2\rangle \delta(t-t')
\end{equation}
can derived from temporal coupled mode theory assuming a constant pump power $\langle |\tilde{s}_\text{in}|^2 \rangle$ coupled at rate $\Gamma_c$ to a cavity with composite amplitude decay rate $\Gamma_l$. The corresponding resonant frequency autocorrelation can then be used to compute the noise spectral density
\begin{equation}
S_{\omega\omega}^\text{SPM}(\omega) = \left(\frac{3\chi^{(3)}}{4\epsilon_0n^4V_T}\right)^2\left(\frac{2\Gamma_c}{\Gamma_l^2+\omega^2} \right)^2 \hbar\omega_0^3 \langle |\tilde{s}_\text{in}|^2 \rangle.
\end{equation}

Similar to the multi-photon absorption case, we evaluate this result at the nonlinear threshold $\langle \Delta\omega_\text{Kerr}\rangle /2\Gamma_l=1$ for a conservative estimate of the associated noise. The final result, considering $\Gamma_c\approx \Gamma_l$ due to the efficient vertical coupling afforded by a superimposed grating (Section~\ref{sec:samples}), is
\begin{equation}
\boxed{S_{\omega\omega}^\text{SPM,threshold}(\omega) = \left(\frac{3\chi^{(3)}}{4\epsilon_0n^4V_TQ_l}\right)\left(\frac{2\Gamma_l}{\Gamma_l^2+\omega^2} \right) \hbar\omega_0^3.}
\end{equation}
Even at the threshold power, Fig.~\ref{fig:NoiseComparison} shows that the SPM noise is over an order of magnitude weaker than TRN.

\section{Comparison of TRN in Various Materials}
Surprisingly, the $Q_\text{eff}^\text{max}/V_\text{eff}$ limits shown in Fig.~3 of the main text for several common materials lie within an order of magnitude. As shown in Table~\ref{table:params}, this observed invariance can be attributed to a inverse relationship between the thermo-optic coefficient and thermal diffusivity in common materials. Yet this relationship is not fundamental: aluminum nitride, for example, is shown to outperform all other plotted materials by over an order of magnitude due to its simultaneously large thermal conductivity and small thermo-optic coefficient. This realization demonstrates the importance of material choice when designing state-of-the-art high-$Q/V_\text{eff}$ resonators.

\renewcommand{\arraystretch}{1}
\begin{table}
\footnotesize
\centering
\begin{tabular}{@{} C{0.7in}C{0.5in}C{0.8in}C{0.5in}C{0.9in}C{1.2in} @{}} \toprule
\textbf{Material} & \textbf{Index} $n$ & \textbf{TO coeff.} $\alpha_\text{TO}$ [K$^{-1}$] & \textbf{Density} $\rho$ [g/cm$^{3}$] & \textbf{Heat capacity} $c_V$ [J/g$\cdot$K] & \textbf{Thermal diffusivity} $D_T$ [cm$^2$/s] \\
\midrule
\textbf{Si} & 3.48 & \color{red} $1.8\times 10^{-4}$ & 2.32 & 0.7 & \color{ForestGreen} 0.8 \\
\textbf{GaAs} & 3.38 & \color{red} $2.35\times 10^{-4}$ & 5.32 & 0.35 & \color{ForestGreen} 0.31 \\
\textbf{InP} & 3.16 & \color{red} $2\times 10^{-4}$ & 4.81 & 0.31 & \color{ForestGreen} 0.37 \\
\textbf{Si$_3$N$_4$} & 1.99 & \color{ForestGreen} $2.5\times 10^{-5}$ & 4.65 & 0.7 & \color{red} 0.02 \\
\textbf{LiNbO$_3$} & 2.21 & \color{ForestGreen} $3.2\times 10^{-5}$ & 5.32 & 0.63 & \color{red} $7\times 10^{-3}$ \\
\textbf{AlN} & 2.19 & \color{ForestGreen} $3\times 10^{-5}$ & 3.23 & 0.6 & \color{ForestGreen} 1.47 \\
\bottomrule
\end{tabular}
\caption{Material properties used to calculate the thermal noise limits in Fig.~3 of the main text. Aluminum nitride is the only material listed with a favorable thermo-optic coefficient \textit{and} thermal diffusivity.}
\label{table:params}
\end{table}

\section{Effects of TRN on All-Optical Qubits}

\subsection{Derivation of Qubit Coupling Strengths}

\newcommand\note[1]{\textcolor{red}{[#1]}}
\newcommand\dV{{\rm d}^3\vec{r}}
\renewcommand{\d}{{\rm d}}

This section derives the figures of merit for qubit operation in nonlinear optical cavities.  For more information, see Refs.~\cite{Mabuchi2012,Krastanov2020}.  The procedure is to first derive the classical equations of motion for fields in nonlinear oscillators and then to quantize them, deriving the Hamiltonian and the single-photon coupling strength.  In classical cavity electrodynamics, a cavity field can be expressed as a sum of resonant modes:
\begin{equation}
    E(\vec{x}, t) = \sum_\omega C_\omega \bigl(A_\omega(t) E_\omega(\vec{x}) e^{-i\omega t} + \mbox{c.c.}\bigr),\ \ \ \ 
	C_\omega = \sqrt{\hbar\omega/2\epsilon_0}
\end{equation}
The modes $E_\omega$ satisfy the Helmholtz equation $\nabla\times (\nabla\times E_\omega) = (n^2\omega^2/c^2) E_\omega$.  This is a generalized eigenvalue equation and the resulting solutions can be orthogonalized: $\int{n^2 E_{\omega'}^* E_\omega \dV} = c^2\int{B_{\omega'}^* B_\omega \dV} = \delta_{\omega'\omega}$.  With this normalization, we find that the electromagnetic energy density in the cavity is $U = \sum_\omega \hbar\omega |A_\omega|^2$.  Therefore, $A_\omega$ is the normalized field operator, where $|A_\omega|^2$ gives the number of photons in the mode $E_\omega$.

Nonlinear interactions can be treated as perturbations because the nonlinearity is weak on the order of a single optical cycle.  The Helmholtz equation acquires a nonlinear polarization $P = \epsilon_0 (\chi^{(2)}:E^2 + \chi^{(3)}:E^3 + \ldots)$, which can be integrated to give perturbations to the equations of motion for $A_\omega$ \cite{BoydBook}:
\begin{equation}
    \nabla\times (\nabla\times E) + \frac{n^2}{c^2} \frac{\partial^2 E}{\partial t^2} = -\frac{1}{c^2} \frac{\partial^2(P/\epsilon_0)}{\partial t^2}
    \ \ \ \Rightarrow\ \ \ 
    \frac{\d A_\omega}{\d t} = \frac{i\omega}{2 C_\omega} \int{E_\omega(\vec{x})^* \Bigl[\frac{P(\vec{x}, t)}{\epsilon_0}\Bigr]_\omega e^{i\omega t}\dV}
\end{equation}

\subsubsection{Kerr ($\chi^{(3)}$) Interaction}

In the $\chi^{(3)}$ case, we have a cavity with a single resonant mode $E_\omega$.  The polarization term due to the Kerr interaction is $P = \epsilon_0 \chi^{(3)}:(C_\omega A_\omega E_\omega e^{-i\omega t}  + \mbox{c.c.})^3$.  This gives rise to the equation of motion $\dot{A}_\omega = -i\chi |A_\omega|^2 A_\omega$, where:
\begin{equation}
    \chi = -\frac{3\hbar\omega^2 \chi^{(3)}}{4n^4\epsilon_0} \frac{1}{V_{\rm Kerr}},\ \ \ \ 
	V_{\rm Kerr} \equiv \frac{\bigl(\int{n^2 |E_{\omega}|^2\dV}\bigr)^2}{\int_*{n^4 |E_{\omega}|^4 \dV}}
\end{equation}
Quantizing the field to satisfy the commutation relations $[\hat{A}_\omega, \hat{A}_\omega^\dagger] = 1$ this equation of motion can be generated from the Hamiltonian:
\begin{equation}
    H_{\rm Kerr} = \frac{1}{2}\chi \hat{A}_\omega^\dagger \hat{A}_\omega^\dagger \hat{A}_\omega \hat{A}_\omega
\end{equation}
As an open quantum system, the field interacts with a bath through Lindblad dissipation terms, in this case $L = \sqrt{2\Gamma} A_\omega$, where $\Gamma = \omega/2Q$.  The figure of merit for strong coupling is:
\begin{equation}
    {\rm FOM}_{\chi^{(3)}} = \frac{\chi}{2\Gamma} = \underbrace{\frac{3\pi \hbar c}{2n\epsilon_0} \frac{\chi^{(3)}}{\lambda^4}}_{K_\chi} \frac{Q}{\tilde{V}_{\rm Kerr}}
    \label{eqn:chi3constant}
\end{equation}

\subsubsection{Second-order ($\chi^{(2)}$) Interaction}

In this case, we have two fields at frequencies $(\omega, 2\omega)$.  The polarization term is: $P = \epsilon_0 \chi^{(2)}:(C_\omega A_\omega E_\omega e^{-i\omega t} + C_{2\omega} A_{2\omega} E_{2\omega} e^{-2i\omega t} + \mbox{c.c.})^2$.  This gives rise to the following equations:
\begin{equation}
    \dot{A}_{2\omega} = -\frac{1}{2}\epsilon A_\omega^2,\ \ \ 
	\dot{A}_\omega = \epsilon A_{2\omega} A_\omega^* \label{eq:shgeom}
\end{equation}
where
\begin{equation}
    \epsilon = \frac{\omega\sqrt{\hbar\omega/\epsilon_0}}{n^3 V_{\rm shg}^{1/2}} \chi^{(2)},\ \ \ \ 
	V_{\rm shg} = \frac{\bigl(\int{n^2 |E_{2\omega}|^2\dV}\bigr) \bigl(\int{n^2 |E_{\omega}|^2\dV}\bigr)^2}{\bigl|\int_*{n^3 E_{2\omega}^* E_\omega E_\omega \dV}\bigr|^2}
\end{equation}
in the case that $\vec{E}$ and $\vec{P}$ are aligned (otherwise $\epsilon$ is reduced by a geometric factor).  The integral $\int{(\ldots)\dV}$ is taken over all space, while $\int_*{(\ldots)\dV}$ is restricted to the nonlinear material.

As before, we can quantize the fields $\hat{A}_\omega$, $\hat{A}_{2\omega}$ and derive a Hamiltonian corresponding to Eqs.~(\ref{eq:shgeom}).  As an open quantum system, there will also be Lindblad dissipation terms $\Gamma_1 = \omega/2Q_1$, $\Gamma_2 = 2\omega/2Q_2$:
\begin{equation}
    H = -i\epsilon \bigl(\hat{A}_{2\omega}^\dagger \hat{A}_\omega\hat{A}_\omega -  \hat{A}_\omega^\dagger \hat{A}_\omega^\dagger \hat{A}_{2\omega}\bigr),\ \ \ 
    L_1 = \sqrt{2\Gamma_1}\,A_\omega,\ \ \ 
    L_2 = \sqrt{2\Gamma_2}\,A_{2\omega}
\end{equation}
The figure of merit for strong coupling again is expressed as a ratio of the coupling rate $\epsilon$ to the loss rate.  Here there are two loss channels, and a common approach is to take the geometric mean of the two (a choice motivated by the limit $Q_2 \ll Q_1$, in which the $\chi^{(2)}$ interaction can be adiabatically eliminated to a $\chi^{(3)}$ one with $\chi/\Gamma \propto \epsilon^2/\Gamma_1\Gamma_2$).  Thus we set the figure of merit to be:
\begin{equation}
    {\rm FOM}_{\chi^{(2)}} = \frac{\epsilon}{2 \bar{\Gamma}} = \frac{\epsilon}{2 \sqrt{\Gamma_1\Gamma_2}} = \underbrace{\sqrt{\frac{\pi\hbar c}{n^3\epsilon_0}} \frac{\chi^{(2)}}{\lambda^2}}_{K_\epsilon} \frac{\sqrt{Q_1 Q_2}}{\tilde{V}_{\rm shg}^{1/2}}
    \label{eqn:chi2constant}
\end{equation}
In the main text, we assume $Q_1=Q_2=Q$ such that ${\rm FOM}_{\chi^{(2)}} \propto Q/V_\text{shg}^{1/2}$.

\subsection{Parameters}

The parameters used to generate Fig.~4 of the main text are included in Table~\ref{table:qubitParams}.

\renewcommand{\arraystretch}{1.2}
\begin{table}[h]
\footnotesize
\centering
\begin{tabular}{@{} L{2in}C{0.5in}C{1.5in}C{2in} @{}} \toprule
\textbf{Parameter} & \textbf{Symbol} & \textbf{Value} & \textbf{Source} \\
\midrule
$\chi^{(3)}$ Nonlinear Index & $n_2$ & $1.2\times 10^{-13}$ cm$^2$/W & \cite[Sec.~11]{HamerlyThesis} \\
$\chi^{(3)}$ FOM Constant & $K_\chi$ & $8.7\times 10^{-11}$ & Calculated (Eqn.~\ref{eqn:chi3constant}) \\
$\chi^{(2)}$ (DC $E$-Field Induced) & $\chi^{(2)}$ & 40 pm/V & \cite{Timurdogan2017} \\
$\chi^{(2)}$ FOM Constant & $K_\epsilon$ & $1.3\times 10^{-7}$ & Calculated (Eqn.~\ref{eqn:chi2constant}) \\
Thermo-optic Coefficient & $\alpha_\text{TO}^\text{Si}$ & $1.8\times 10^{-4}$ K$^{-1}$ & \cite{Komma2012} \\ 
Thermal Diffusivity & $D^\text{Si}$ & 0.8 cm$^2$/s & \cite{Cuffe2015} \\
\bottomrule
\end{tabular}
\caption{Silicon material properties assumed to calculate the qubit figures of merit at $\lambda_0=2.3~\upmu$m and $T=300$K.}
\label{table:qubitParams}
\end{table}

\bibliography{ThermorefractiveNoise}